\documentclass[final,authoryear,3p,times]{elsarticle}

\usepackage{graphicx,color}
\usepackage{enumerate}
\usepackage[colorlinks]{hyperref}
\usepackage[table,xcdraw]{xcolor}
\usepackage{amsmath,amssymb,amsthm,bm,amsbsy}
\usepackage{multicol}
\usepackage{ulem}
\usepackage{afterpage}
\usepackage{calc}
\usepackage{ifthen}
\usepackage{algorithm}
\usepackage{algpseudocode}
\usepackage{caption}
\usepackage{subcaption}
\usepackage{relsize}
\usepackage{graphicx}
\usepackage{xr}
\algtext*{Indent}
\algtext*{EndIndent}

\newcommand{\boldface}[1]{\boldsymbol{#1}}  
\newcommand{\bfa}{\boldface{a}}
\newcommand{\bfb}{\boldface{b}}

\newcommand{\bft}{\boldface{t}}
\newcommand{\bfu}{\boldface{u}}
\newcommand{\bfv}{\boldface{v}}
\newcommand{\bfw}{\boldface{w}}

\newcommand{\bfy}{\boldface{y}}
\newcommand{\bfz}{\boldface{z}}
\newcommand{\bfA}{\boldface{A}}

\newcommand{\bfC}{\boldface{C}}

\newcommand{\bfF}{\boldface{F}}

\newcommand{\bfI}{\boldface{I}}

\newcommand{\bfP}{\boldface{P}}
\newcommand{\bfQ}{\boldface{Q}}
\newcommand{\bfR}{\boldface{R}}

\newcommand{\bfX}{\boldface{X}}

\newcommand{\bfepsilon}{\boldsymbol{\varepsilon}}

\newcommand{\bftheta}{\boldsymbol{\theta}}

\newcommand{\bfmu}{\boldsymbol{\mu}}

\newcommand{\bfSigma}{\boldsymbol{\Sigma}}

\newcommand{\calB}{\mathcal{B}}
\newcommand{\calC}{\mathcal{C}}
\newcommand{\calD}{\mathcal{D}}

\newcommand{\calG}{\mathcal{G}}

\newcommand{\calI}{\mathcal{I}}

\newcommand{\calN}{\mathcal{N}}

\newcommand{\calP}{\mathcal{P}}

\newcommand{\calU}{\mathcal{U}}

\newcommand{\calX}{\mathcal{X}}

\newcommand{\Rset}{\mathbb{R}}

\newlength{\boxwidth}
\def\dd{\;\!\mathrm{d}}

\def\btheorem{\begin{theorem}}
\def\etheorem{\end{theorem}}
\def\blemma{\begin{lemma}}
\def\elemma{\end{lemma}}
\def\bproposition{\begin{proposition}}
\def\eproposition{\end{proposition}}
\def\bcorollary{\begin{corollary}}
\def\ecorollary{\end{corollary}}
\def\bdefinition{\begin{definition}}
\def\edefinition{\end{definition}}
\def\bexample{\begin{example}}
\def\eexample{\end{example}}
\def\bremark{\begin{remark}}
\def\eremark{\end{remark}}

\newcommand{\be}{\begin{equation}}
\newcommand{\ee}{\end{equation}}
\newcommand{\beq}{\begin{eqnarray}}
\newcommand{\eeq}{\end{eqnarray}}
\newcommand{\bem}{\begin{multline}}
\newcommand{\eem}{\end{multline}}
\newcommand{\ba}{\begin{align}}
\newcommand{\ea}{\end{align}}

\renewcommand{\figurename}{Figure }

\newcommand{\customCaption}[4]{Model discovery with the hidden #1 benchmark model \eqref{eq:#2} using #3 data #4 for different noise levels. (a) Marginal posterior distribution of the material parameters $\bftheta$ indicated via violin plots. The features are labeled according to the indices defined in Table~\ref{tab:features}. (b) Average activity of each feature in the posterior distribution. (c)-(h) Strain energy density along the different deformation paths in \eqref{eqn:strain_paths} for the (true) hidden model and the discovered models sampled from the MCMC chain; see Section~\ref{sec:results} for details.}

\newcommand{\customCaptionStatic}[2]{\customCaption{#1}{#2}{quasi-statics}{}}

\newcommand{\customCaptionStaticSupp}[2]{\customCaption{#1}{#2}{quasi-statics}{and suppressed true features}}

\newcommand{\customCaptionDynamic}[2]{\customCaption{#1}{#2}{dynamics}{}}

\newcommand{\RR}[1]{\textcolor{black}{#1}}

\journal{}

\begin{document}

\begin{frontmatter}

\title{{Bayesian-EUCLID: discovering hyperelastic material laws with uncertainties}}

\author[a]{Akshay Joshi}
\author[a]{Prakash Thakolkaran}
\author[a]{Yiwen Zheng}
\author[b]{Maxime Escande}
\author[b]{Moritz Flaschel}
\author[b]{Laura De Lorenzis}
\author[a]{Siddhant Kumar\corref{cor1}}

\cortext[cor1]{Email: Sid.Kumar@tudelft.nl}
\address[a]{Department of Materials Science and Engineering, Delft University of Technology, 2628 CD Delft, The Netherlands}
\address[b]{Department of Mechanical and Process Engineering, ETH Z\"{u}rich, 8092 Z\"{u}rich, Switzerland}

\begin{abstract}
{Within the scope of our recent approach for Efficient Unsupervised Constitutive Law Identification and Discovery (EUCLID),} we propose an unsupervised Bayesian learning framework for discovery of parsimonious and interpretable constitutive laws with quantifiable uncertainties. {As in deterministic EUCLID, we do not resort to stress data, but only to realistically measurable full-field displacement and global reaction force data;} as opposed to calibration of an a priori assumed model, we start with a constitutive model ansatz based on a large catalog of candidate functional features{; we leverage domain knowledge} by including features based on {existing, both physics-based and phenomenological}, constitutive models. {In the new Bayesian-EUCLID approach}, we use a hierarchical Bayesian model with sparsity-promoting priors and Monte Carlo sampling to efficiently solve the parsimonious model selection task and discover physically consistent constitutive equations in the form of multivariate multi-modal probabilistic distributions. We demonstrate and validate the ability to accurately and efficiently recover isotropic and anisotropic  hyperelastic models like the Neo-Hookean, Isihara, Gent-Thomas, Arruda-Boyce, Ogden, and Holzapfel models in both elastostatics and elastodynamics. The discovered constitutive models are reliable under both epistemic uncertainties {-- i.e. uncertainties on the true features of the constitutive catalog -- and aleatoric uncertainties} -- {which arise from the noise in the displacement field data, and are automatically estimated by the hierarchical Bayesian model.}

\end{abstract}


\end{frontmatter}

\section{Introduction}
\label{sec:introduction}

Owing to the empirical nature of constitutive/material models, data-driven methods (and more recently, their hybridization with integration of physics knowledge) are rapidly pushing the boundaries where classical modeling methods have fallen short. In general, the state-of-the-art approaches either $surrogate$ or completely $bypass$ material models \citep{flaschel_unsupervised_2021}. \textit{Surrogating} material models involve learning a mapping between strains and stresses using techniques ranging from piece-wise interpolation \citep{crespo_wypiwyg_2017,sussman_model_2009} to Gaussian process regression \citep{rocha_onthefly_2021,fuhg_local_2022} and artificial neural networks \citep{ghaboussi_knowledgebased_1991,fernandez_anisotropic_2021,klein_polyconvex_2022,vlassis_sobolev_2021,kumar_inverse-designed_2020,bastek_inverting_2022,zheng_data_2021,Mozaffar2019,Bonatti2021,Vlassis2020,kumar_ml_mech_review}; the latter are particularly attractive  because of their ability  to efficiently and accurately learn from large  and high-dimensional data. In contrast, the model-free data-driven approach \citep{kirchdoerfer_data-driven_2016,Ibaez2017,Kirchdoerfer_dynamics_2017,conti_data-driven_2018,nguyen_data-driven_2018,eggersmann_model-free_2019,carrara_data-driven_2020,karapiperis_data-driven_2021} \textit{bypasses} constitutive relations by mapping a material point's deformation to an appropriate stress state (subject to compatibility constraints) directly from a large dataset of stress-strain pairs.  Recent approaches  \citep{ibanez_hybrid_2019,gonzalez_learning_2019} also proposed adding data-driven corrections to the existing constitutive models. A common challenge in both approaches is the lack of interpretability, as they either substitute the constitutive model with a black-box or bypass it, thereby precluding any physical understanding of the {material} behavior. Interpretability or a physical intuition of the material model is critical to identifying where the model fails and can aid better design and use of materials such as composites \citep{Liu2021}. Recent works have began addressing this issue \citep{Asad2022,Liang2022}.

In addition to lacking interpretability, both approaches of $surrogating$ and $bypassing$ material models in data-driven constitutive modeling are rooted in a supervised learning setting and require {a large number} of strain-stress pairs. This presents a two-fold limitation, especially in the context of experimental data. \textit{(i)}~Probing the entire high-dimensional strain-stress space with viable mechanical tests, e.g., uni-/bi-axial tensile or bending tests is infeasible. \textit{(ii)}~Stress \textit{tensors} cannot be measured experimentally (force measurements are only boundary-averaged projections of stress tensors), which is prohibitive to learning full tensorial constitutive models. {While} exhaustive and tensorial stress-strain data can be artificially generated using multiscale simulations \citep{yuan_toward_2008,wang_multiscale_2018}, {the computational cost associated to the generation of large datasets} for complex material systems {is currently still prohibitive}. It is thus important to be able to learn the material behavior directly from data that {are} realistically available through mechanical testing.

Full-field displacement data, {e.g. obtained from digital image correlation (DIC), combined with applied force data from load cells,} have been the mainstay of {modern} material model calibration \RR{\citep{DIC_Hild}}. This is traditionally performed via the finite element model updating (FEMU) \citep{Marwala2010} or virtual fields method (VFM) \citep{Pierron_2012}. While FEMU requires iterative optimization of model parameters until simulated and measured displacement fields match, VFM solves for {the unknown} material parameters directly by satisfying {the} weak form of momentum balance with measured full-field displacement data. In the model-free data-driven realm, recent works \citep{Leygue2018,dalemat_measuring_2019,Cameron2021} have demonstrated estimation of stress fields from full-field displacement data. In the model-based realm, physics-informed neural networks (PINNs) and their variations have shown promising results -- by first learning the forward solution, i.e., displacement fields, to the mechanical boundary boundary value problem as a function of the material parameters and then estimating the unknown parameters via gradient-based optimization \citep{huang_learning_2020,tartakovsky_learning_2018,haghighat_deep_2020,Chen2021}. However, all the aforementioned methods (including FEMU, VFM, and PINNs) are limited to \textit{a priori} assumed constitutive models (e.g., with known deformation modes) with only a few unknown parameters. Such a restricted model ansatz is applicable only to specific materials and stress states, and cannot be generalized beyond the calibration/test data.

In light of the above challenges, previous work \citep{flaschel_unsupervised_2021} by some of the authors and independently by \citet{Wang2021} presented a method to neither \textit{surrogate} nor \textit{bypass}, but rather automatically \textit{discover} interpretable constitutive models of hyperelastic materials from full-field displacement data and global reaction forces. {More recently, the approach was extended to plasticity \citep{flaschel_plasticity} and denoted as Efficient Unsupervised Constitutive Law Identification and Discovery (EUCLID).} {EUCLID} is based on sparse regression -- originally developed for automated discovery of physical laws in nonlinear dynamic systems \citep{schmidt_distilling_2009,brunton_discovering_2016}.  It formulates the constitutive model as parameterized sum of a large catalog of candidate functional features chosen based on physical knowledge and decades of prior phenomenological modeling experience. The problem is posed as a parsimonious model selection task, thereby minimizing the bias of pre-assumed models with limited expressability (in, e.g., FEMU, VFM, PINNs). {Importantly, EUCLID is \textit{unsupervised}, i.e., it does not use stress labels but only realistically measurable full-field displacement data and global reaction forces.  The lack of stress labels for model discovery is bypassed  by leveraging the physical constraint that the displacement fields must satisfy linear momentum conservation}.

\begin{figure}[t]
	\centering
	\includegraphics[width=1.0\textwidth]{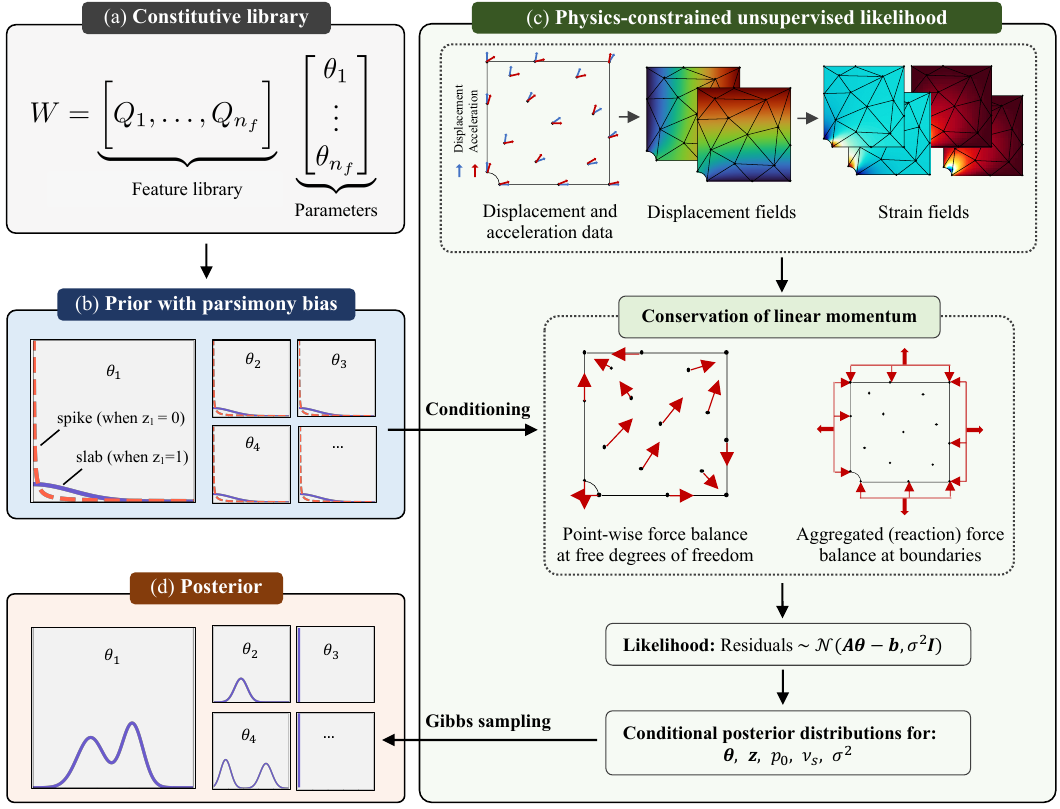}
	\caption{Schematic of Bayesian-EUCLID for unsupervised discovery of hyperelastic constitutive laws with uncertainties. (a) A large library of constitutive features (inspired from physics-based and phenomenological models) is chosen for the hyperelastic strain energy density. (b) Sparsity-promoting spike-slab priors are placed on the material parameters to induce bias towards parsimonious constitutive models. (c) The likelihood of the observed data (consisting of displacement data -- including accelerations, if available -- and reaction forces) is unsupervised and based on satisfying the physical constraint of linear momentum balance. Conditioned on the prior, the force residuals are modeled using a Gaussian likelihood. (d) Using Bayes' rule and Gibbs sampling, physically admissible, interpretable, and parsimonious constitutive models are discovered in the form of multi-modal posterior distributions with quantifiable epistemic and aleatoric uncertainties.}\label{fig:schematic}
\end{figure}

{In this paper, we bring EUCLID to the next level by developing a Bayesian paradigm for automated discovery of hyperelastic constitutive laws \textit{with quantifiable uncertainties}. The approach remains unsupervised, i.e. it requires only full-field displacement and global reaction force data. The physical constraint of linear momentum conservation leads to a residual} which defines the likelihood. We use a hierarchical Bayesian model with sparsity-promoting spike-slab priors \citep{nayek_spike-and-slab_2021} and Monte Carlo sampling to efficiently solve the parsimonious model selection task and discover constitutive models in the form of multivariate multi-modal posterior probability distributions. {Figure~\ref{fig:schematic} summarizes the schematic of the method.} The discovered constitutive models are reliable under both epistemic uncertainties, i.e. uncertainties on the true features of the constitutive model catalog, and aleatoric uncertainties, i.e. those due to noise in the displacement field data \citep{hullermeier_aleatoric_2021} which is automatically estimated by the hierarchical Bayesian model.  In addition to the multi-modal solutions and uncertainty quantification, Bayesian-EUCLID provides an increase in computational speed and data efficiency (in terms of the amount of data required) of two orders of magnitude in comparison to deterministic EUCLID  \citep{flaschel_unsupervised_2021}. The current work also extends the previous method to \textit{(i)} anisotropic hyperelasticity as well as \textit{(ii)} elastodynamics; the latter also leveraging inertial data for model discovery. \RR{For the scope of the current work, we assume the anisotropy to be in-plane with a priori known directions. This assumption is necessary as the 2-D nature of the DIC displacement data does not allow for accurate characterization of out-of-plane anisotropy.}

In the following, we present the {new} unsupervised Bayesian learning framework in Section~\ref{sec:method}.  Section~\ref{sec:benchmarks} presents {several} benchmarks based on well-known physical and phenomenological constitutive models,  before Section~\ref{sec:conclusion} concludes the study. {A pseudo-code is included in \ref{sec:AppPSEUD} {to enable a}  better understanding of the numerical implementation.}

\section{{Bayesian-EUCLID:} Unsupervised discovery of hyperelastic constitutive laws with uncertainties}
\label{sec:method}

\subsection{Available data}
\label{sec:available_data}

Consider a two-dimensional domain with reference configuration $\Omega\in\Rset^2$ (with boundary $\partial\Omega$) subjected to displacement-controlled {loading}\footnote{In mechanical testing under load-controlled {conditions, reaction forces are replaced by the known applied loads.}} on $\partial \Omega_u\subseteq\partial\Omega$. The domain {geometry} and {the} boundary conditions are designed to generate diverse and heterogeneous strain states upon loading (see Section~\ref{sec:data_generation}). Under quasi-static loading, the observed data consist of snapshots of displacement measurements $\{\bfu^{a,t} \in \Rset^2: a=1,\dots,n_n; t=1,\dots,n_t\}$ (obtained via, e.g., DIC)  at $n_n$ reference points $\calX = \{\bfX^a \in \Omega :a=1,\dots,n_n\}$ and $n_t$ time steps. We also consider experiments with dynamic loading, wherein the point-wise acceleration data may be inferred from the displacement history via the second-order central difference as
\be\label{eq:acc}
\ddot{\bfu}^{a,t} = \frac{\bfu^{a,t+\delta t}-2\bfu^{a,t}+\bfu^{a,t-\delta t}}{(\delta t)^2} + \text{O}\left((\delta t)^2\right).
\ee
Here, $\delta t \ll 1$ is the temporal resolution of the DIC. Note that the temporal resolution of the data snapshots (i.e., $t=1,\dots,n_t$) is assumed to be significantly coarser than the DIC frame rate. The observed data also consist of  reaction force measurements $\{R^{\beta,t}:\beta=1,\dots,n_\beta;t=1,\dots,n_t\}$ from $n_\beta$ load cells at some boundary segments. Note that a reaction force is a boundary-aggregate measurement; the point-wise traction distribution on a displacement-constrained boundary is practically immeasurable and therefore, not considered to be known here. In addition, the number of reaction force measurements is far smaller than the number of points for which the displacement is tracked, i.e., $n_\beta \ll n_n$ (which further adds to the challenge of unavailable stress labels).  Assuming the domain $\Omega$ is homogeneous and hyperelastic, the inverse problem is to determine a constitutive model that best fits the observed data. It is noteworthy that the aforementioned problem involves inferring the constitutive model of the material only from realistically available data, unlike supervised learning approaches which use experimentally inaccessible stress tensors to learn/bypass the material model. For the scope of this work, we consider a two-dimensional plane strain setting, while noting that the proposed method may be extended to three-dimensions with appropriate techniques (e.g., Digital Volume Correlation). Specific details pertaining to data generation are presented in Section \ref{sec:data_generation}.

\subsection{Field approximations from point-wise data}

We mesh the reference domain with linear triangular finite elements using the point set $\calX$. The point-wise   displacements are then interpolated as
\be\label{eq:displacement_approximation}
\bfu^t(\bfX) = \sum_{a=1}^{n_n} N^a(\bfX) \bfu^{a,t},
\ee
where $N^a:\Omega\rightarrow \Rset$ denotes the shape function associated with the $a^\text{th}$ node. The deformation gradient field is then approximated as 
\begin{equation}
    \bfF^t (\bfX) = \bfI + \sum_{a=1}^{n_n} {\bfu}^{a,t} \otimes \nabla N^a(\bfX),
\end{equation}
where $\nabla(\cdot) = \partial (\cdot)/\partial \bfX$ denotes gradient in reference coordinates. Note that $\nabla N^a(\bfX)$ is constant within each element due to the use of linear shape functions. For the sake of brevity, we drop the superscript $(\cdot)^t$ in the subsequent discussion while tacitly implying that the formulation applies to snapshots at every time step.

\subsection{Constitutive {model} library}
\label{sec:energy_library}

The constitutive response of hyperelastic materials derives from a  strain energy density $W(\bfF)$ such that the first Piola–Kirchhoff stress tensor is given by $\bfP(\bfF) = \partial W/\partial \bfF$.  The strain energy density must satisfy certain  constraints for physical admissibility {and to guarantee the existence of minimizers} \citep{Morrey1952,hartmann_polyconvexity_2003,Schroder2010}. The coercivity condition  requires $W(\bfF)$ to become infinite when the material undergoes infinitely large deformation or compression to close to zero volume, i.e.,
\be
    W(\bfF) \geq c_1 \bigg(\|\bfF\|^p+\|\text{Cof}(\bfF)\|^q+ \det\left(\bfF\right)^r\bigg) +c_0 \qquad \forall\qquad  \bfF \in \text{GL}_{+}(3) \ 
\ee
with constants $c_0, c_1>0, \ p\geq2, \ q\geq p/(p-1), \ r>1$, and
\be
    W(\bfF) \to +\infty \qquad \text{for}\qquad \ \det(\bfF)\to 0^+.
\ee
Here, $\text{Cof}(\cdot)$ and $\det(\cdot)$ denote the cofactor and determinant, respectively{, whereas $\text{GL}_{+}(3)$ indicates the set of invertible second-order tensors with positive determinant}. {A further requirement on $W(\bfF)$ is} quasiconvexity \citep{Morrey1952,Schroder2010} w.r.t.~$\bfF$, i.e.,
\be
\int_\calB W(\bar\bfF + \nabla \bfw)\dd V \geq  W(\bar \bfF)\int_\calB \dd V \qquad \forall \qquad \calB\subset\Rset^3, \ \ \bar\bfF\in\text{GL}_{+}(3), \ \ \bfw\in C^\infty_0(\calB) \ \  \text{(i.e., $\bfw=0$ on $\partial\calB$)}.
\ee
However, due to the analytical/computational intractability of the quasiconvexity condition \citep{Kumar2019nme,Voss2022}, this is usually replaced  by that of polyconvexity (the latter implies the former) which automatically guarantees material stability (see \cite{ball_convexity_1976,Schroder2010} for a detailed discussion on polyconvexity). $W(\bfF)$ is polyconvex if and only if there exists a convex function $\calP:\Rset^{3\times3}\times\Rset^{3\times3}\times\Rset\rightarrow\Rset$ (in general non-unique) such that
\be
W(\bfF) = \calP(\bfF,\text{Cof}\bfF,\det \bfF).
\ee
In addition, the stress must vanish at the reference configuration, i.e., $\bfP(\bfF=\bfI)=\boldsymbol{0}$. The constitutive model must also satisfy objectivity, i.e., $W(\bfF)=W(\bfR\bfF) \ \forall \bfR\in \text{SO}(3)$.  Reformulating the strain energy density as $W(\bfC)$ (where $\bfC=\bfF^T\bfF$ denotes the right Cauchy–Green deformation tensor) satisfies the objectivity condition identically but may violate the polyconvexity condition as the off-diagonal terms of $\bfC$ are non-convex in $\bfF$ \citep{klein_polyconvex_2022}. To this end, it is common practice to model the strain energy density as a function of  invariants  of $\bfC$ that are convex in $\bfF$. Following arguments of \citet{spencer_constitutive_1984} on material symmetry, the energy density for anisotropic hyperelastic materials with fiber directions $\{\bfa_1,\bfa_2,\dots\}$ is expressed in terms of both isotropic and anisotropic strain invariants (the latter corresponding to each fiber direction) as
\be
W(\bfC,\{\bfa_1,\bfa_2,\dots\}) \equiv W\bigg(I_1, I_2, I_3, J_4, J_5, J_6, J_7,\dots\bigg),
\ee
where
\be
\begin{aligned}
\text{Isotropic invariants:}& \quad I_1(\bfC) = \text{tr}(\bfC), \quad I_2(\bfC) = \frac{1}{2}\left[\text{tr}(\bfC)^2-tr(\bfC^2)\right], \quad I_3(\bfC) = \text{det}(\bfC)\\
\text{Anisotropic invariants (fiber 1):}& \quad J_4(\bfC,\bfa_1) = \bfa_1\cdot\bfC\cdot\bfa_1, \quad J_5(\bfC,\bfa_1) = \bfa_1\cdot\bfC^2\cdot\bfa_1 \\
\text{Anisotropic invariants (fiber 2):}& \quad J_6(\bfC,\bfa_1) = \bfa_2\cdot\bfC\cdot\bfa_2, \quad J_7(\bfC,\bfa_2) = \bfa_2\cdot\bfC^2\cdot\bfa_2 \\
\vdots \qquad \qquad  & \qquad \qquad \vdots
\end{aligned}.
\ee
Without loss of generality,  the remainder of this work concerns two fiber directions\footnote{{An anisotropic hyperelastic material with arbitrary number of symmetrical distributed fiber directions can be equivalently modeled with two mutually perpendicular fiber families of appropriate stiffnesses \citep{DONG2021104377}.}}; however, the framework can be extended to more {fiber families}. Additionally, only $J_4$ and $J_6$ terms are considered in modelling the energy density of the anisotropic material \citep{holzapfel_new_2000}. {This is because $J_4$ and $J_6$, being square of the stretches projected in the direction of the fibers, offer a more intuitive basis to model the energy density function.}

The objective of the inverse problem is to identify an appropriate strain energy density {function} from the observed data (displacement field and reaction force measurements). We consider the following ansatz:
\be
W(I_1,I_2,I_3,J_4,J_6) = \bfQ (I_1,I_2,I_3,J_4,J_6) \cdot \bftheta,
\label{eqn:WQT}
\ee
where $\bfQ:\Rset^5\rightarrow \Rset^{n_f}$ denotes a large library of $n_f$ isotropic and anisotropic features. Here, $\bftheta\in \Rset_{+}^{{n_f}}$ is a vector of  positive scalar coefficients representing unknown material parameters to be estimated. If the features are coercive and polyconvex, the positivity constraint on the entries of $\bftheta$  automatically ensures that $W(\bfF)$ is also coercive and polyconvex. The feature library can contain essentially any mathematical function of the input strain invariants; we derive inspiration from existing phenomenological and physics-based models (see \citet{Marckmann2006,dal_performance_2021} for detailed reviews). For the scope of this work, we consider compressible hyperelasticity with the following feature library:
\be
\begin{aligned}
    \boldsymbol{Q}(I_1,I_2,I_3,J_4,J_6) \ =\  & \underbrace{\bigl[(\Tilde{I}_1-3)^i(\Tilde{I}_2-3)^{j-i}: \ i \in \{{0},\dots,j \} \ \text{and} \ j \in \{1,\dots,N_\text{MR} \} \bigr]^T}_{\text{Generalized {Mooney-Rivlin} features}} \\
    & \oplus \ \underbrace{\left[(J-1)^{2i}:i \in \{1,\dots,N_\text{vol} \}\right]^T}_\text{Volumetric energy features} \ \oplus \ \underbrace{[\log(\Tilde{I}_2/3)]}_\text{logarithmic feature} \oplus \underbrace{\text{AB}(\Tilde{I}_1)}_\text{Arruda-Boyce feature} \\
    & \oplus \ \underbrace{\left[\text{OG}_i(\Tilde{I}_1,J):i \in \{1,\dots,N_\text{Ogden}\}\right]}_\text{Ogden features} \\
    & \oplus \ \underbrace{\left[(\Tilde{J}_4-1)^i : i \in \{{2},\dots,N_\text{aniso}\}\right]
    \ \oplus\ \left[(\Tilde{J}_6-1)^i : i \in \{{2},\dots,N_\text{aniso}\}\right]}_\text{Anisotropy features},
\end{aligned}\label{eq:library}
\ee
where $\Tilde{I}_1 = J^{-2/3}I_1,\  \Tilde{I}_2 = J^{-4/3}I_2,\  J = \det(\bfF) = I_3^{1/2}, \ \Tilde{J}_4 = J^{-2/3}J_4, \text{ and } \Tilde{J}_6 = J^{-2/3}J_6$ are the invariants {of} the unimodular\footnote{\RR{The unimodular anisotropic invariants used here may give rise to non-physical effects under certain deformations \citep{HELFENSTEIN20102056}. However, the Bayesian-EUCLID framework is agnostic to the choice of features.}} Cauchy-Green deformation tensor ${\Tilde{\bfC}}= J^{-2/3}\bfC$. The chosen feature library automatically satisfies the conditions of objectivity, stress-free reference configuration, coercivity, and polyconvexity (with only one exception discussed below). In the following, we explain the different types of features in the library.
\begin{itemize}
    \item The generalized Mooney-Rivlin features consist of monomials based on the strain invariants and are a superset of several well-known models such as Neo-Hookean \citep{Treloar1943}, Isihara \citep{isihara_statistical_1951}, Haines-Wilson \citep{haines_strain-energy_1979}, Biderman \citep{biderman1958calculation}, and many more. Note that the terms containing $(\Tilde{I}_2-3)^{j-i}$ with $(j-i)\geq 1$ do not satisfy polyconvexity \citep{hartmann_polyconvexity_2003,Schroder2010}. However, they are commonly used in many material models \citep{isihara_statistical_1951,haines_strain-energy_1979,dal_performance_2021} and do not exhibit problems due to physical inadmissibility in practical deformation ranges. Thus, these functions are also included in the feature library.
    \item The logarithmic term is inspired by the classical Gent-Thomas model \citep{gent_forms_1958}. 
    \item The Arruda-Boyce \citep{arruda_three-dimensional_1993} feature, based on statistical mechanics of polymeric chains,  is given by
    \be\label{eq:feature_AB}
        \text{AB}(\Tilde{I}_1) = 10\sqrt{N_{{c}}}\left[\beta_{{c}} \lambda_{{c}} + \sqrt{N_{{c}}} \ \log\left(\frac{\beta_{{c}}}{\sinh(\beta_{{c}})}\right)\right] - c_\text{AB}  \quad \text{with} \quad \beta_{{c}} = \mathcal{L}^{-1}\left(\frac{\lambda_{{c}}}{\sqrt{N_{{c}}}}\right),
    \ee
    where $\lambda_{{c}} = \sqrt{\Tilde{I_1}/3}$ represents the stretch in the polymeric chains arranged along the diagonals of a cubic representative volume element; $N_{c}$ denotes the number of elements in the polymeric chain with $\sqrt{N_{c}}$ being a measure of maximum possible chain elongation. Without loss of generality, $N_{c}$ is chosen to be 28 for the scope of this work. $\mathcal{L}^{-1}$ denotes the inverse Langevin function which in this work is numerically approximated as \citep{bergstrom_large_1999}
    \be
        \mathcal{L}^{-1}(x) = \begin{cases}
        1.31 \tan(1.59 x) + 0.91 x \quad &\text{for} \quad |x|<0.841,\\ (\text{sgn}(x)-x)^{-1} \quad &\text{for} \quad 0.841 \leq |x|< 1,
        \end{cases}
    \ee
    where $\text{sgn}(\cdot)$ denotes the signum function.
    The factor `10' is pre-multiplied to ensure that the coefficient of the Arruda-Boyce feature has the same order of magnitude as the other features. The constant {$c_\text{AB} = 15.16$} is set such that the feature vanishes at  $\bfF=\bfI$, i.e., $\text{AB}|_{\bfF=\bfI} = 0$. The Arruda-Boyce model also encompasses its simpler approximation --  the Gent model \citep{Gent1996} (including the limited chain stretching behavior phenomena), hence we do not include the latter in the feature library.
    \item The Ogden features \citep{ogden_large_1972} are given by
    \be\label{eq:feature_OG}
        \text{OG}_i(\Tilde{I}_1,J) = \frac{2}{\alpha_i}\left( \Tilde{\lambda}_1^{\alpha_{i}}+\Tilde{\lambda}_2^{\alpha_{i}}+\Tilde{\lambda}_3^{\alpha_{i}}\right)  \qquad \text{with} \qquad \Tilde{\lambda}_k = J^{-1/3}\lambda_k, \ k=1,2,3,
    \ee
    where ${\lambda}_1,{\lambda}_2,{\lambda}_3$ denote the principal stretches and $\alpha_i$ denotes the exponent corresponding to the $i^{th}$ Ogden feature. In plane strain, the Ogden features can be reformulated purely in terms of $\Tilde{I}_1$ and $J$. Using the characteristic polynomial of $\bfC$ whose eigenvalues are the squares of the principal stretches, it can be shown {that}
    \be
    \Tilde{\lambda}_1, \Tilde{\lambda}_2 = \left(\frac{\left(\Tilde{I}_1 - J^{-\frac{2}{3}}\right)\pm \sqrt{\left(\Tilde{I}_1 - J^{-\frac{2}{3}}\right)^2 - 4J^{\frac{2}{3}}}}{2}\right)^{1/2} \qquad \text{and} \qquad \Tilde{\lambda}_3 = J^{-1/3}.
    \ee
    Given that the feature coefficients in $\bftheta$ are positive, $\alpha_i\geq1$ automatically ensures polyconvexity and coercivity \citep{ogden_large_1972,ciarlet,hartmann_polyconvexity_2003}. The choice of $\{\alpha_i,i=1,\dots,N_\text{Ogden}\}$ is part of feature engineering and can be decided both with or without consideration of the physics of the material. 
    \item The anisotropic features -- similar to isotropic generalized Mooney-Rivlin features -- are based on monomials of $J_4$ and $J_6$ \citep{Wang2021}. {Although polynomial terms up to fourth order are considered here for illustrative purposes, higher order terms can be used to obtain a more accurate surrogate for complicated anisotropic models like the Holzapfel model \citep{holzapfel_new_2000}.} For the scope of this work, we assume that the true fiber directions $\bfa_1$ and $\bfa_2$ are known a priori. \RR{For example, in the Holzapfel model benchmark in Section~\ref{sec:benchmarks}, $\bfa_1 = \left(\cos(30^{\circ}), \sin(30^{\circ}),0\right)^T$ and $\bfa_2 = \left(\cos(30^{\circ}), -\sin(30^{\circ}), 0\right)^T$ are known, i.e, the fibers are assumed to be oriented at $\pm30^{\circ}$.} The feature library does not include the linear $(J_4-1)$ and $(J_6-1)$ terms, as these energy features do not have zero derivatives for $\boldsymbol{F}=\boldsymbol{I}$, and thus induce non-zero stresses in the reference configuration.
    \item While the previous features account for the deviatoric deformations, the volumetric features -- given by monomials of $(J-1)$ -- {are used to account for }the energy due to volumetric deformations. 
\end{itemize} 

Compared to the previous work by \citet{flaschel_unsupervised_2021} and \citet{Wang2021}, the library includes a more diverse set of features beyond isotropic and 
polynomial features; the positivity constraint on $\bftheta$ ensures that all solutions found are automatically physically admissible. We note that the method is not limited to the feature library chosen here and can be extended to include any additional features based on prior knowledge about the material system.

\subsection{Model prior with parsimony bias}
\label{sec:prior}

{It is not desirable to have all the $n_f$ features active in the constitutive model  for two reasons.} \textit{(i)}~A model {that utilizes many} features is overly complex and, consequently, likely to overfit the observed data and show poor generalization (analogous to fitting a high-degree polynomial to very few data points). \textit{(ii)} The {degree of physical interpretability decreases} as the model becomes too complex (tending towards black-box models).  {To address these issues, the idea of parsimony is adopted as a regularization for model learning, and for inducing physical interpretability as bias \citep{brunton_discovering_2016,Kutz2022}.}

In this light, we pose the inverse problem of identifying the constitutive response as a model selection task -- the objective is {not only to} estimate $\bftheta$ but also the appropriate subset of features {leading to an optimal compromise between accuracy and parsimony}. A brute-force approach is not feasible as  the number of unique families of possible constitutive models is $(2^{n_f}-1)$, which increases exponentially with number of features $n_f$. {The problem} has been classically addressed via sparse regression \citep{brunton_discovering_2016} -- which is based on iterative minimization of an objective function in a non-probabilistic setting with lasso, ridge, or similar parsimony-inducing regularization. {The resulting discovered constitutive models are of deterministic nature} \citep{flaschel_unsupervised_2021,Wang2021}. 

\begin{figure}
	\centering
	\includegraphics[width=0.5\textwidth]{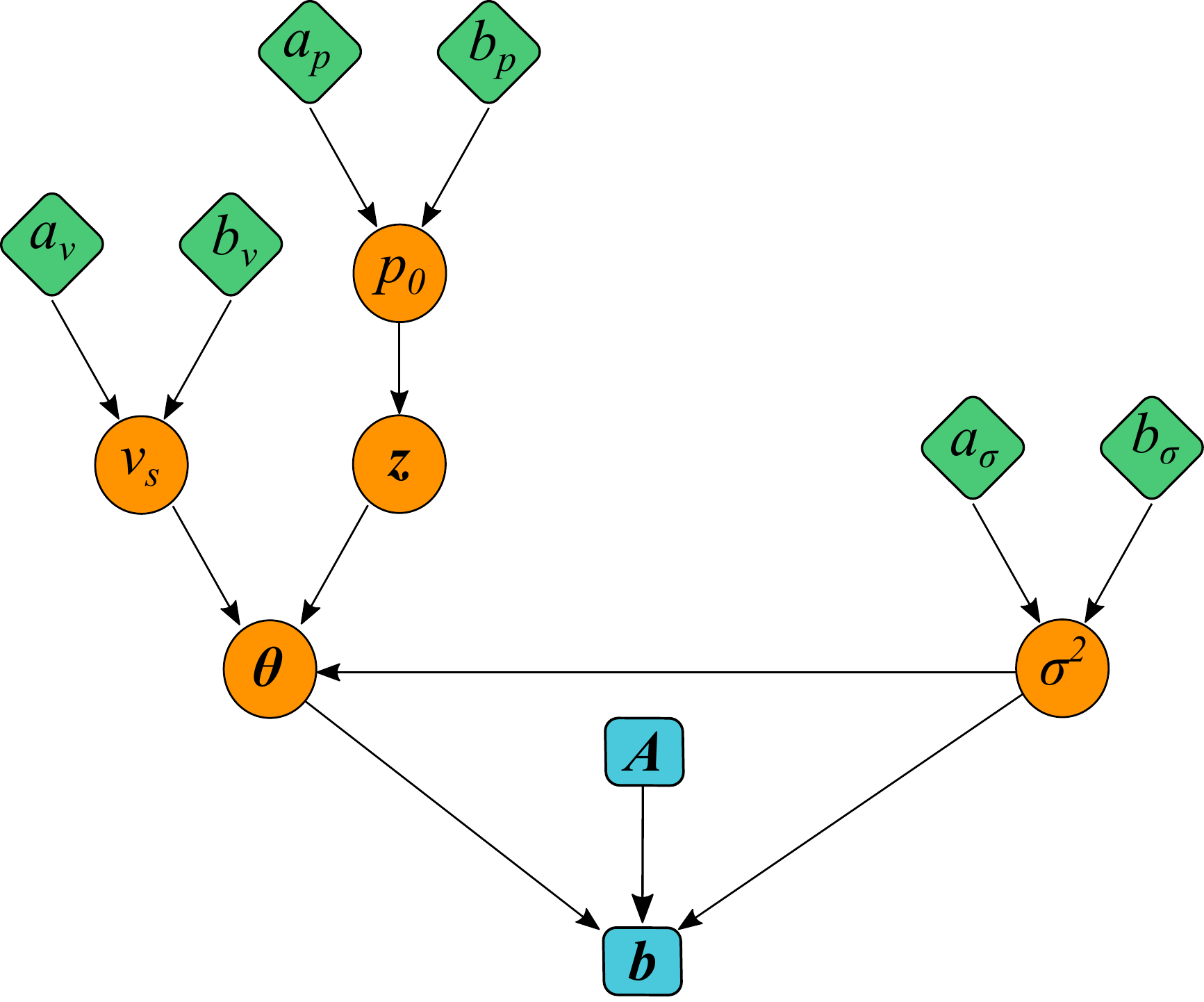}
	\caption{Schematic of the hierarchical spike-slab Bayesian model (adapted from \citet{nayek_spike-and-slab_2021}). The orange circles represent the random variables for which the prior distributions are prescribed. The green diamonds denote known hyperparameters. The observations (displacements, accelerations, and reaction forces) and conservation of linear momentum are embedded in $\bfA$ and $\bfb$ which are used to estimate the posterior probabilities for the unknown random variables.}\label{fig:DAG}
\end{figure}

{In this work, we replace deterministic sparse regression with a probabilistic framework able} to simultaneously perform model selection and uncertainty quantification. We use a modification of the hierarchical Bayesian model with discontinuous spike-slab priors \citep{Ishwaran2005} introduced by \citet{nayek_spike-and-slab_2021} (albeit in a supervised setting) and described as follows; see \figurename\ref{fig:DAG} for schematic. \RR{Spike-slab priors and their variants have a strong peak (i.e., spike) at zero and extremely shallow tails (i.e., slab). A random variable may be sampled from either the spike or slab part of the distribution. When the random variable is sampled from the slab, it can take large values with uniform-like distribution. However, due to the strong peak, the random variable is more likely to be sampled from the spike (i.e., to be zero-valued) -- representing our bias towards parsimonious solutions.\footnote{While other priors such as horseshoe \citep{Carvalho2010} and Laplacian \citep{Kabn2007} also promote sparsity, discontinuous spike-slab priors have been shown to be more robust and efficient in finding parsimonious solutions \citep{nayek_spike-and-slab_2021}.} A schematic of the  spike-slab priors is shown in \figurename \ref{fig:schematic}b.}

To set up the discontinuous spike-slab priors on the solution vector $\bftheta$, we introduce a random indicator variable $\bfz\in[0,1]^{n_f}$ and auxiliary random variables $\nu_s\geq 0$ and $\sigma^2\geq 0$  such that
\be
p(\theta_i\ |\ z_i, \nu_s, \sigma^2) = \begin{cases}
p_\text{spike}(\theta_i) = \delta(\theta_i) \qquad &\text{for} \qquad z_i=0\\
p_\text{slab}(\theta_i\ |\ \nu_s,\sigma^2) = \calN_{+}(0,\nu_s \sigma^2) \qquad &\text{for} \qquad z_i=1
\end{cases},
\qquad i=1,\dots,n_f.
\ee
When $z_i=0$, the $i^\text{th}$ corresponding feature is deactivated (i.e., $\theta_i=0$) and the probability density of that feature is given by a Dirac delta (\textit{spike}) centered at zero. If $z_i=1$, then the corresponding  feature is activated and $\theta_i$ is sampled from a truncated normal distribution \citep{Botev2016} $\calN_{+}$ with zero mean and variance of $\nu_s \sigma^2$ (\textit{slab}). The truncated normal distribution  is derived from a normal distribution where the samples are restricted to non-negative values only, i.e., $\theta_i\geq0$ (necessary to satisfy physical admissibility; see Section~\ref{sec:energy_library}). With the prior assumption that the components of $\boldsymbol{\theta}$ are independent and identically distributed (i.i.d.), the joint prior distribution of $\bftheta$ is written as
\be
    p \ (\bftheta \ | \ \boldsymbol{z}, \nu_s, \sigma^2) = p_{\text{slab}} \ (\bftheta_r\ |\ \nu_s,\sigma^2) \prod_{i:z_i=0} p_{\text{spike}} \ (\theta_i) \qquad \text{with} \qquad p_{\text{slab}} \ (\bftheta_r \ |\ \nu_s,\sigma^2) = \calN_{+} \  (\boldsymbol{0},\sigma^2\nu_s\bfI_r),
\ee
where $\bftheta_r\in\Rset^r_+$ is a reduced vector consisting of only slab/active components of $\bftheta$ and $\bfI_r$ denotes the $r\times r$ identity matrix. We use the algorithm by \cite{Botev2016} for sampling from the multivariate truncated normal distribution. All features (activated or not) are modeled as an i.i.d. Bernoulli trial with probability $p_0\in[0,1]$, i.e.,
\be\label{eq:z_bernoulli_likelihood}
z_i  \ | \ p_0 \thicksim \text{Bern}(p_0).
\ee
This represents the prior belief that each feature is activated with a probability of $p_0$. The prior probability  distributions for the variables $\nu_s$, $\sigma^2$, and $p_0$ are modeled as 
\be
\begin{aligned}
    &\nu_s \thicksim \calI\calG(a_{\nu},b_{\nu})  & \qquad \text{(Inverse gamma distribution)} \\
    &\sigma^2 \thicksim \calI\calG(a_{\sigma},b_{\sigma})  & \qquad  \text{(Inverse gamma distribution)}\\
    &p_0 \thicksim \text{Beta} (a_p,b_p)  & \qquad  \text{(Beta distribution)},
\end{aligned}
\ee
where $(a_{\nu},b_{\nu})$, $(a_{\sigma},b_{\sigma})$, and $(a_p,b_p)$ are the respective parameters of the hyper-priors. Note that while both $\nu_s$ and $\sigma^2$ scale the variance of the slab, $\sigma^2$ also scales the variance of the likelihood as discussed in the following section. {For both $\nu_s$ and $\sigma^2$, the prior is modeled with an inverse gamma distribution since it is an analytically tractable  conjugate prior for the unknown variance of a normal distribution. Similarly, the beta distribution is chosen as an informative  prior for $p_0$ because it is analytically conjugate to itself for a Bernoulli likelihood -- as is the case in \eqref{eq:z_bernoulli_likelihood}.}

\subsection{Physics-constrained and unsupervised likelihood}
\label{sec:likelihood}

{To obtain the posterior probabilistic estimates of the material model (characterized by $\bftheta$), we consider the `likelihood'  of the  observed data  (which include the displacement, acceleration and reaction force data) satisfying physical laws -- in this case, the conservation of linear momentum.} The weak form of the linear momentum balance is given by
\be
    \int_{\Omega}  \bfP : \nabla \bfv  \dd V - \int_{\partial \Omega_t} \hat{\bft} \cdot   \bfv  \dd S + \int_{\Omega}  \rho_0  \ddot{\bfu} \cdot \bfv \dd V = \boldsymbol{0} \quad \forall \quad \text{admissible}\ \ \bfv,
    \label{eqn:weakform}
\ee
where $\hat{\boldsymbol{t}}$ denotes the traction acting on $\partial \Omega_t=\partial\Omega\setminus\partial \Omega_u$ and $\rho_0$ is the density in the reference configuration. {Note that the boundary tractions on $\partial \Omega_t$} are zero for displacement-controlled loading but non-zero otherwise. The stress tensor is given in indicial notation by
\be
P_{ij} = \frac{\partial W(\bfF)}{\partial F_{ij}} = \frac{\partial \bfQ^T (I_1,I_2,I_3,J_4,J_6)}{\partial F_{ij}} \bftheta,\label{eqn:PWF}
\ee
Here we use automatic differentiation \citep{baydin2018} to compute the feature derivatives in $\partial \bfQ^T/\partial F_{ij}$.  {The test function $\textit{\textbf{v}}$ is any sufficiently regular function that vanishes on the Dirichlet boundary $\partial\Omega_u$.} Formulating the problem in the weak form mitigates the high noise sensitivity introduced via double spatial derivatives in the collocation or strong form of linear momentum balance \citep{flaschel_unsupervised_2021}.

Let $\calD= \{(a,i) : a=1,\dots,n_n; i=1,2\}$ denote the set of all nodal degrees of freedom in the discretized reference domain. $\calD$ is further split into two subsets: $\calD^\text{free} \subset \calD$ and $\calD^\text{fix}=\calD\setminus \calD^\text{free}$ consisting of free and fixed (via Dirichlet constraints) degrees of freedoms. Using the same shape functions as displacements, we approximate the test functions (Bubnov-Galerkin approximation) as \be
v_i (\bfX) = \sum_{a=1}^{n_n} N^a(\bfX) v^a_i,  \qquad\text{with}\qquad v^a_i=0 \ \ \text{if} \ \ (a,i)\in \calD^\text{fix}
\ee
for admissibility. Substituting this into \eqref{eqn:weakform} together with \eqref{eq:displacement_approximation} yields 
\be
    \sum_{a=1}^{n_n} v^a_i \left[ 
    \int_{\Omega}  \left(\frac{\partial \bfQ^T}{\partial F_{ij}} \bftheta \right) \nabla_j N^a (\bfX) \dd V 
    - \int_{\partial \Omega_t}  \hat{t}_i   N^a(\bfX) \dd S 
    + \sum_{b=1}^{n_n}\ddot{u}^b_i\int_{\Omega}\rho_0 N^a (\bfX) N^b (\bfX) \dd V \right] = 0.
\ee
The above equation is true for all arbitrary admissible test functions if and only if
\be\label{eq:consistent_balance}
    \int_{\Omega}  \left(\frac{\partial \bfQ^T}{\partial F_{ij}} \bftheta \right) \nabla_j N^a (\bfX) \dd V 
    - \int_{\partial \Omega_t}  \hat{t}_i   N^a(\bfX) \dd S
    + \sum_{b=1}^{n_n}\ddot{u}^b_i\underbrace{\int_{\Omega}\rho_0 N^a (\bfX) N^b (\bfX) \dd V}_{\text{consistent mass matrix}}  = 0   \qquad \forall \qquad (a,i)\in\calD^\text{free}.
\ee
The consistent mass matrix (as indicated above) is approximated with the lumped nodal mass given by 
\be
M^{ab} = \begin{cases} m^a \quad &\text{if} \quad a=b \\
0 \quad &\text{otherwise}
\end{cases} \qquad \text{with} \qquad m^a = \int_\Omega \rho_0 N^a (\bfX) \dd V.
\ee
Therefore, \eqref{eq:consistent_balance} reduces to
\be\label{euclid_free}
F^{\text{int},a}_i(\bftheta) - F^{\text{ext},a}_i + F^{\text{m},a}_i = 0 \qquad \forall \qquad (a,i)\in\calD^\text{free},
\ee
where 
\be
\underbrace{F^{\text{int},a}_i(\bftheta) = \int_{\Omega}  \left(\frac{\partial \bfQ^T}{\partial F_{ij}} \bftheta \right) \nabla_j N^a (\bfX) \dd V}_{\text{internal/constitutive force}}, \qquad 
\underbrace{F^{\text{ext},a}_i = \int_{\partial \Omega_t}  \hat{t}_i   N^a(\bfX) \dd S}_{\text{external force}}, \qquad \text{and} \qquad
\underbrace{F^{\text{m},a}_i = \vphantom{\int_{\partial \Omega_t}} m^a \ddot{u}^a_i}_{\text{inertial force}}
\ee
are interpreted as internal/constitutive, external, and inertial forces, respectively, acting on the degree of freedom denoted by $(a,i)$. Note that the inertial forces vanish for quasi-static loading. The force balance in \eqref{euclid_free} represents a linear system of equations for the unknown material parameters $\bftheta$, which can be assembled in the form
\be\label{eq:Abfree_superset}
\bfA^\text{free} \bftheta = \bfb^\text{free} \qquad \text{with} \qquad \bfA^\text{free}\in\Rset^{|\calD^\text{free}|\times n_f} \quad \text{and}\quad \bfb^\text{free}\in\Rset^{|\calD^\text{free}|}.
\ee
For computational efficiency, we randomly subsample $n_\text{free}<|\calD^\text{free}|$ degrees of freedom from \eqref{eq:Abfree_superset} such that $\bfA^\text{free}\in\Rset^{n_\text{free}\times n_f}$ and $\bfb^\text{free}\in\Rset^{n_\text{free}}$.

Additionally, the reaction forces are known at $n_\beta$ Dirichlet boundaries. For $\beta=1,\dots,n_\beta$, let $\calD^{\text{fix},\beta}\subseteq \calD^\text{fix}$ (with $\calD^{\text{fix},\beta} \cap \calD^{\text{fix},\beta'} = \emptyset$ for $\beta\neq\beta'$) denote the subset of degrees of freedom for which the sum of reaction forces $R^\beta$ is known. Each observed reaction force must balance the aggregated internal and inertial forces, i.e.,
\be
\sum_{(a,i)\in \calD^{\text{fix},\beta}} \left( F^{\text{int},a}_i (\bftheta) + F^{\text{m},a}_i \right)\ = \ \sum_{(a,i)\in \calD^{\text{fix},\beta}} F^{\text{ext},a}_i \ = \ R^\beta, \qquad \forall \qquad \beta=1,\dots,n_\beta,
\ee
which can further be assembled into the form
\be\label{euclid_fix}
\bfA^\text{fix} \bftheta = \bfb^\text{fix} \qquad \text{with} \qquad \bfA^\text{fix}\in\Rset^{n_\beta\times n_f} \quad \text{and}\quad \bfb^\text{fix}\in\Rset^{n_\beta}.
\ee

Recall that we dropped the superscript $(\cdot)^t$ for the sake of brevity while implying that the above formulation applies to snapshots at every time step. The force balance constraints \eqref{euclid_free} and \eqref{euclid_fix} across all $t=1,\dots,n_t$ snapshots are concatenated into a joint system of equations as
\be\label{eqn:finaltheta}
\bfA \bftheta = \bfb \qquad \text{with} \qquad
\bfA = \begin{bmatrix}
\bfA^{\text{free},1}\\
\vdots\\
\bfA^{\text{free},n_t}\\
\lambda_r \bfA^{\text{fix},1}\\
\vdots\\
\lambda_r \bfA^{\text{fix},n_t}\\
\end{bmatrix} 
\quad \text{and} \quad 
\bfb = \begin{bmatrix}
\bfb^{\text{free},1}\\
\vdots\\
\bfb^{\text{free},n_t}\\
\lambda_r \bfb^{\text{fix},1}\\
\vdots\\
\lambda_r \bfb^{\text{fix},n_t}\\
\end{bmatrix},
\ee
where $\lambda_r>0$ is a hyperparameter that controls the relative importance between the force balance of the free and fixed degrees of freedom. \RR{The choice of all the hyperparameters is reported and discussed in \ref{sec:protocols}.}

The term $\bfb$ is directly related to the nodal accelerations and the measured reaction forces, which may have uncertainties. Additionally, observational noises in the nodal displacements give rise to commensurate uncertainties in $\bfA$. Thus, \eqref{eqn:finaltheta} is better written as
\begin{equation}\label{eqn:finalthetanoise}
    \bfb = \bfA\bftheta + \bfepsilon,
\end{equation}
where $\bfepsilon$ is the residual of the momentum balance equations and is indicative of the uncertainty in the observations as well as of model inadequacies. Given a constitutive model (represented by $\bftheta$) sampled from the prior (Section~\ref{sec:prior}) and the observed data (now in the form of $\bfA$ and $\bfb$), we place an i.i.d. normal distribution likelihood on the residual as
\be\label{eq:likelihood}
\bfb\  | \ \bftheta, \sigma^2, \bfA  \thicksim \calN(\bfA\bftheta,\sigma^2 \bfI_N),
\ee
where $\bfI_N$ denotes the $N\times N$ identity matrix with $N=\left(n_{\text{free}}+n_\beta\right) n_t$.

\subsection{Model discovery via posterior estimation}

The next step in the Bayesian learning process is to evaluate the joint posterior probability distribution for the random variables in the hierarchical model, i.e., $\{\bftheta, \bfz, p_0, \nu_s, \sigma^2\}$, given the observed data and momentum balance constraints embodied by $\bfA$ and $\bfb$. Using Bayes' theorem, the posterior is given by
\be\label{eq:posterior}
p(\bftheta,\bfz,p_0,\nu_s,\sigma^2 \ |\  \bfA, \bfb) \propto \underbrace{p(\bfb\ |\ \bftheta,\bfz,p_0,\nu_s,\sigma^2, \bfA)}_{\text{physics-constrained likelihood}} \ \underbrace{p(\bftheta,\bfz,p_0,\nu_s,\sigma^2)}_{\text{spike-slab model prior}}.
\ee
Note that the model prior is independent of the observed data in $\bfA$ and hence, its conditioning on $\bfA$ has been removed. Analytical sampling  from \eqref{eq:posterior} directly is difficult due to the presence of the spike-slab priors and requires the use of Markov Chain Monte Carlo (MCMC) methods. For the purpose of this work, we use a Gibbs sampler \citep{casella_explaining_1992} to empirically estimate the model posterior distribution. The Gibbs sampling procedure requires knowledge of the conditional posterior distributions for all the random variables.  \citet{nayek_spike-and-slab_2021} derived analytical expressions for all the conditional posterior distributions which we briefly summarize here.

\begin{itemize}
    \item Conditional posterior distribution of $\bftheta$: 
    \be\label{eq:cond_0}
        \bftheta_r \ |\   \bfz,p_0,\nu_s,\sigma^2, \bfA, \bfb\ \thicksim\ \calN_{+}(\bfmu, \sigma^2 \bfSigma), \qquad \text{with} \qquad \bfSigma = \left(\bfA_r^T \bfA_r + \nu_s^{-1} \bfI_r\right)^{-1} \quad\text{and}\qquad \bfmu = \bfSigma\bfA_r^T\bfb.
    \ee
    $\bfA_r\in \Rset^{N\times s_z}$ is obtained by concatenating the columns $\{i=1,\dots,n_f\}$ of $\bfA$ for which $z_i=1$; the distribution only (but jointly) applies to the corresponding active features, i.e., $\bftheta_r$. $\bfI_r$ is the $s_z\times s_z$ identity matrix where  $s_z=\sum_{i=1}^{n_f} z_i$ denotes the number of active features. If $z_i=0$, the corresponding $\theta_i$ is set to zero. Note that \citet{nayek_spike-and-slab_2021} did not consider any constraint on the values of $\bftheta$, in which case the above conditional posterior is given by an unconstrained multivariate normal distribution. In the context of our work, the positivity constraint on $\bftheta$ replaces the same with a constrained multivariate normal distribution $\calN_+$ (see \cite{Botev2016} for sampling methodology) without affecting the remaining distributions.
    
    \item Conditional posterior distribution of $\sigma^2$: 
    \be\label{eq:cond_1}
         \sigma^2 \ |\ \bftheta,\bfz,p_0,\nu_s, \bfA, \bfb \ \thicksim\ \calI\calG  \left( a_\sigma + \frac{N}{2}, b_\sigma + \frac{1}{2}\left(\bfb^T\bfb-\bfmu^T\Sigma^{-1}\bfmu\right) \right)
    \ee    
    
    \item Conditional posterior distribution of $\nu_s$: 
    \be\label{eq:cond_2}
        \nu_s\ |\ \bftheta,\bfz,p_0,\sigma^2, \bfA, \bfb \ \thicksim\ \calI\calG\left( a_\nu + \frac{s_z}{2}, b_\nu+ \frac{\bftheta_r^T\bftheta_r}{2\sigma^2} \right)
    \ee
    
    \item Conditional posterior distribution of $p_0$: 
    \be\label{eq:cond_3}
         p_0 \ |\ \bftheta, \bfz,\nu_s,\sigma^2, \bfA, \bfb\  \thicksim\ \text{Beta}\left(a_p+s_z, b_p+n_f-s_z\right)
    \ee
    
    \item Conditional posterior distribution of $\bfz$ (component-wise):
    \be\label{eq:cond_4}
        z_i \ | \ \bftheta,{\bfz_{-i}},p_0,\nu_s,\sigma^2  \bfA, \bfb \ \thicksim\ \text{Bern}(\xi_i)\qquad\text{with}\qquad \xi_i = p_0 \left[p_0 + \frac{p(\bfb\ |\ z_i=0, \bfz_{-i}, \nu_s,\bfA)}{p(\bfb\ |\ z_i=1, \bfz_{-i}, \nu_s,\bfA)}(1-p_0)\right]^{-1}.
    \ee    
    Note that the components of $\bfz$ are sampled in random order at each step of the Gibbs sampler. $\bfz_{-i}$ denotes the reduced vector $\bfz$ without the $i^\text{th}$ component, i.e., {without} $z_i$.  The marginal likelihood $p(\bfb\ |\ \bfz, \nu_s,\bfA)$, obtained by integrating out $\bftheta$ and $\sigma^2$ in the likelihood \eqref{eq:likelihood}, is given by
    \be\label{eq:cond_5}
        p(\bfb\ |\ \bfz,\nu_s,\bfA) =
        \frac{\Gamma(a_\sigma + 0.5 N)}{(2\pi)^{N/2} (\nu_s)^{s_z/2}}
        \frac{(b_\sigma)^{a_\sigma}}{\Gamma(a_\sigma)}
        \frac{
        \left[
            \det\left(
                \left(\bfA_r^T\bfA_r + \nu_s^{-1}\bfI_r\right)^{-1}
            \right)
        \right]^{1/2}
        }
        {
            \left[
                b_\sigma 
                + 
                0.5\bfb^T 
                \left(
                    \bfI_N - \bfA_r \left(
                        \bfA_r^T\bfA_r + \nu_s^{-1}\bfI_r
                    \right)^{-1} \bfA_r^T 
                \right)
                \bfb
            \right]^{a_\sigma+0.5N}
        }
    \ee
    where $\Gamma(\cdot)$ denotes the Gamma function.
\end{itemize}
The Gibbs sampler repeatedly and successively samples each variable in $\{\bftheta, \bfz, p_0, \nu_s, \sigma^2\}$ using the respective conditional posterior distributions \eqref{eq:cond_0}-\eqref{eq:cond_4}. This produces the following Markov chain of length $N_G$:
\begin{equation}\label{eq:chain}
    \boldsymbol{\theta}^{(0)}\rightarrow \sigma^{2^{\mathlarger{(0)}}} \rightarrow \nu_s^{(0)}\rightarrow p_0^{(0)}\rightarrow \boldsymbol{z}^{(0)}\rightarrow \dots \dots \rightarrow \boldsymbol{\theta}^{(N_G)}\rightarrow {\sigma^2}^{(N_G)} \rightarrow\nu_s^{(N_G)}\rightarrow p_0^{(N_G)}\rightarrow \boldsymbol{z}^{(N_G)},
\end{equation}
which empirically approximates the joint posterior distribution $p(\bftheta,\bfz,p_0,\nu_s,\sigma^2 \ |\  \bfA, \bfb)$.  Note that the Markov chain states are only recorded after discarding the first $N_\text{burn}$ burn-in samples to avoid the effects of a bad starting point. For the same reason, {we also generate $N_\text{chains}$ independent Markov chains (each with different starting points selected randomly)} which are concatenated after discarding the burn-in states. {A pseudo-code for  MCMC sampling of the posterior probability distribution is described in \ref{sec:AppPSEUD}.}

The average activity of the $i^{th}$ feature ($i=1,\dots,n_f$) of the energy {density function} library is defined as
\be\label{eq:avg_activity}
z^{\text{avg}}_i = \frac{1}{N_G\times N_\text{chains}} \sum_{k=1}^{N_G\times N_\text{chains}}z_i^{(k)},
\ee
where $z_i^{(k)}\in\{0,1\}$ is the activity of the $i^\text{th}$ feature in the $k^\text{th}$ sample of the Markov chain. Note that the average activity is a marginalization over the  variables in the joint posterior distribution. Therefore,  $z^{\text{avg}}_i=1$ and $z^{\text{avg}}_i=0$ imply that the corresponding feature is always active or inactive, respectively; whereas any  value between zero and one implies that the joint distribution is likely multi-modal with the corresponding feature being both active and inactive in different modes.

\section{Benchmarks}
\label{sec:benchmarks}

\begin{figure}
	\centering
	\includegraphics[width=0.7\textwidth]{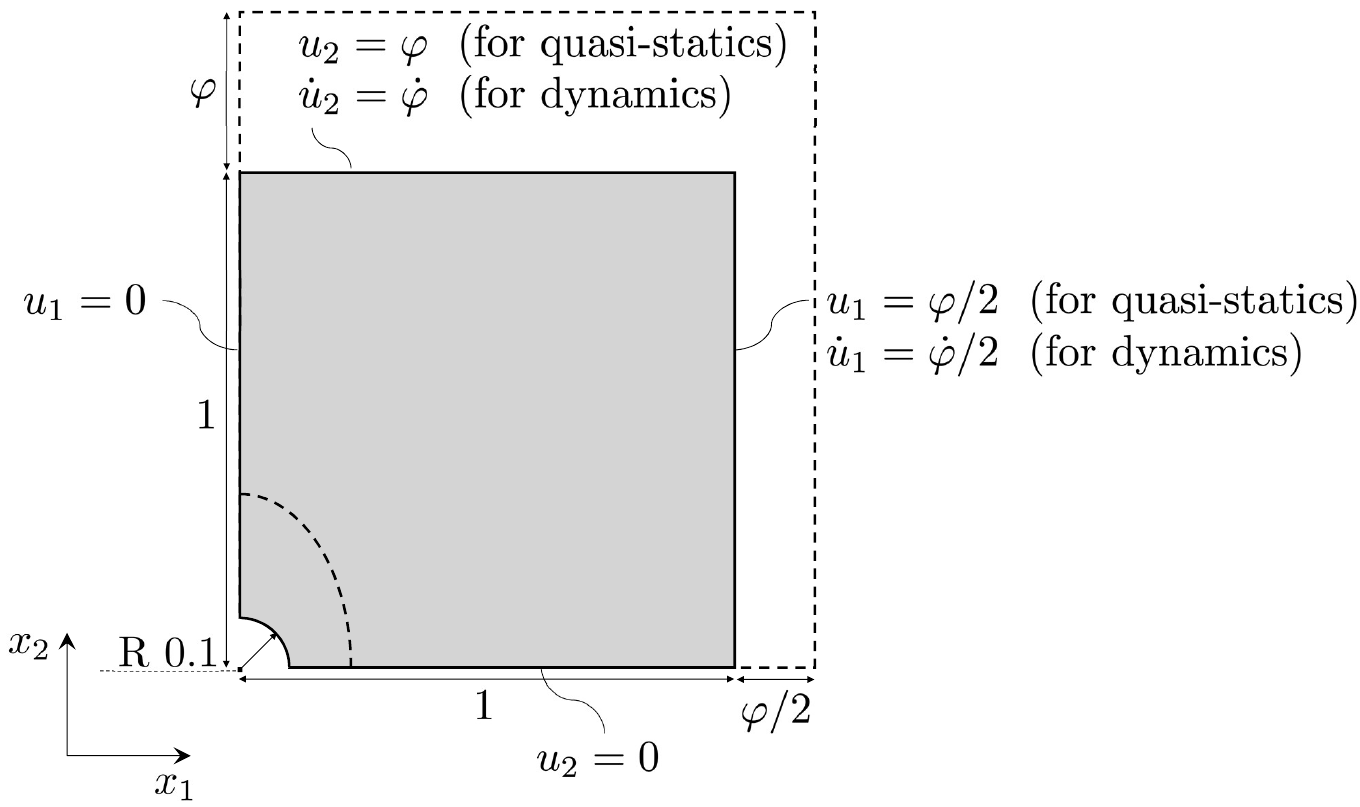}
	\caption{{Geometry of a single quadrant of a plate with a central hole used in the data generation simulations. Also shown are the boundary conditions for the virtual experiment. In the quasi-static case (no accelerations), for every load step, the displacements are incremented by a value of $\varphi$ along $x_2$ direction and $\varphi/2$ along the $x_1$ direction. In the dynamic case (with accelerations), the displacement increment rate is  $\dot{\varphi}$ along the $x_2$ direction and $\dot{\varphi}$/2 along the $x_1$ direction. All dimensions are consistent in units with respect to the other parameters. All lengths and displacements are normalized with respect to the side length of the {undeformed} specimen.}}\label{fig:geometry}
\end{figure}

\subsection{Data generation}
\label{sec:data_generation}

For the scope of this work, we use the finite element method (FEM) to generate synthetic data which would otherwise be obtained from DIC experiments. For benchmarking, we consider the same mechanical testing as that of \cite{flaschel_unsupervised_2021} where a hyperelastic square specimen with a finite-sized hole is deformed under displacement-controlled asymmetric biaxial tension (see \figurename\ref{fig:geometry} for schematic). This testing configuration subjects the material to a sufficiently diverse range of localized strain states necessary for discovering a generalizable constitutive model. Due to two-fold symmetry in the problem, only a quarter of the plate is simulated, with appropriate boundary conditions to enforce the symmetry. \RR{For the scope of the current work dealing with only in-plane behavior of the material, we assume plane strain conditions.} The data is simulated using linear triangular elements. Both static and dynamic simulations are performed. For the static simulations, as shown in \figurename\ref{fig:geometry}, the boundary displacements are prescribed by the loading parameter $\varphi$ for $L_\text{static}$ consecutive load steps. For the dynamic simulations, the boundary velocities are prescribed by the constant loading rate $\dot\varphi$ with time step size $\delta t\ll 1$ for $L_\text{dynamic}$ consecutive steps and explicit time integration. A total of $n_t$ equispaced snapshots consisting of nodal displacements, nodal accelerations (only for the dynamic case with $n_t\ll L_\text{dynamic}$), and four boundary reaction forces (normal to the top, bottom, left, and right boundaries of the specimen) are recorded. The accelerations are inferred from the displacement fields using the central difference scheme \eqref{eq:acc} (see Section~\ref{sec:available_data}). All  parameters related to the data generation are summarized in \ref{sec:protocols}.

To emulate real measurements, artificial noise is added to the displacement data. The noise level realistically depends on the imaging setup and pixel accuracy and is independent of the applied displacements. Therefore, we apply the same absolute noise level (noise floor) to all the nodal displacement data irrespective of the local displacement field. The synthetically generated nodal displacements are noised as 
\begin{equation}
    u_i^{a,t} = u_i^{\text{fem,a,t}} +  u_i^{\text{noise,a,t}} \quad \text{with} \quad  u_i^{\text{noise,a,t}} \thicksim \calN(0,\sigma_u^2) \quad \forall \quad (a,i)\in\calD,\ t\in\{1,\dots,n_t\}.
\end{equation}
$u_i^{\text{fem,a,t}}$ denotes the $i^\text{th}$ displacement component at node $a$ as generated from the FEM simulation to which the noise $u_i^{\text{noise,a,t}}$  (sampled from the i.i.d. Gaussian distribution with zero mean and $\sigma_u$ standard deviation) is added. 
To demonstrate the efficacy of the proposed Bayesian learning framework, we study two different noise levels (normalized with the specimen length): 
\begin{itemize}
    \item low noise: $\sigma_u=10^{-4}$
    \item high noise: $\sigma_u=10^{-3}$
\end{itemize}
which are representative of modern DIC setups \citep{Schreier2009,Bonnet2005}. \RR{In the current work, for illustrative purposes, we assume uncorrelated Gaussian noise in the displacements. To account for spatio-temporally correlated noise or non-Gaussian noise that may be encountered while imaging \citep{WuDufanXRAY}, the likelihood (\ref{eq:likelihood}) must accordingly be adapted to represent non-Gaussian multivariate distributions.} Following the benchmarks of \citet{flaschel_unsupervised_2021}, we also spatially denoise each displacement field snapshot using Kernel Ridge Regression (KRR), as would be done in the case of experimental measurements. Further details regarding the noise and denoise protocols can be found in \cite{flaschel_unsupervised_2021}. Note that $\sigma_u$ represents the displacement noise and is different from $\sigma$ which represents the noise in satisfying the momentum balance likelihood \eqref{eq:likelihood}.

The following material models are used for benchmarking the Bayesian constitutive model discovery.
\begin{enumerate}
    \item Neo-Hookean solid:
            \begin{equation}\label{eq:NH2}
                W(\bfF) = 0.5 (\Tilde{I}_1-3) + 1.5 (J-1)^2.
            \end{equation}
    \item Isihara solid \citep{isihara_statistical_1951}:
            \begin{equation}\label{eq:IH}
                W(\bfF) = 0.5 (\Tilde{I}_1-3) + (\Tilde{I}_2-3) + (\Tilde{I}_1-3)^2 + 1.5 (J-1)^2.
            \end{equation}
    \item Gent-Thomas model \citep{gent_forms_1958}:
            \begin{equation}\label{eq:GT}
                W(\bfF) = 0.5 (\Tilde{I}_1-3) + (\Tilde{I}_1-3)^2 + \log(\Tilde{I}_2/3) + 1.5 (J-1)^2.
            \end{equation}
    \item Haines-Wilson model \citep{haines_strain-energy_1979}:
            \begin{equation}\label{eq:HW}
                W(\bfF) = 0.5 (\Tilde{I}_1-3) + (\Tilde{I}_2-3) + 0.7 (\Tilde{I}_1-3) (\Tilde{I}_2-3) + 0.2 (\Tilde{I}_1-3)^3 + 1.5 (J-1)^2.
            \end{equation}
    \item Arruda-Boyce model \citep{arruda_three-dimensional_1993}:
            \begin{equation}\label{eq:AB}
                W(\bfF) = 0.25 \text{AB}(\Tilde{I}_1) + 1.5 (J-1)^2,
            \end{equation}
            with $\text{AB}$ given by \eqref{eq:feature_AB}.
    \item Ogden model \citep{ogden_large_1972} with 1 term:
            \begin{equation}\label{eq:OG}
                W(\bfF) = 0.65\ \text{OG}_1(\Tilde{I}_1,J) + 1.5 (J-1)^2,
            \end{equation}
            where $\text{OG}_1$ is given by \eqref{eq:feature_OG} with $\alpha_1=1.3$.
    \item Ogden model with 3 terms:
            \begin{equation}\label{eq:OG3}
                W(\bfF) = 0.4\  \text{OG}_1(\Tilde{I}_1,J) + 0.0012\ \text{OG}_2(\Tilde{I}_1,J) + 0.1\ \text{OG}_3(\Tilde{I}_1,J) + 1.5 (J-1)^2,
            \end{equation}
            where $\text{OG}_1$, $\text{OG}_2$, $\text{OG}_3$ are given by \eqref{eq:feature_OG} with $\alpha_1=1.3$, $\alpha_2=5$, and $\alpha_3=2$, respectively.
    \item Anisotropic Holzapfel model \citep{holzapfel_new_2000} with two {fiber families} at $+30^{\circ}$ and $-30^{\circ}$ orientations:
            \begin{equation}\label{eq:HZ}
                W(\bfF) = 0.5 (\Tilde{I}_1-3) + (J-1)^2 + \frac{k_{1h}}{2k_{2h}}\left[\exp\left(k_{2h}(\Tilde{J}_4-1)^2\right)+\exp\left(k_{2h}(\Tilde{J}_6-1)^2\right)-2\right]
            \end{equation}
    with the fiber directions $\bfa_1 = \left(\cos(30^{\circ}), \sin(30^{\circ}),0\right)^T$, $\bfa_2 = \left(\cos(30^{\circ}), -\sin(30^{\circ}), 0\right)^T$ and constants $k_{1h} = 0.9$, $k_{2h}=0.8$. We highlight that the Holzapfel model is specifically chosen to test the generalization capability of the proposed framework since its characteristic features are not available in the chosen feature library.
\end{enumerate}
All the aforementioned models {are} assumed to have weak compressibility by adding quadratic volumetric penalty terms to the energy density.

\subsection{Model discovery with quasi-static data} 
\label{sec:results}

The  subsequent results are based on quasi-static data with the feature library listed in Table~\ref{tab:features}. When explicitly stated, certain features in the library will be excluded intentionally to test the generalization capability of the proposed method. All protocols and parameters pertaining to the Bayesian learning are presented in \ref{sec:protocols}. 

The model discovery results on the eight benchmarks \eqref{eq:NH2}-\eqref{eq:HZ} -- for both low noise ($\sigma_u=10^{-4}$) and high noise ($\sigma_u=10^{-3}$) cases  are summarized in Figures \ref{fig:NH2}-\ref{fig:HZ}. In each figure, the \textit{violin plot} (top-left) shows the marginal posterior distribution (along the vertical axis) for each component of $\bftheta$ using the Markov chain \eqref{eq:chain}. The features are labeled according to the indices defined in Table~\ref{tab:features}. {Each of the overlaid grey lines represents} {a coefficient vector $\bftheta$ sampled from the joint posterior distribution. Together, they complement the information conveyed by the accompanying violin plots, which present the posterior distribution for each component of $\bftheta$. Darker regions in the line plot arise when many samples from the MCMC chain possess similar values for the coefficient vector $\bftheta$, and thus indicate the most probable values for $\bftheta$.} The features present in the true model are highlighted in cyan background, with red horizontal dashes denoting the true value of the coefficients in $\bftheta$. The \textit{activity plot} (bottom left) shows the average posterior activity of each feature, where the height of the $i^\text{th}$ bar equals $z_i^\text{avg}$ (see \eqref{eq:avg_activity}). For each sample from the MCMC chain, we compute the strain energy density 
\be
W^{(k)}(\bfF) = \bfQ^T \bftheta^{(k)}, \qquad k=1,\dots,(N_G\times N_\text{chains})
\ee
along six different deformation paths parameterized by $\gamma\in[0,1]$ as
\be
\begin{aligned}
\boldsymbol{F}^{\text{UT}}(\gamma) &= \begin{bmatrix}
1 + \gamma & 0\\
0 & 1
\end{bmatrix}, \ \ \
\boldsymbol{F}^{\text{UC}}(\gamma) = \begin{bmatrix}
\frac{1}{1 + \gamma} & 0\\
0 & 1
\end{bmatrix}, \ \ \ 
\boldsymbol{F}^{\text{BT}}(\gamma) = \begin{bmatrix}
1 + \gamma & 0\\
0 & 1+ \gamma
\end{bmatrix}, \\
\boldsymbol{F}^{\text{BC}}(\gamma) &= \begin{bmatrix}
\frac{1}{1 + \gamma} & 0 \\
0 & \frac{1}{1 + \gamma}
\end{bmatrix}, \ \ \
\boldsymbol{F}^{\text{SS}}(\gamma) = \begin{bmatrix}
1 & \gamma\\
0 & 1
\end{bmatrix}, \ \ \
\boldsymbol{F}^{\text{PS}}(\gamma) = \begin{bmatrix}
1 + \gamma & 0\\
0 & \frac{1}{1 + \gamma}
\end{bmatrix}.
\end{aligned}
\label{eqn:strain_paths}
\ee
The deformation paths correspond to uniaxial tension (UT), uniaxial compression (UC), biaxial tension (BT), biaxial compression (BC), simple shear (SS), and pure shear (PS), respectively. To visualize the accuracy of the discovered models, the true energy density (denoted by the red line) is plotted alongside the mean (denoted by the black line) and $95\%$ percentile bounds (shaded grey), i.e., the region between the $2.5$ and $97.5$ percentile samples across the chain of energy densities  for each deformation gradient. The percentile bounds represent the uncertainty in the energy density prediction. \RR{To quantify the prediction accuracy, the coefficient of determination (denoted by $R^2$) between the mean of the predicted energy densities and the true energy density is also shown for each deformation path. As a general definition, given two series of true and predicted values -- $\bfy_\text{true}$ and $\bfy_\text{pred}$,  respectively, $R^2$ as a measure for goodness of fit is computed as
\be
R^2 =  1 - \frac{\sum_i (y_\text{true, i}-y_\text{pred, i})^2}{\sum_i (y_\text{true, i}-\overline{\bfy_\text{true}})^2}
\ee
where $R^2=1$ denotes that the predictive model perfectly matches the true values, while  $R^2\leq 0$ corresponds to the predictive model being worse than or equivalent to constantly predicting the mean of the true values.}

\begin{table}[ht]
\centering
\begin{tabular}{c|c|c|c}
\hline
\rowcolor[HTML]{9B9B9B} 
\textbf{Index no.} & \textbf{Feature} & \textbf{Index no.} & \textbf{Feature} \\ \hline
\cellcolor[HTML]{C0C0C0}1     & $(\Tilde{I}_1-3)$            &\cellcolor[HTML]{C0C0C0}14     & $(\Tilde{I}_2-3)^4$  \\ \hline
\cellcolor[HTML]{C0C0C0}2     & $(\Tilde{I}_2-3)$            &\cellcolor[HTML]{C0C0C0}15    & $(J-1)^2$  \\ \hline
\cellcolor[HTML]{C0C0C0}3     & $(\Tilde{I}_1-3)^2$          &\cellcolor[HTML]{C0C0C0}16     & $\log(\Tilde{I}_2/3)$  \\ \hline
\cellcolor[HTML]{C0C0C0}4     & $(\Tilde{I}_1-3)(\Tilde{I}_2-3)$        &\cellcolor[HTML]{C0C0C0}17     & $\text{AB}(\Tilde{I}_1)$       \\ \hline
\cellcolor[HTML]{C0C0C0}5     & $(\Tilde{I}_2-3)^2$          &\cellcolor[HTML]{C0C0C0}18     & $\text{OG}_1(\Tilde{I}_1,J)$        \\ \hline
\cellcolor[HTML]{C0C0C0}6     & $(\Tilde{I}_1-3)^3$          &\cellcolor[HTML]{C0C0C0}19     & $\text{OG}_2(\Tilde{I}_1,J)$           \\ \hline
\cellcolor[HTML]{C0C0C0}7     & $(\Tilde{I}_1-3)^2(\Tilde{I}_2-3)$      &\cellcolor[HTML]{C0C0C0}20     & $\text{OG}_3(\Tilde{I}_1,J)$       \\ \hline
\cellcolor[HTML]{C0C0C0}8     & $(\Tilde{I}_1-3)(\Tilde{I}_2-3)^2$      &\cellcolor[HTML]{C0C0C0}21     & $(\Tilde{J}_4-1)^2$     \\ \hline
\cellcolor[HTML]{C0C0C0}9     & $(\Tilde{I}_2-3)^3$          &\cellcolor[HTML]{C0C0C0}22     & $(\Tilde{J}_4-1)^3$      \\ \hline
\cellcolor[HTML]{C0C0C0}10    & $(\Tilde{I}_1-3)^4$          &\cellcolor[HTML]{C0C0C0}23    & $(\Tilde{J}_4-1)^4$         \\ \hline
\cellcolor[HTML]{C0C0C0}11    & $(\Tilde{I}_1-3)^3(\Tilde{I}_2-3)$      &\cellcolor[HTML]{C0C0C0}24    & $(\Tilde{J}_6-1)^2$      \\ \hline
\cellcolor[HTML]{C0C0C0}12    & $(\Tilde{I}_1-3)^2(\Tilde{I}_2-3)^2$    &\cellcolor[HTML]{C0C0C0}25    & $(\Tilde{J}_6-1)^3$    \\ \hline
\cellcolor[HTML]{C0C0C0}13    & $(\Tilde{I}_1-3)(\Tilde{I}_2-3)^3$      &\cellcolor[HTML]{C0C0C0}26    & $(\Tilde{J}_6-1)^4$     \\ \hline
\end{tabular}
\caption{List of features used in the library $\boldsymbol{Q}$.}
\label{tab:features}
\end{table}

\subsubsection{Model accuracy and aleatoric uncertainties}
\label{sec:aleatoricunc}
For all the benchmarks, the energy density along the six deformation paths is predicted with high accuracy \RR{(indicated by good $R^2$ scores)} and confidence, the latter being specifically indicated by the narrow percentile bounds containing the true energy density. As expected, the \RR{$R^2$ scores and percentile bounds are lower and wider, respectively,} for the high noise case than for the low noise case. \RR{Across all benchmarks, the approach correctly and interpretably detects the presence/absence of anisotropy in the all the constitutive models.} In the anisotropic Holzapfel benchmark (\figurename\ref{fig:HZ}), both isotropic and anisotropic features are correctly identified, the latter given by fourth-order polynomial terms as approximation to the true exponential Holzapfel features in \eqref{eq:HZ} (not included in the feature library).  

The data used for model discovery are based on a single test consisting of asymmetric biaxial tension of the specimen. Consequently, the loading  mainly activates the tensile strain states,  while the shear strains are only activated indirectly due to and in proximity of the hole in the specimen. This is automatically reflected by the Bayesian learning in the energy density predictions in the form of lower uncertainty (narrower percentile bounds; \RR{higher $R^2$ scores}) along uni-/bi-axial tension and compression deformation paths vs.~high uncertainty (wider percentile bounds; \RR{lower $R^2$ scores}) along simple and pure shear deformation paths.\footnote{The specimen shape may be designed to optimally activate a diverse range of strain states which will further increase the confidence and accuracy of the predictions. However, this is beyond the scope of the current work.} The hierarchical Bayesian approach enables the automatic estimation of the aleatoric uncertainty (arising due to random noise in the observed data) by placing a hyper-prior on the likelihood variance $\sigma^2$. Note that $\sigma^2$ is the variance in satisfying the physics-based constraint of momentum balance, and is different from the displacement field noise variance $\sigma_u^2$. However, both are correlated -- specifically, $\sigma^2$ increases with $\sigma_u^2$ -- as evident from the marginal posterior distribution of $\sigma^2$ {(see \figurename\ref{fig:sig2} and the accompanying discussion in \ref{sec:appSIGS})}.

\begin{figure}[H]
	\centering
	\includegraphics[width=0.4\textwidth]{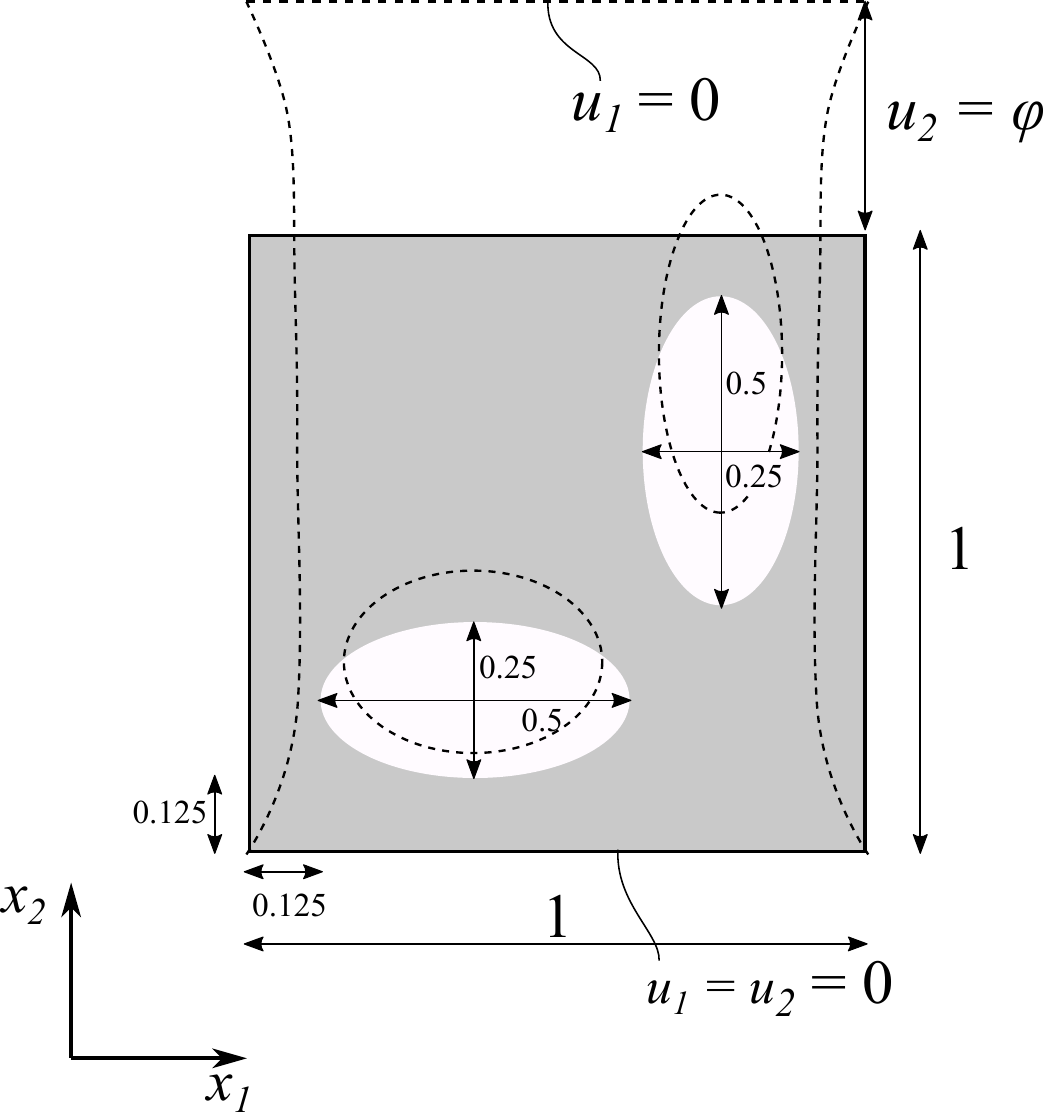}
	\caption{\RR{Geometry of a plate with two asymmetric elliptical holes used for validating the model obtained from Bayesian-EUCLID in a FEM setting. Also shown are the boundary conditions for the virtual experiment. The plate is quasi-statically stretched along the vertical direction as indicated by the load parameter $\varphi$. All lengths and displacements are normalized with respect to the side length of the {undeformed} specimen.}
	}\label{fig:DualHole}
\end{figure}

\RR{To further test the generalization accuracy of the discovered models to unseen strain states, we deploy them in the FEM simulation of a complex specimen different from the one used for model discovery. For this purpose, we consider the uniaxial tensile loading of a square plate with two asymmetric elliptical holes (see \figurename \ref{fig:DualHole} for geometry and boundary conditions) and under plane strain. The simulation uses a mesh of triangular elements with single quadrature points and a total of 4908 nodes. The specimen is quasi-statically extended up to $\varphi = 1$ in 10 equi-spaced loadsteps.  The FEM simulations employing the mean of the predicted strain energy models discovered by the Bayesian-EUCLID framework are validated against the ground-truth model by comparing the values of the strain invariants at all quadrature points of the mesh.  Figures \ref{fig:ValidLN} and \ref{fig:ValidHN} present the FEM-based validation results for the Holzapfel benchmark \eqref{eq:HZ} for the low and high noise cases, respectively. The $R^2$ scores indicate the mismatch in the strain distribution across the specimen with respect to the ground-truth simulation. In the ideal case of $R^2=1$, each point in the plot of the element-wise predicted and ground-truth strain invariants must lie on a straight line with unit slope and zero intercept. As expected, we obtain a good match in the low noise case, while considerable errors are present for the high noise case.}

\subsubsection{Model parsimony}

\begin{figure}[ht]
	\centering
	\begin{subfigure}[ht]{\textwidth}
	    \centering
	    \includegraphics[width=\textwidth]{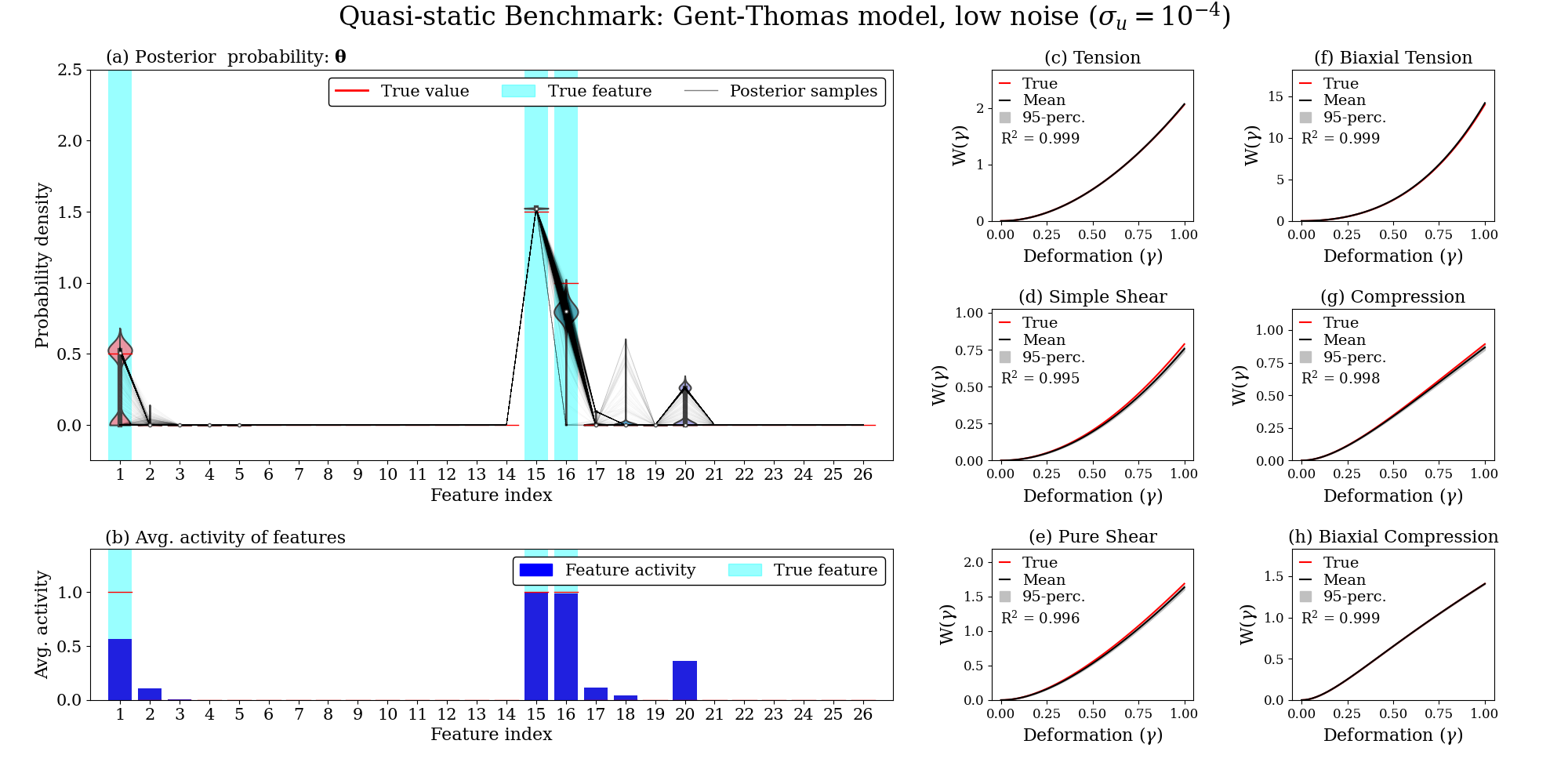}
	\end{subfigure}
	\hrule \vskip 15pt
	\begin{subfigure}[ht]{\textwidth}
	    \centering
	    \includegraphics[width=\textwidth]{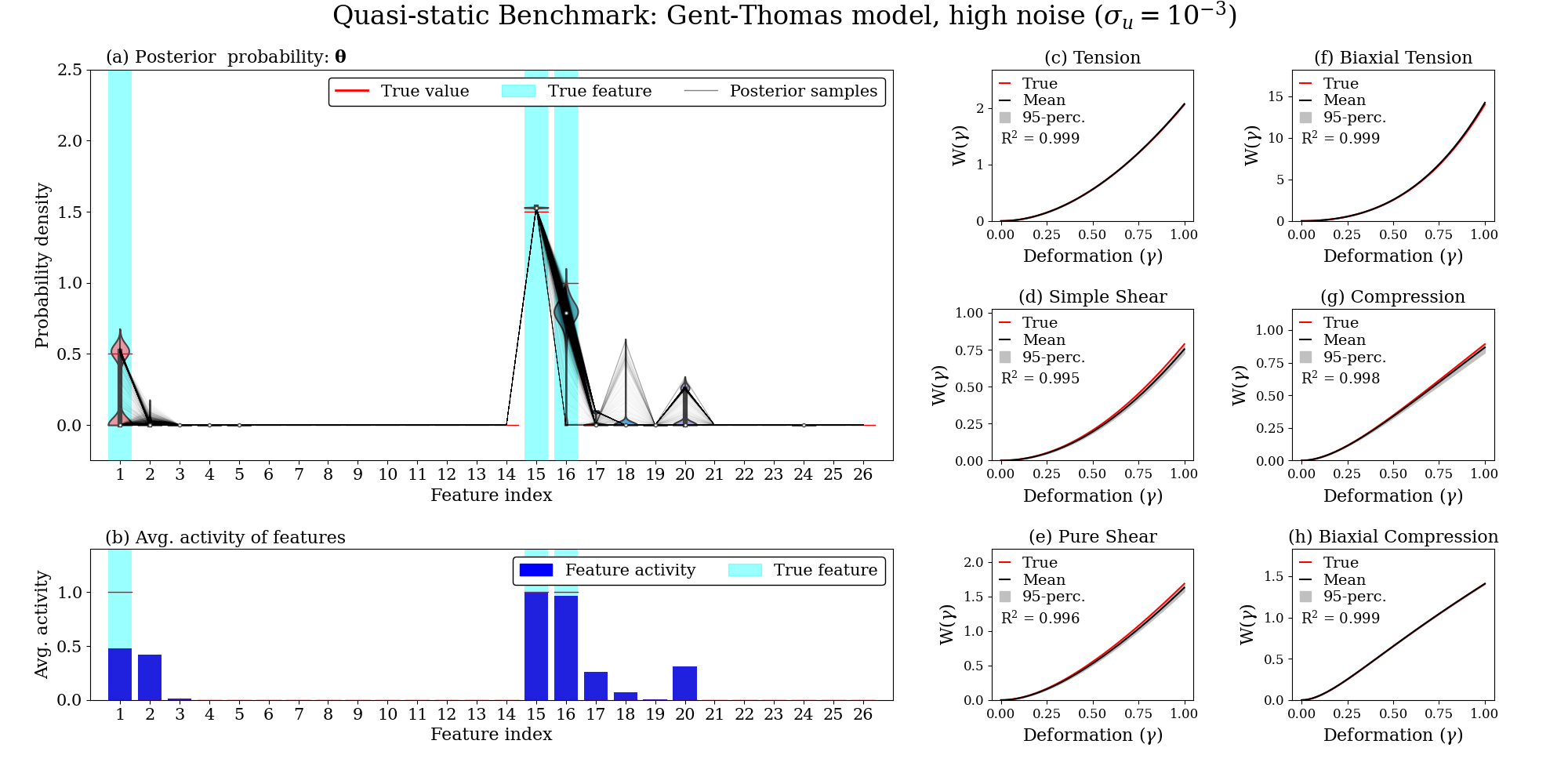}
	\end{subfigure}
    \caption{\customCaptionStatic{Gent-Thomas}{GT}}\label{fig:GT}
\end{figure}

\begin{figure}[ht]
	\centering
	\begin{subfigure}[ht]{\textwidth}
	    \centering
	    \includegraphics[width=\textwidth]{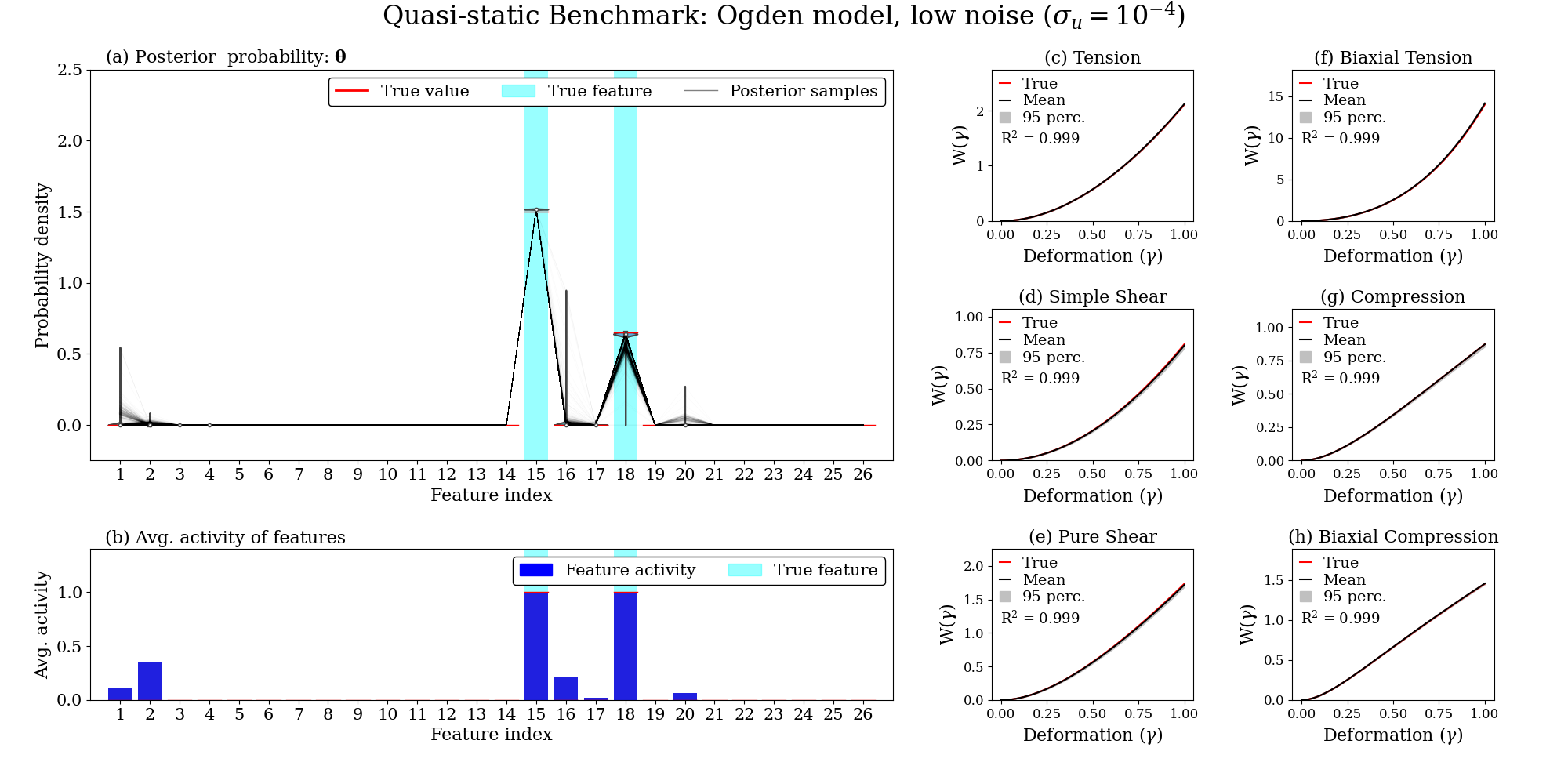}
	\end{subfigure}
	\hrule \vskip 15pt
	\begin{subfigure}[ht]{\textwidth}
	    \centering
	    \includegraphics[width=\textwidth]{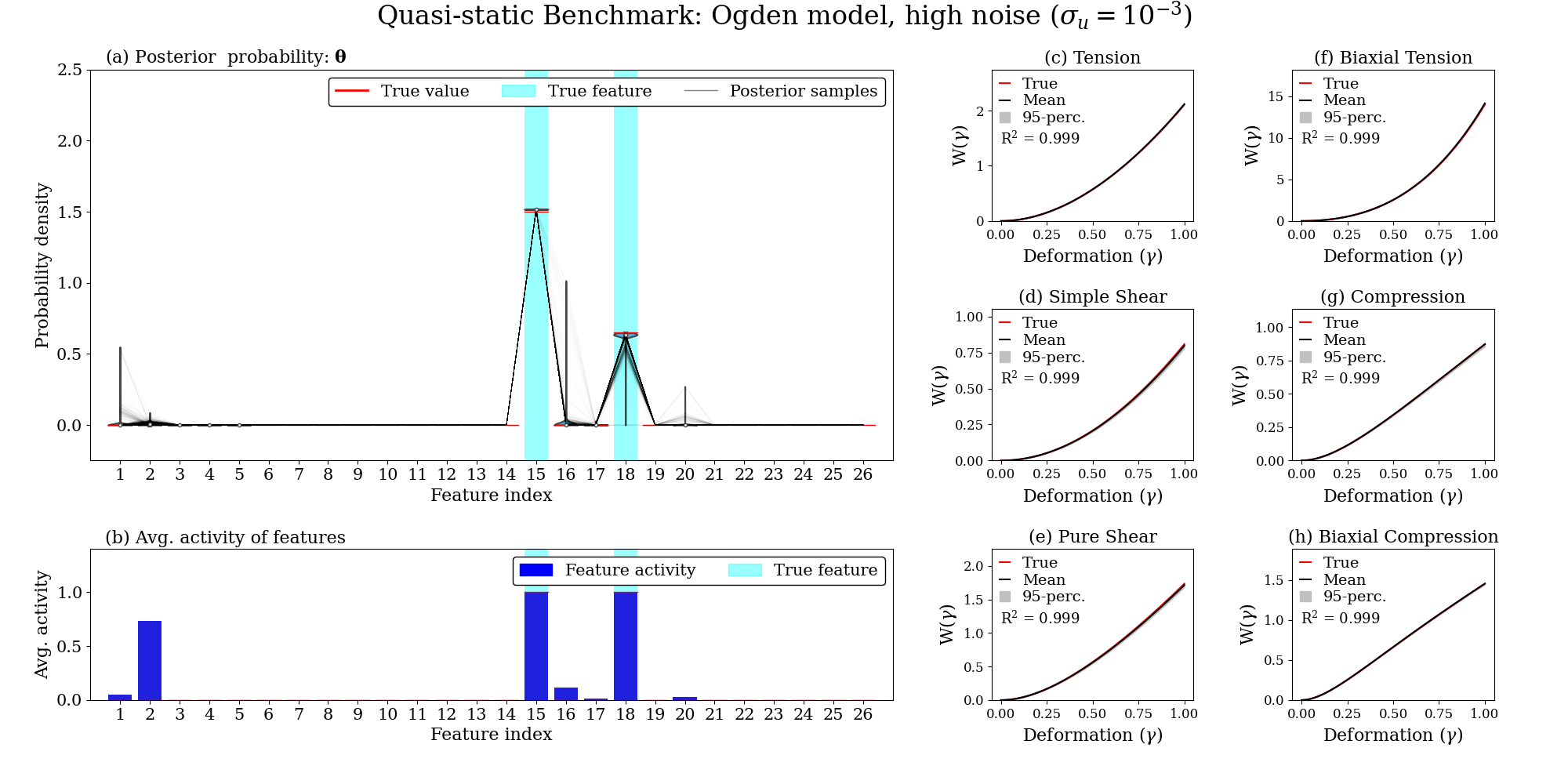}
	\end{subfigure}
    \caption{\customCaptionStatic{Ogden}{OG}}\label{fig:OG}
\end{figure}

\begin{figure}[ht]
	\centering
	\begin{subfigure}[ht]{\textwidth}
	    \centering
	    \includegraphics[width=\textwidth]{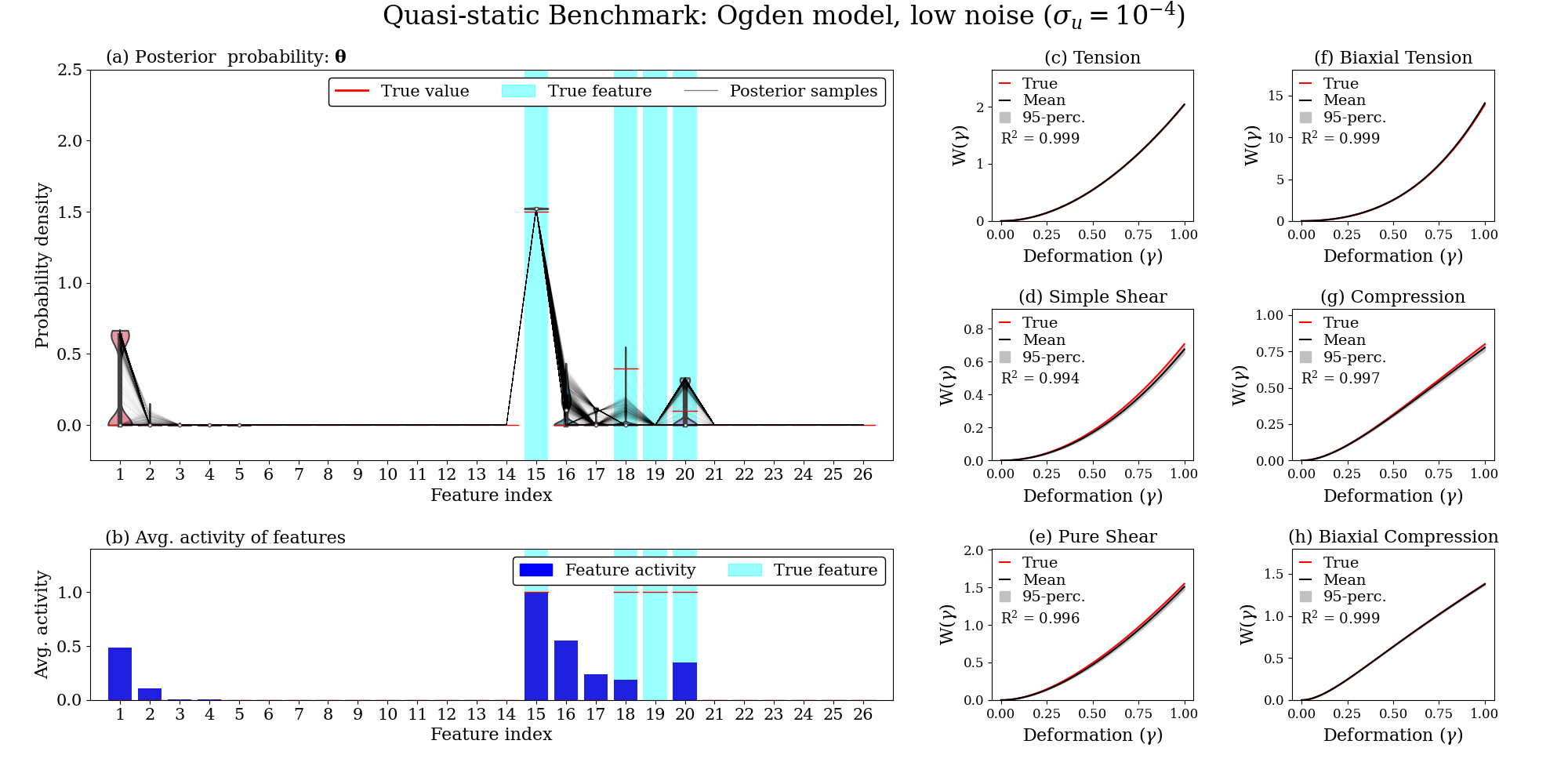}
	\end{subfigure}
	\hrule \vskip 15pt
	\begin{subfigure}[ht]{\textwidth}
	    \centering
	    \includegraphics[width=\textwidth]{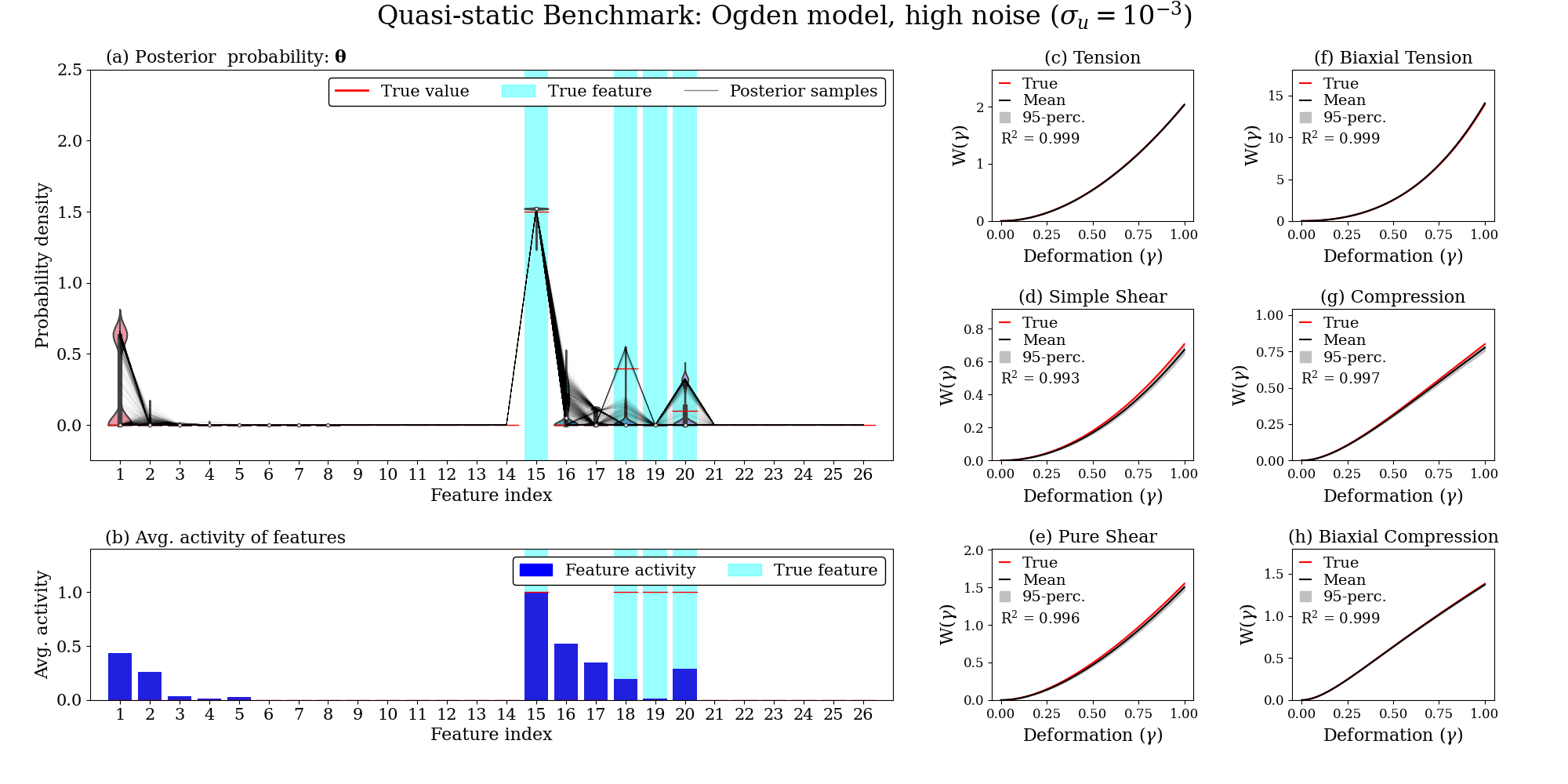}
	\end{subfigure}
    \caption{\customCaptionStatic{3-term Ogden}{OG3}}\label{fig:OG3}
\end{figure}

\begin{figure}[ht]
	\centering
	\begin{subfigure}[ht]{\textwidth}
	    \centering
	    \includegraphics[width=\textwidth]{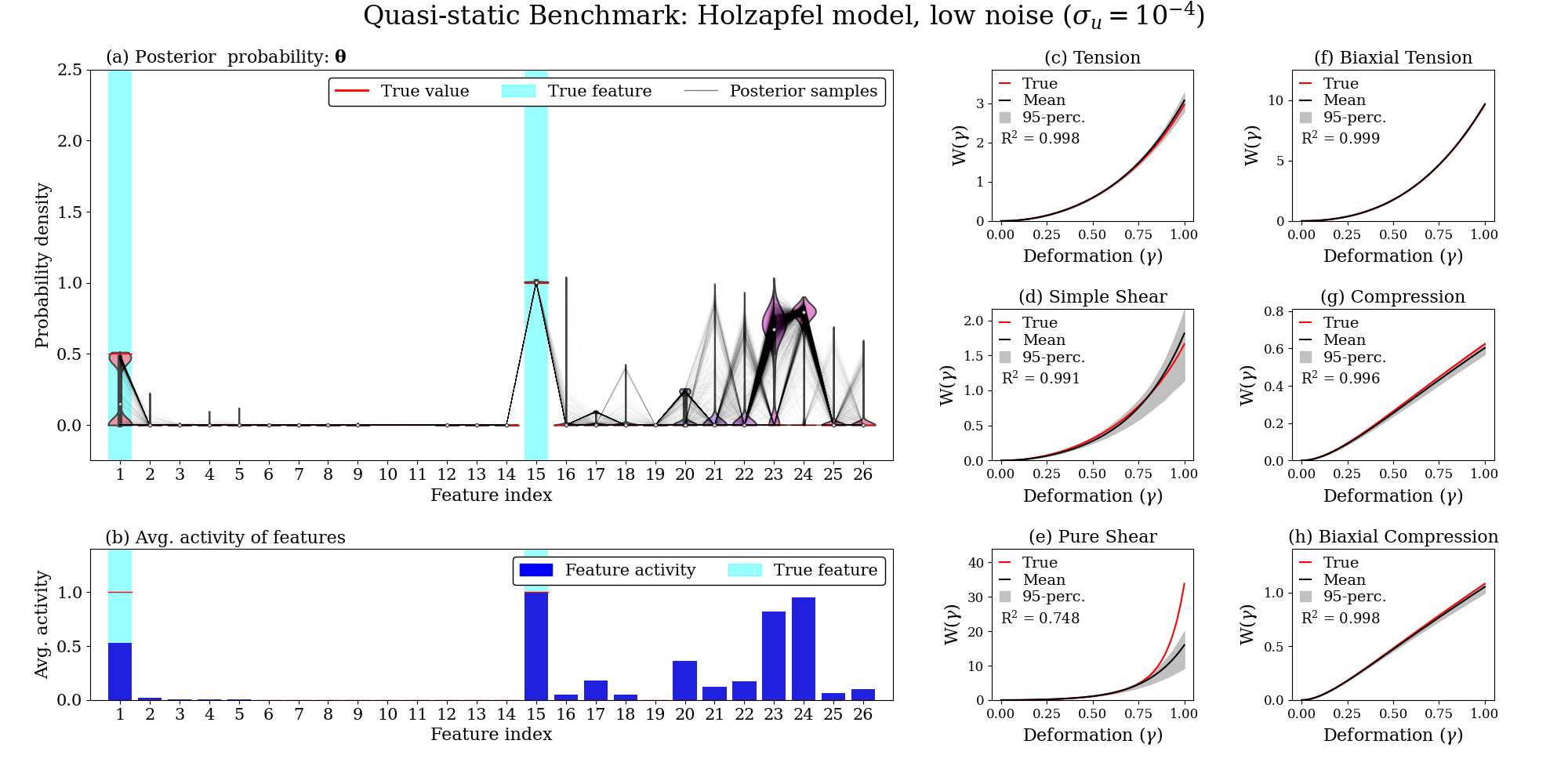}
	\end{subfigure}
	\hrule \vskip 15pt
	\begin{subfigure}[ht]{\textwidth}
	    \centering
	    \includegraphics[width=\textwidth]{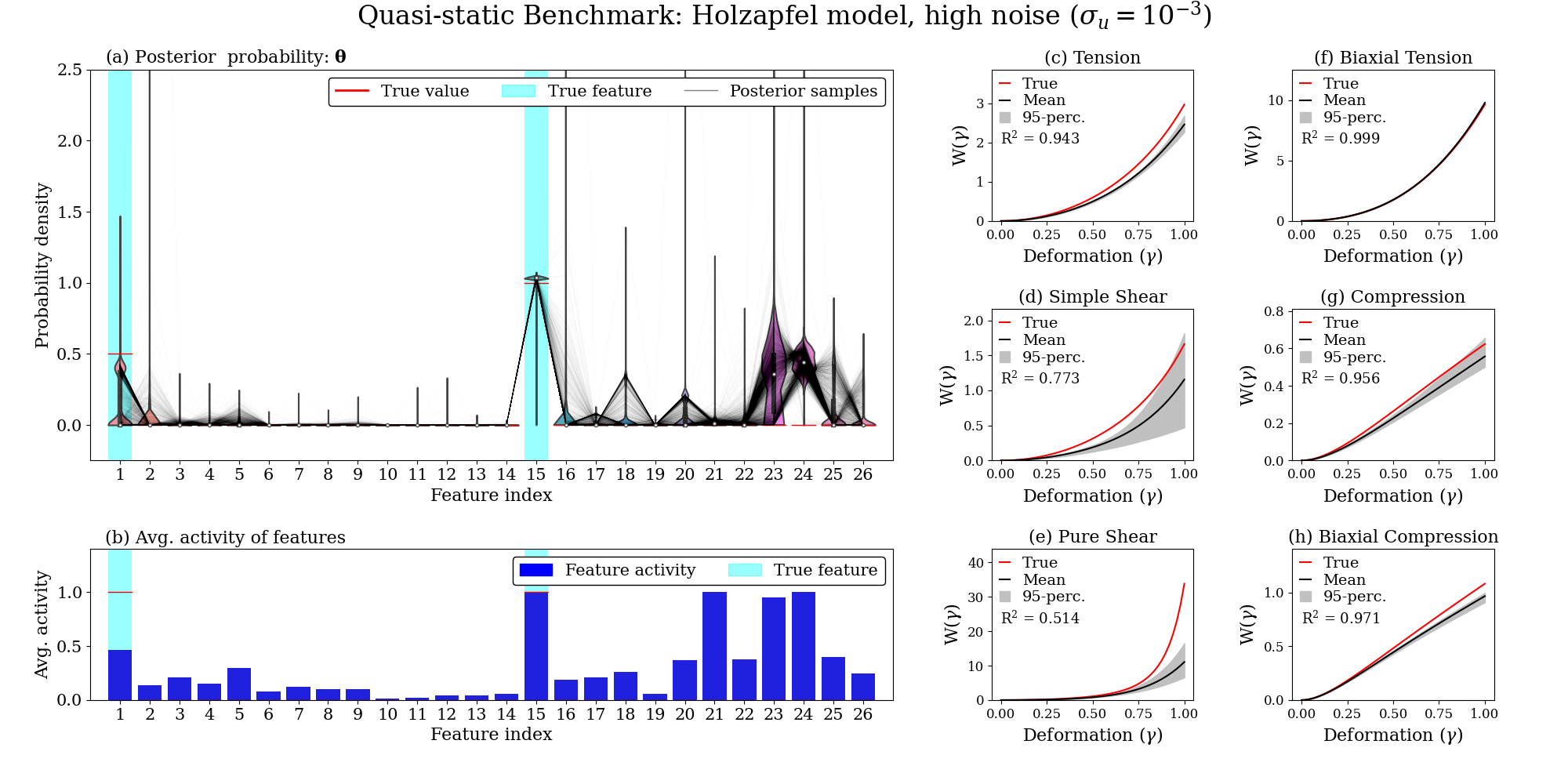}
	\end{subfigure}
    \caption{\customCaptionStatic{anisotropic Holzapfel}{HZ}}\label{fig:HZ}
\end{figure}

The discovered models in all the benchmarks are parsimonious -- containing a combination of only a few terms, as indicated by the zero average activity and/or zero-centered Dirac-delta-type marginal posterior distribution of several features. In general, the parsimony is lower for the high noise case, which is expected as, under increasing uncertainty, overfitting becomes more likely and the effect of any regularization weakens (in this case, parsimony {acts as} a regularization \citep{Kutz2022} via the spike-slab prior). \RR{The results shown in Figures \ref{fig:NH2}-\ref{fig:HZ} clearly indicate that the Bayesian-EUCLID framework is able to detect the presence/absence of anisotropy and identify the relevant volumetric and deviatoric features. Although the number of possible models is exceedingly large (around $2^{26}$ for 26 features in the library), enforcement of parsimony enables the method to discover the simplest possible model that captures the material's behavior. This ability to select and discover relevant material models distinguishes the EUCLID framework from other approaches that seek to characterize materials by assuming energy models a priori.}

\subsubsection{Model multi-modality}

\begin{figure}[ht]
	\centering
	\begin{subfigure}[ht]{\textwidth}
	    \centering
	    \includegraphics[width=\textwidth]{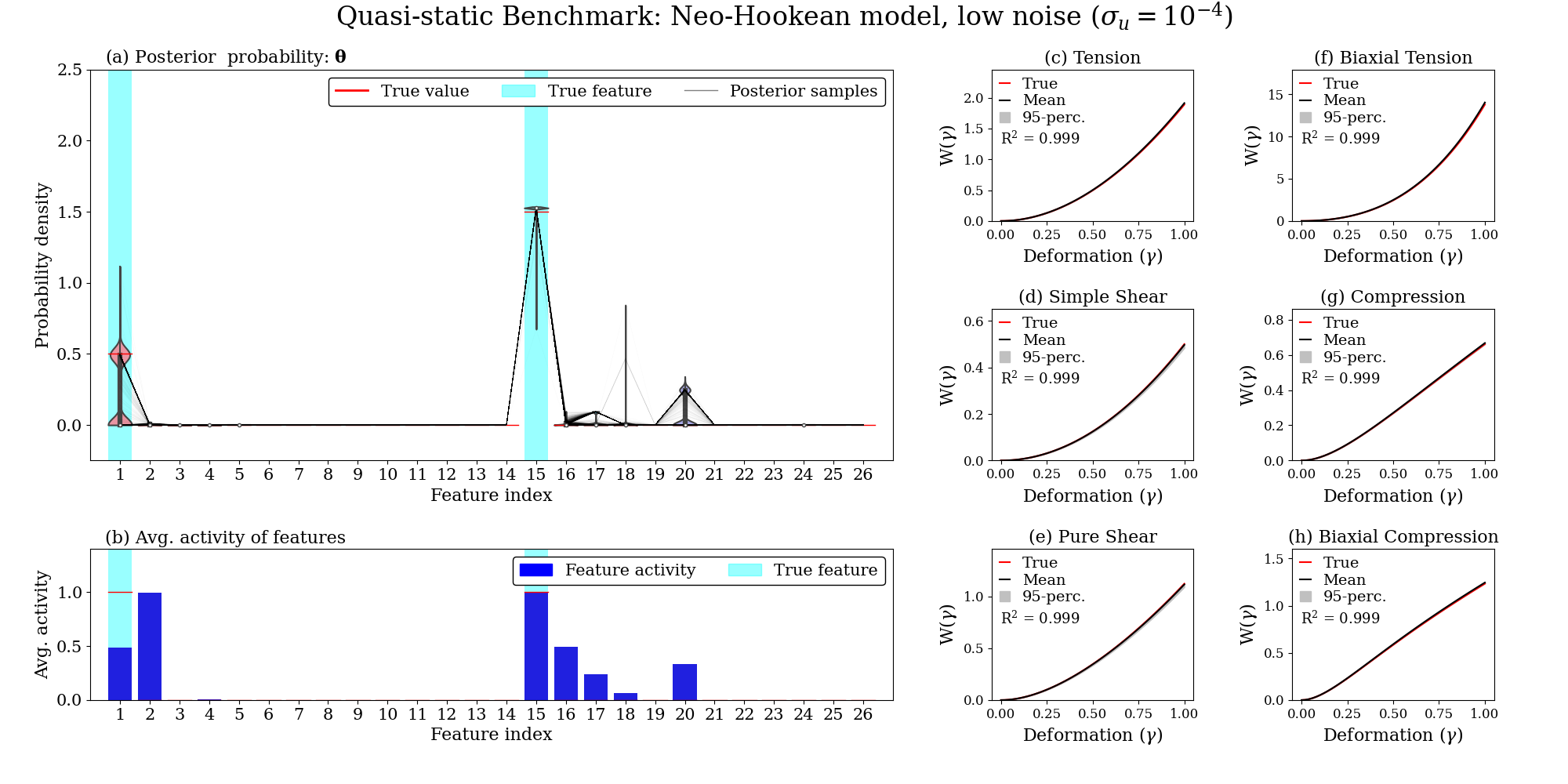}
	\end{subfigure}
	\hrule \vskip 15pt
	\begin{subfigure}[ht]{\textwidth}
	    \centering
	    \includegraphics[width=\textwidth]{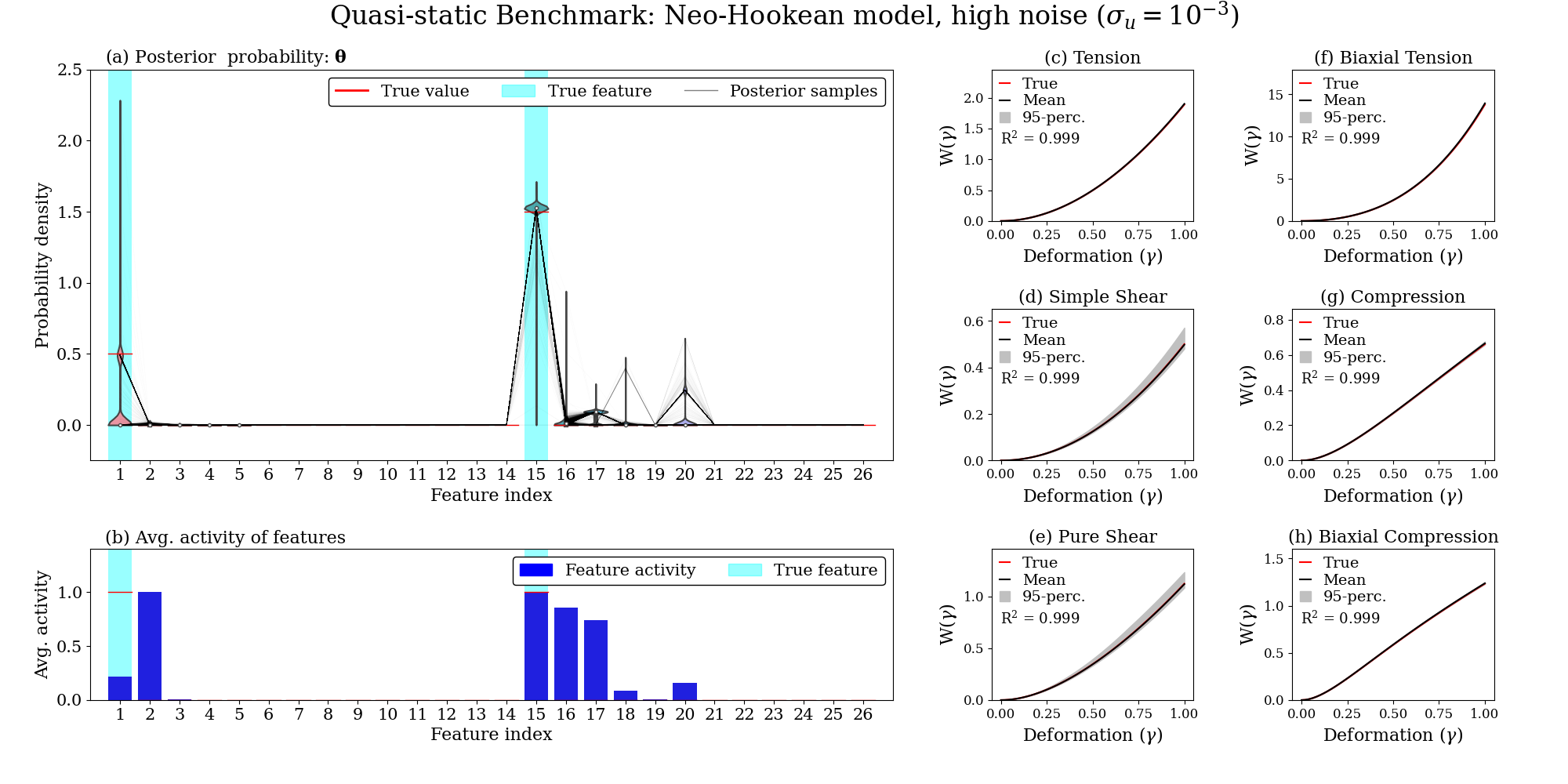}
	\end{subfigure}
    \caption{\customCaptionStatic{Neo-Hookean}{NH2}}\label{fig:NH2}
\end{figure}

\begin{figure}[ht]
	\centering
	\begin{subfigure}[ht]{\textwidth}
	    \centering
	    \includegraphics[width=\textwidth]{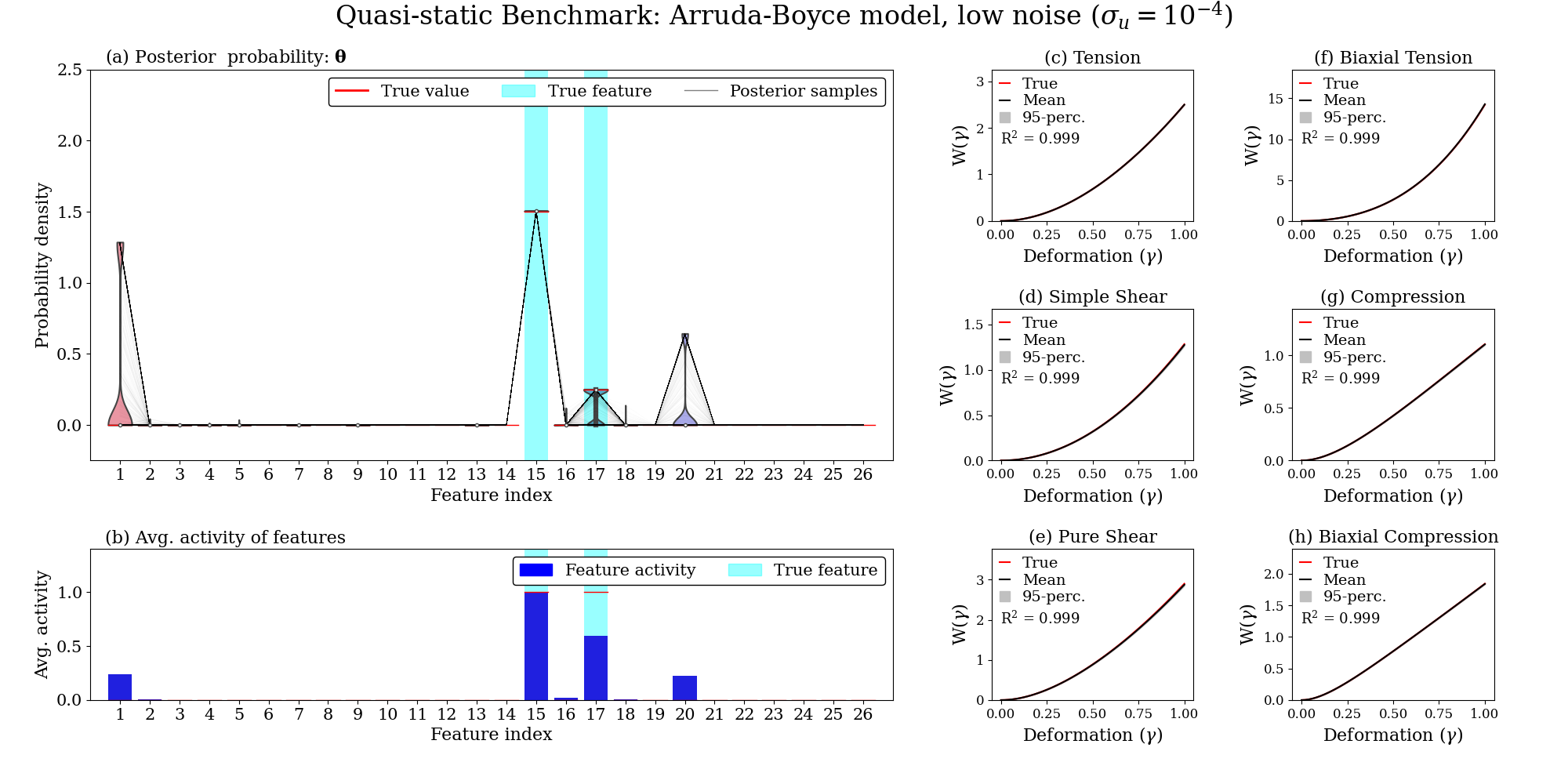}
	\end{subfigure}
	\hrule \vskip 15pt
	\begin{subfigure}[ht]{\textwidth}
	    \centering
	    \includegraphics[width=\textwidth]{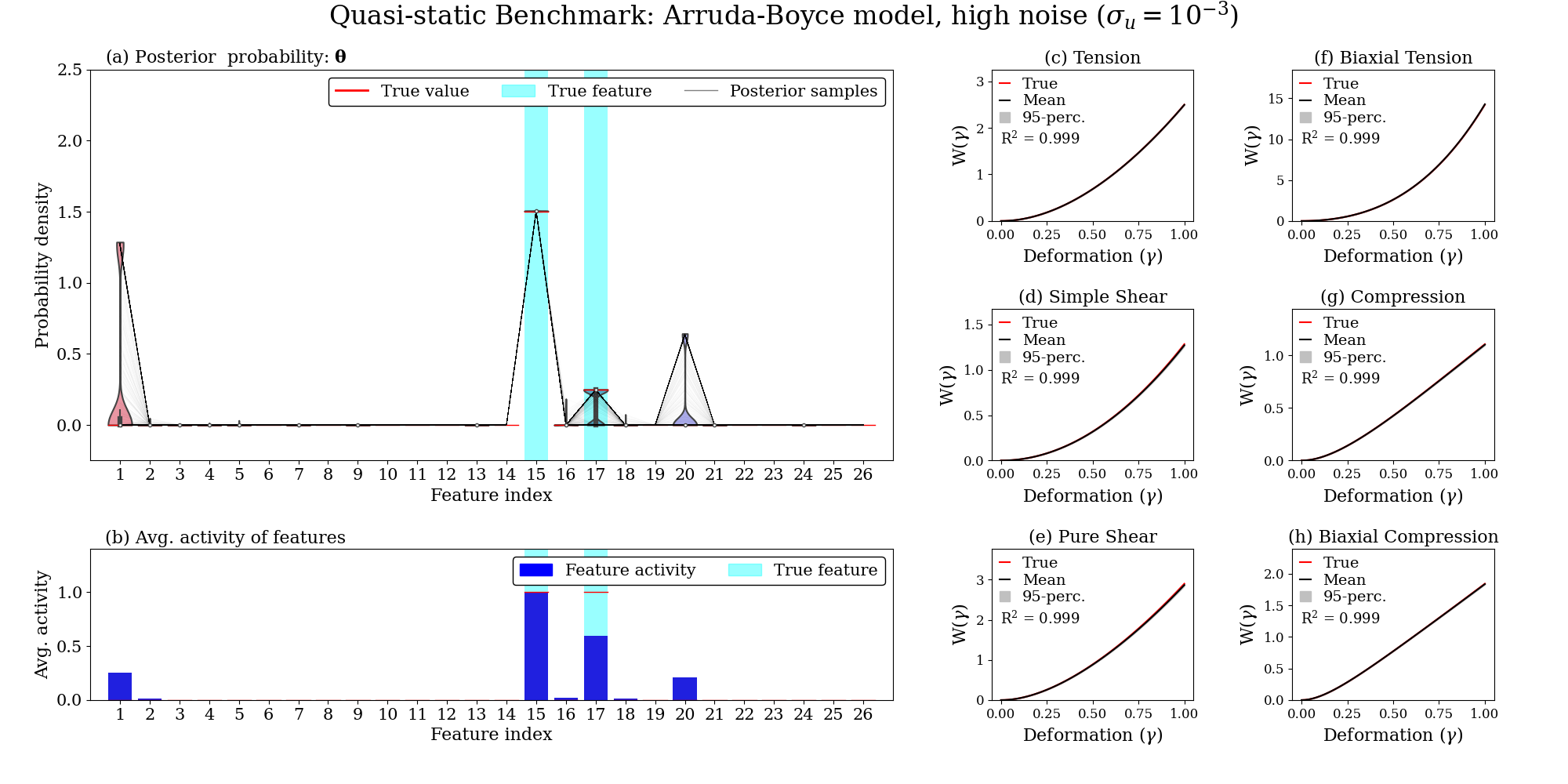}
	\end{subfigure}
    \caption{\customCaptionStatic{Arruda-Boyce}{AB}}\label{fig:AB}
\end{figure}

The marginal posterior {distributions} of the active features are multi-modal, indicating that alternative constitutive models different from the ground truth are also discovered. {E.g., in the Neo-Hookean benchmark (\figurename\ref{fig:NH2}), three dominant models are discovered where the features: $(\Tilde{I}_1-3)$ (Neo-Hookean feature with index: $1$), $\text{AB}$ (Arruda-Boyce feature with index: $17$), and $\text{OG}_3$ (Ogden feature with index: $20$) are primarily active with mutual exclusivity. The same features are also activated mutually exclusively in the Arruda-Boyce benchmark (\figurename\ref{fig:AB}).} In addition, the average activities are also informative of the likelihood of each mode/model. Although the predicted feature coefficients for the alternative models are different from the ground truth, the good agreement between the predicted and true energy densities across all the benchmarks suggests that the alternative models are indeed capable of accurately explaining the observed data and, in general, representing the true constitutive response of the material.

\begin{figure}[ht]
	\centering
	\begin{subfigure}[ht]{\textwidth}
	    \centering
	    \includegraphics[width=\textwidth]{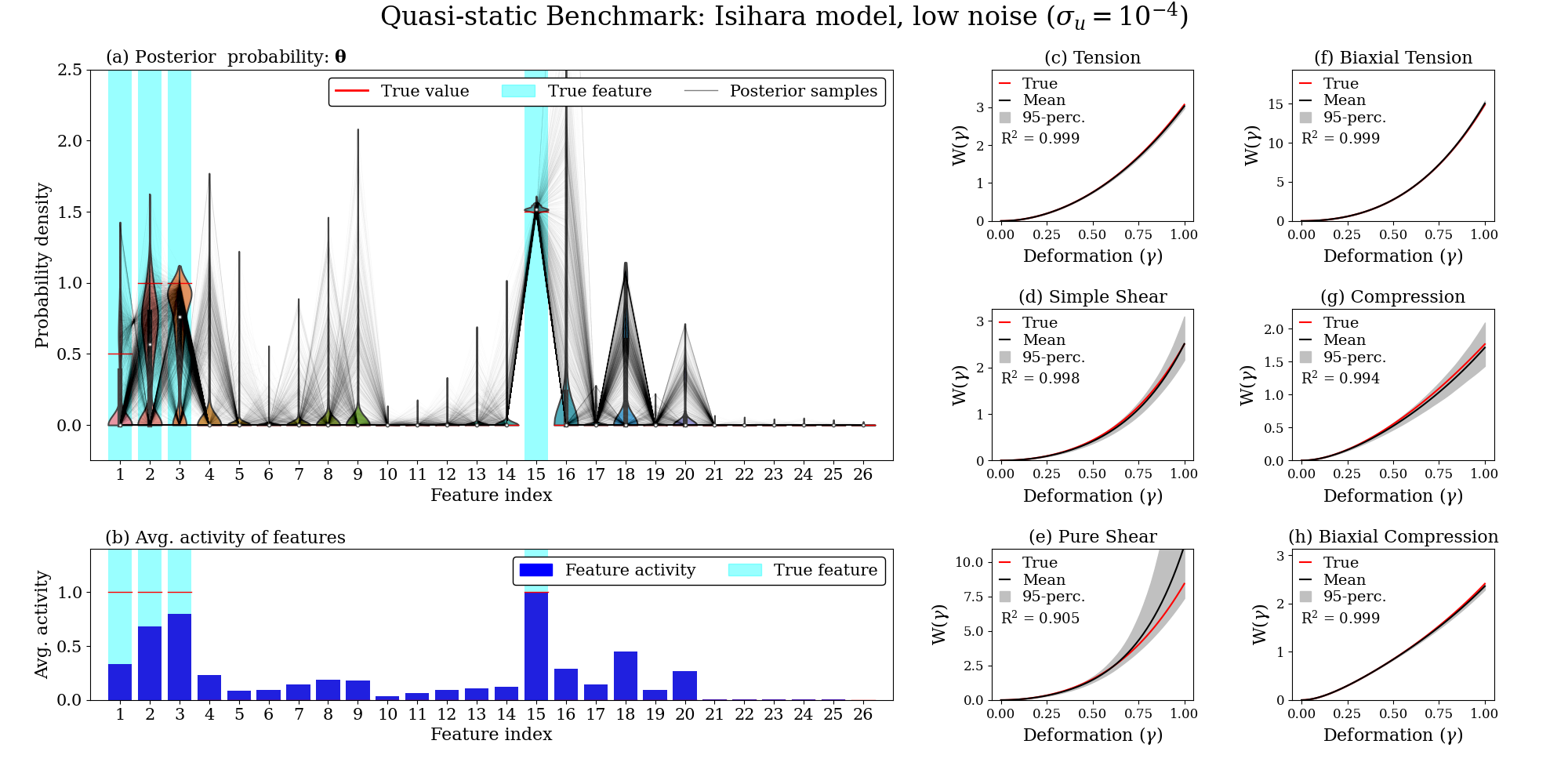}
	\end{subfigure}
	\hrule \vskip 15pt
	\begin{subfigure}[ht]{\textwidth}
	    \centering
	    \includegraphics[width=\textwidth]{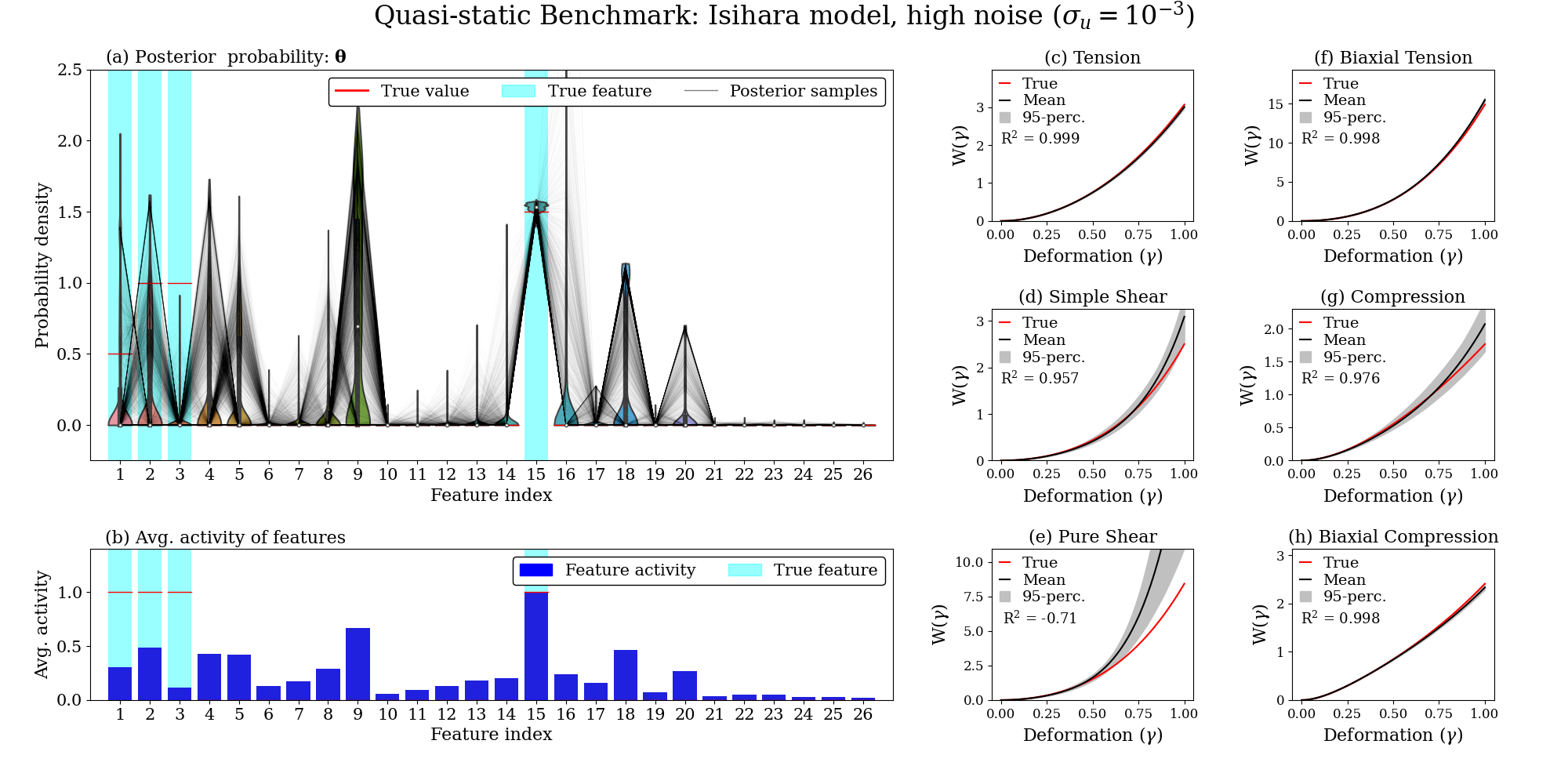}
	\end{subfigure}
    \caption{\customCaptionStatic{Isihara}{IH}}\label{fig:IH}
\end{figure}

\begin{figure}[ht]
	\centering
	\begin{subfigure}[ht]{\textwidth}
	    \centering
	    \includegraphics[width=\textwidth]{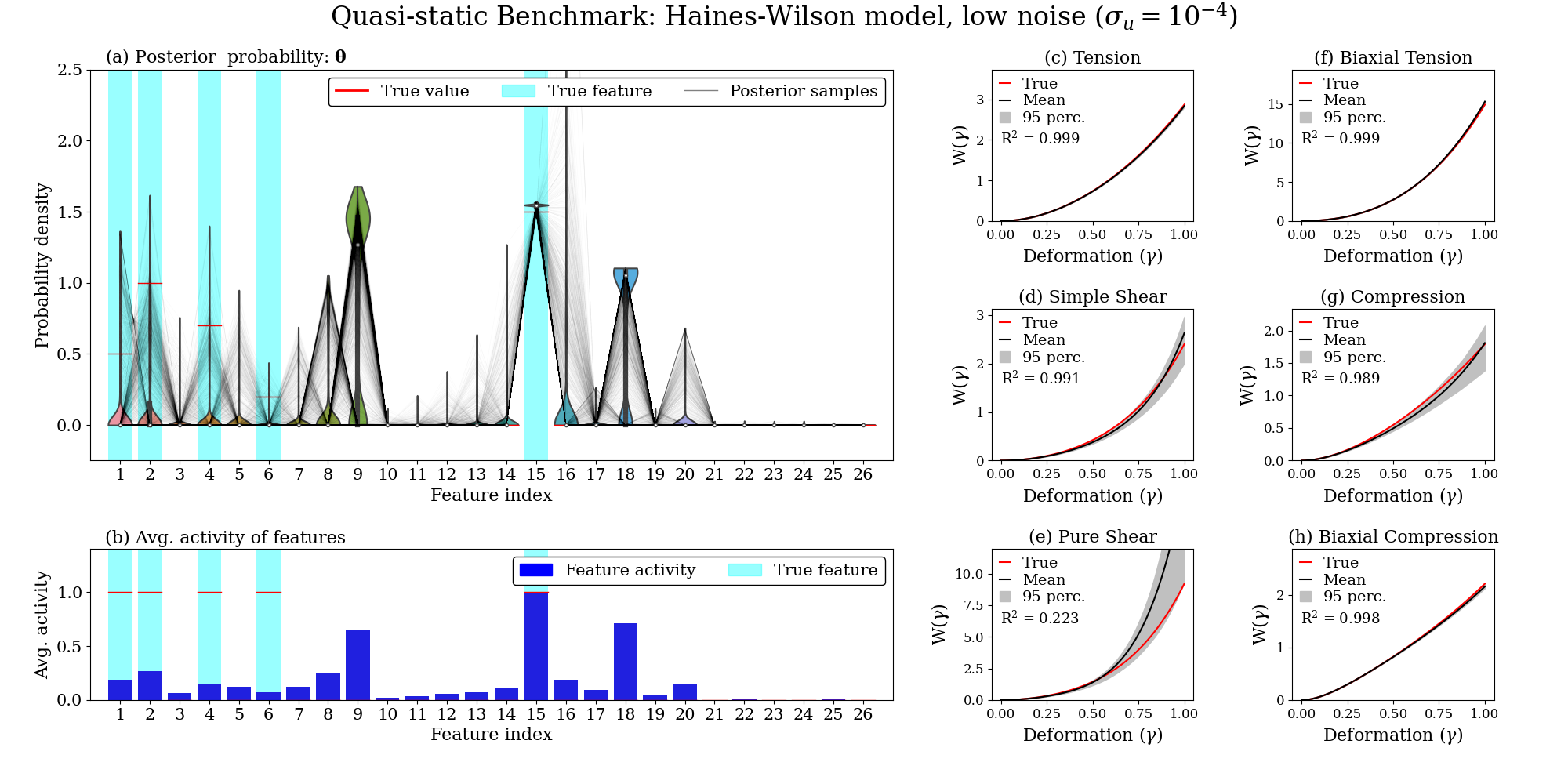}
	\end{subfigure}
	\hrule \vskip 15pt
	\begin{subfigure}[ht]{\textwidth}
	    \centering
	    \includegraphics[width=\textwidth]{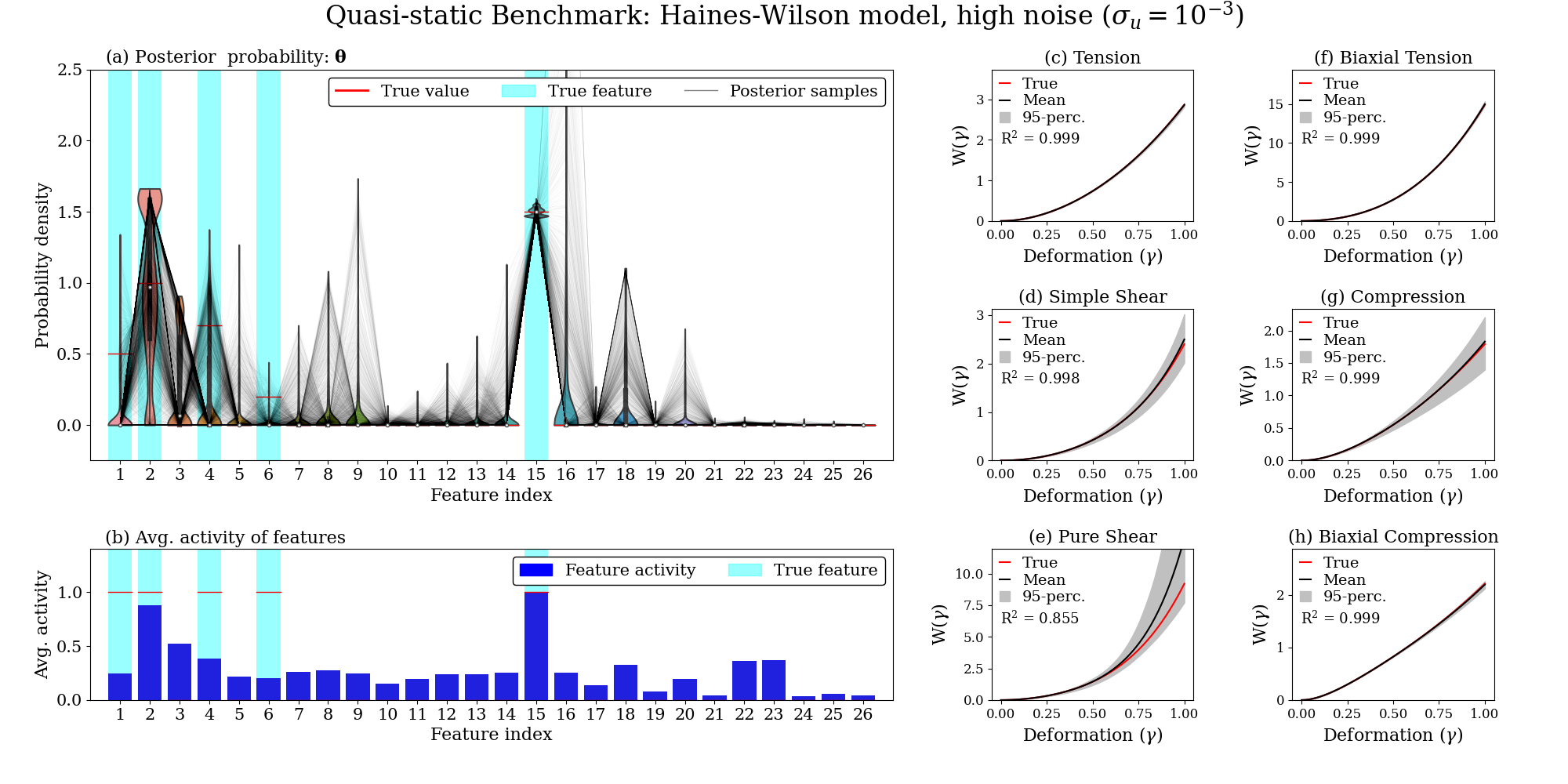}
	\end{subfigure}
    \caption{\customCaptionStatic{Haines-Wilson}{HW}}\label{fig:HW}
\end{figure}

The multi-modality is attributed to {two main} sources. \textit{(i)} Given the observed data, two or more features may have high correlation. In the limit when two features are exactly equal, infinitely many solutions are admissible. The high feature correlations combined with noisy and limited data give rise to multi-modal solutions, as observed in the following representative examples.
\begin{itemize}
    \item The Neo-Hookean feature is a linearization of the Ogden and Arruda-Boyce features. Therefore, all of them are highly correlated for small deformations, which leads to multi-modality in Figures \ref{fig:NH2} and \ref{fig:AB}.
    \item In the Isihara and Haines-Wilson benchmarks (Figures \ref{fig:IH} and \ref{fig:HW}), the true generalized Mooney-Rivlin polynomial features are highly correlated, which also leads to highly multi-modal solutions.
\end{itemize}
\textit{(ii)} In the current set of benchmarks with $n_f=26$, the number of unique combinations of the features, hence the number of possible unique models, is more than $67$~million. Such a big model space is inherently likely to admit more than one model that can satisfy the small amount of observed data, especially in the unsupervised setting where the stress labels are absent. Therefore, it is important to consider all the modes/models until new data are available {to further refine the model estimate.} The new data may come from specially designed specimen geometries with diverse strain distributions that minimize the feature correlations or be obtained from labeled stress-strain pairs via simpler uni-/bi-axial tension{/compression, bending or torsion} tests.

\subsubsection{Model generalization \RR{under} epistemic uncertainties}
\label{sec:EPIS}

\begin{figure}[ht]
	\centering
	\begin{subfigure}[ht]{\textwidth}
	    \centering
	    \includegraphics[width=\textwidth]{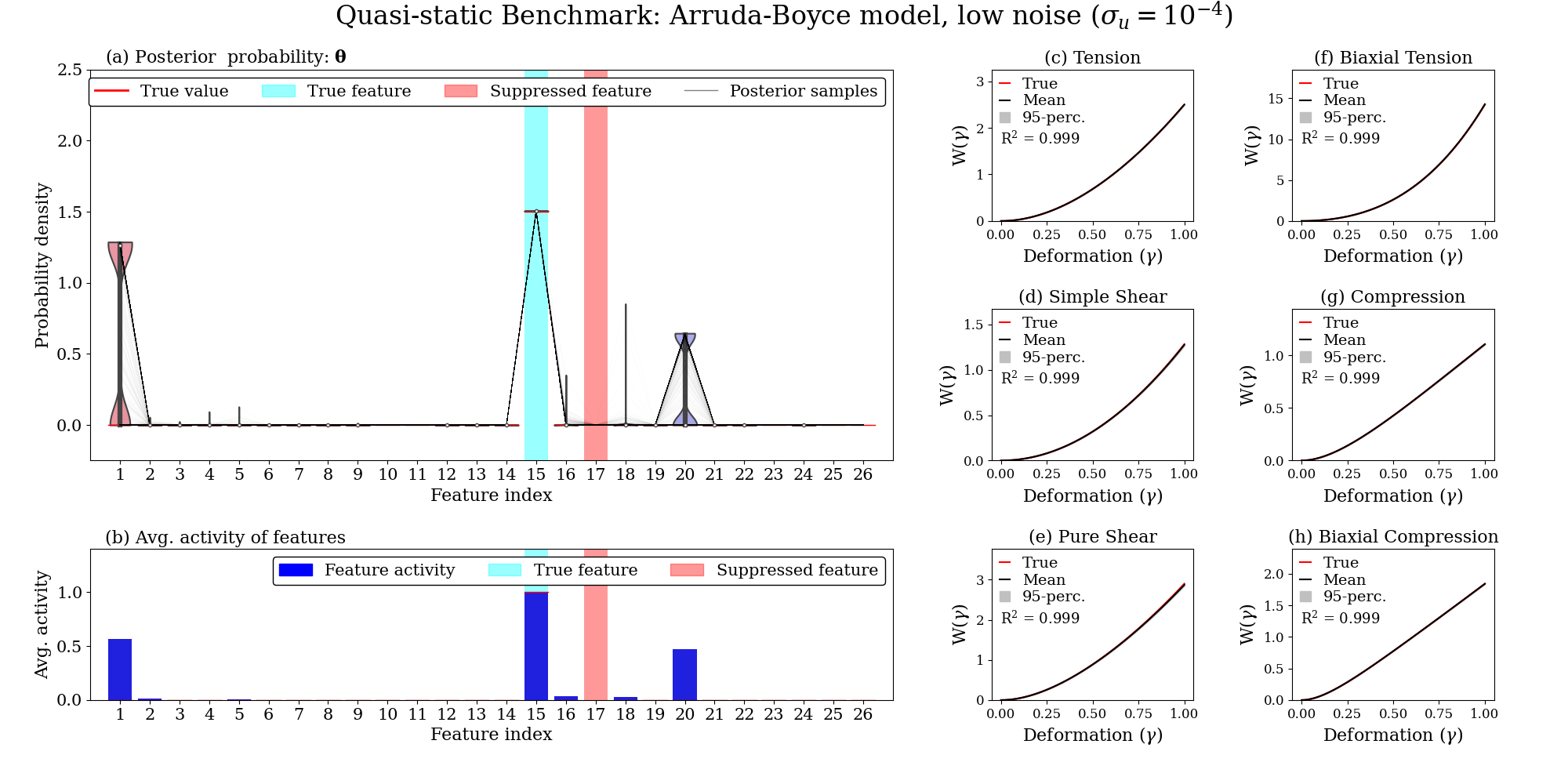}
	\end{subfigure}
	\hrule \vskip 15pt
	\begin{subfigure}[ht]{\textwidth}
	    \centering
	    \includegraphics[width=\textwidth]{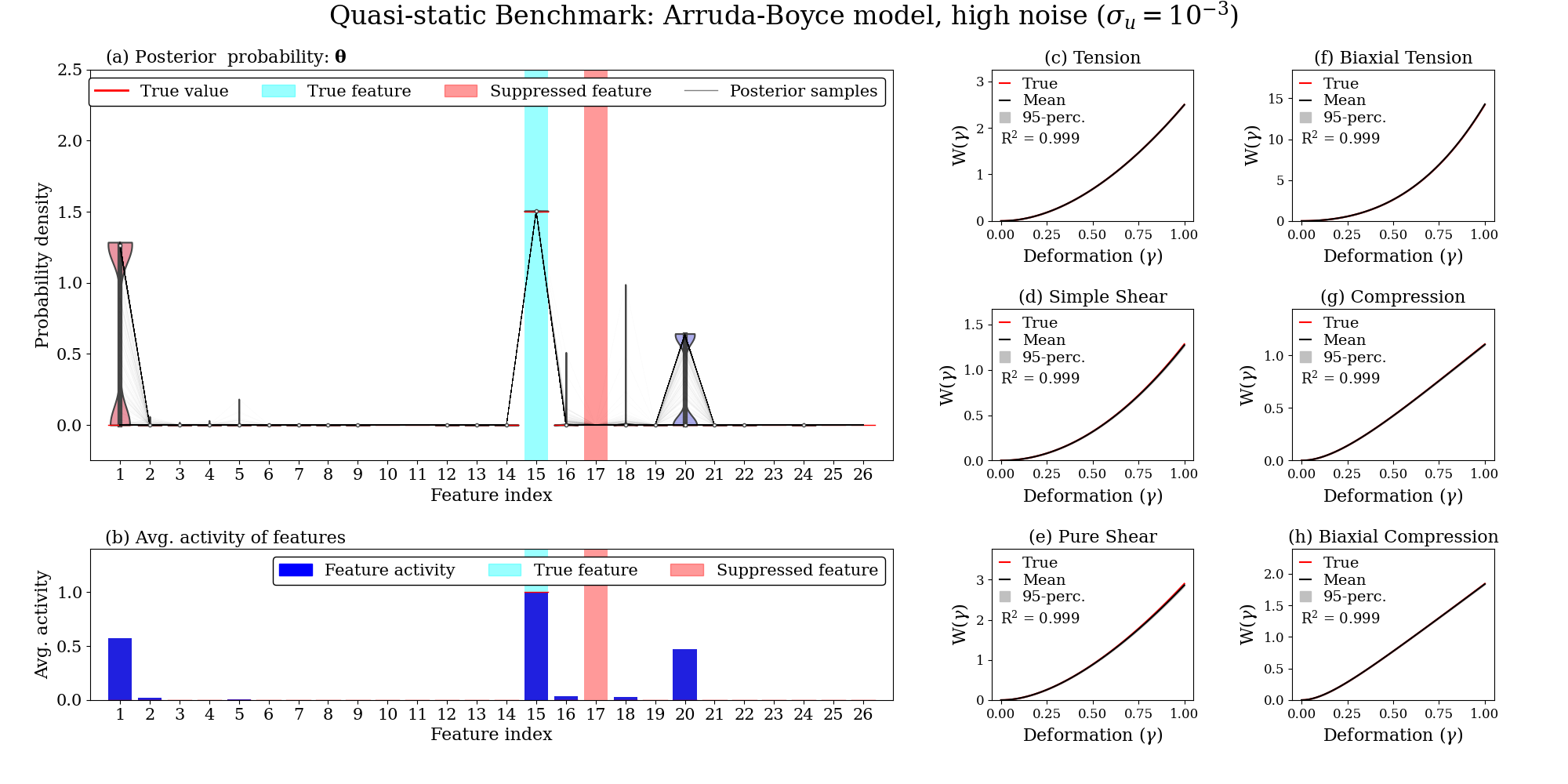}
	\end{subfigure}
    \caption{\customCaptionStaticSupp{Arruda-Boyce}{AB}}\label{fig:AB-supp}
\end{figure}

\begin{figure}[ht]
	\centering
	\begin{subfigure}[ht]{\textwidth}
	    \centering
	    \includegraphics[width=\textwidth]{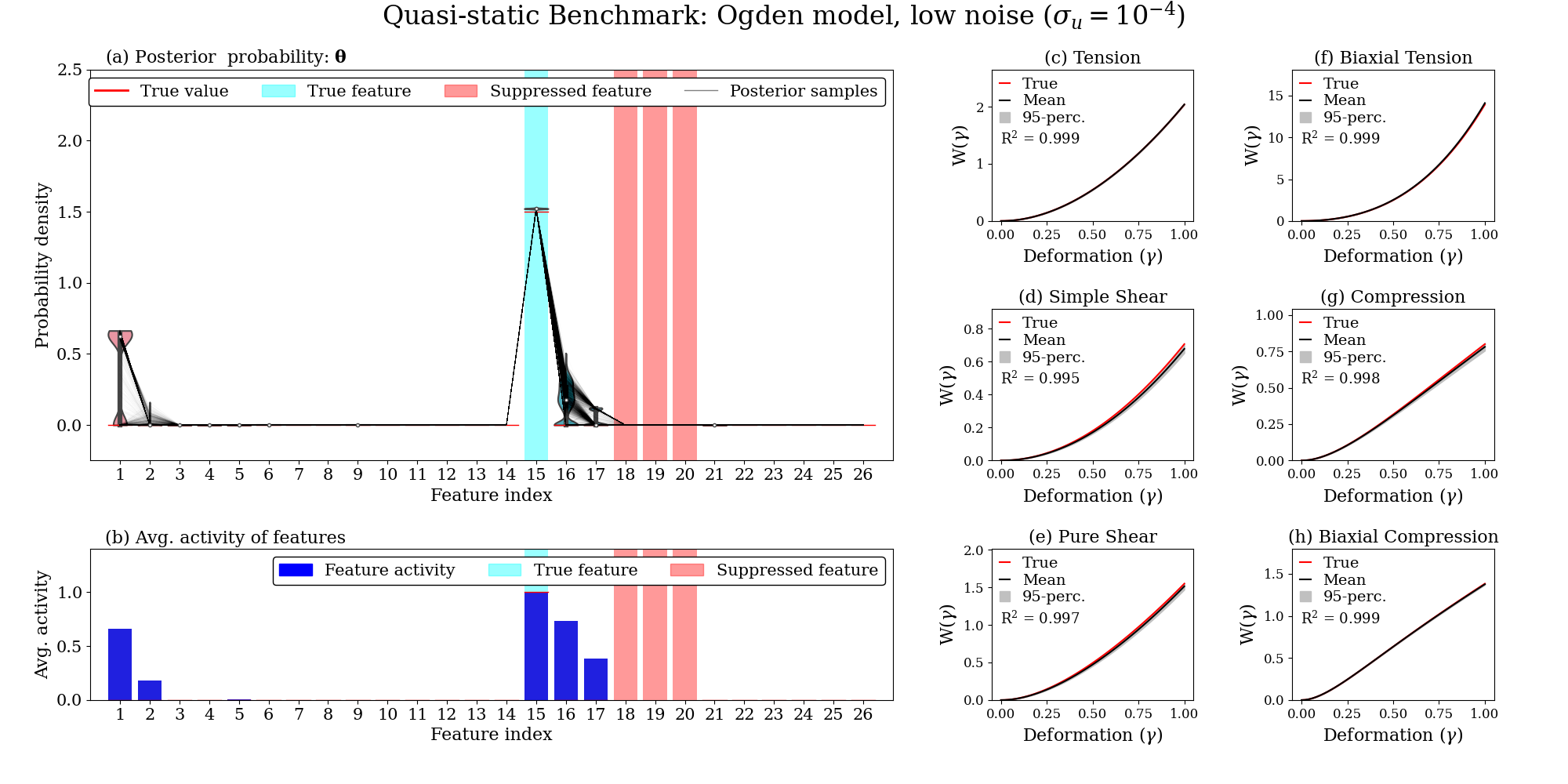}
	\end{subfigure}
	\hrule \vskip 15pt
	\begin{subfigure}[ht]{\textwidth}
	    \centering
	    \includegraphics[width=\textwidth]{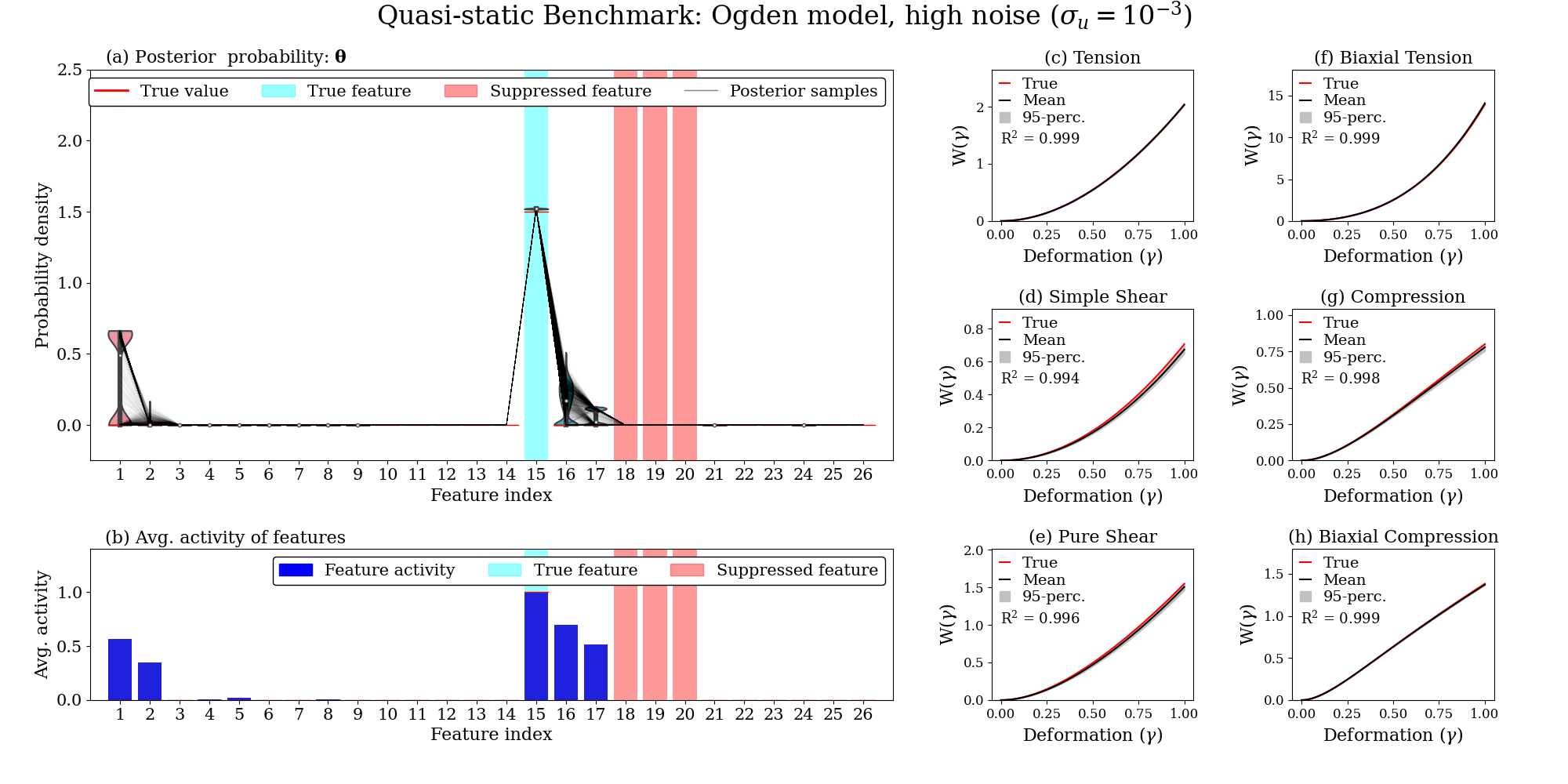}
	\end{subfigure}
    \caption{\customCaptionStaticSupp{3-term Ogden}{OG3}}\label{fig:OG3-supp}
\end{figure}

Epistemic uncertainties arise due to the lack of knowledge about the best model \citep{hullermeier_aleatoric_2021}. In the context of this work, this {translates into} missing prior knowledge in {the creation of} the feature library $\bfQ$. Specifically, the model discovery should be generalizable even when true features are missing in the library.
To this end, we intentionally suppress certain true features in the feature library of Table~\ref{tab:features} and test the generalization capability of the discovered constitutive models. We consider two cases:
\begin{enumerate}
    \item Arruda-Boyce benchmark \eqref{eq:AB} with the true Arruda-Boyce feature (index 17 in Table~\ref{tab:features}) suppressed,
    \item 3-term Ogden benchmark \eqref{eq:OG3} with the true Ogden features (indices 18, 19, and 20 in Table~\ref{tab:features}) suppressed.
\end{enumerate}
The corresponding model discovery results are summarized in Figures~\ref{fig:AB-supp} and \ref{fig:OG3-supp} with the suppressed features highlighted in red background in both the marginal posterior and average activity plots. In the first case, the Neo-Hookean and Ogden features automatically become active to {compensate} for the missing Arruda-Boyce feature. In the second case, the missing Ogden features are  replaced by the Neo-Hookean, Gent-Thomas, and Arruda-Boyce features. In both cases, the predicted energy densities are accurate with highly confident percentile bounds, which demonstrates the robustness of the proposed method under epistemic uncertainties. Additionally, the Holzapfel benchmark \eqref{eq:HZ} is another evidence of the generalization capability in the case of anisotropy, as the true anisotropic Holzapfel features are not part of the feature library in \eqref{eq:library} and Table~\ref{tab:features}. \RR{In \ref{sec:appEPIS}, we further test the generalization capability of the Bayesian-EUCLID framework under epistemic uncertainties such as incorrect assumptions about the fiber directions or the absence of anisotropy features in the feature library.}

Compared to the {deterministic} method of \citet{flaschel_unsupervised_2021}, the  proposed Bayesian method not only enables  multi-modal solutions and uncertainty quantification, but also provides significantly higher data and computational efficiency. The Bayesian method only requires $n_\text{free}=100$ data points per snapshot and computing time on the order of 10-20 minutes on a single average modern processor. In contrast, the previous method by \citet{flaschel_unsupervised_2021} required $n_\text{free}\sim 126,000$ data points (approximately $63,000$ nodes with two degrees of freedom each) per snapshot and  computing time on the order of 10 minutes with 200 parallel processors for similar accuracy. The speedup is partly enabled by the probabilistic framework, that looks at the entire solution space altogether, and partly by automatically enforcing physical admissibility \textit{a priori} as opposed to iteratively searching for models that  empirically satisfy convexity (by checking monotonicity of energy density on some deformation paths).

\begin{figure}[ht]
	\centering
    \includegraphics[width=\textwidth]{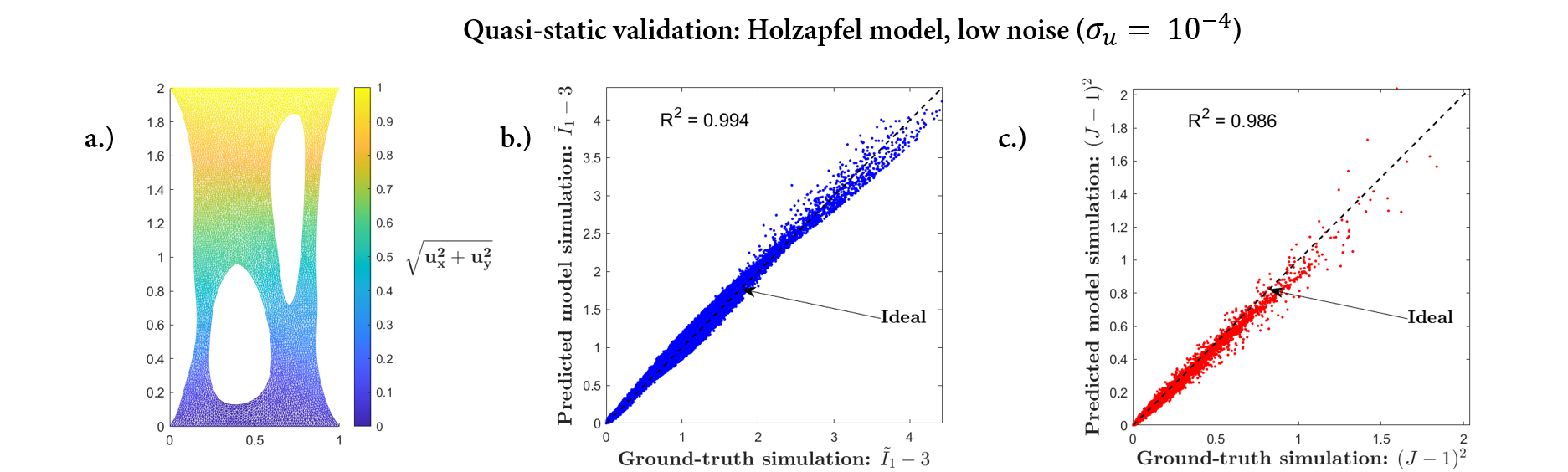}
    \caption{\RR{Quasi-static validation for the hidden Holzapfel benchmark model \eqref{eq:HZ} -- comparing simulation results from the ground-truth model and the mean of the energy density models predicted using the low-noise ($\sigma_u = 10^{-4}$) displacement data. (a) Deformed geometry at $\varphi=1$ obtained using the mean of the predicted models. (b) Predicted vs.~true strain invariant $(\Tilde{I}_1-3)$ across all quadrature points and loadsteps. (c) Predicted vs.~true strain invariant $(J-1)^2$ across all quadrature points and loadsteps.}}\label{fig:ValidLN}
    \vskip 15pt
	\centering
    \includegraphics[width=\textwidth]{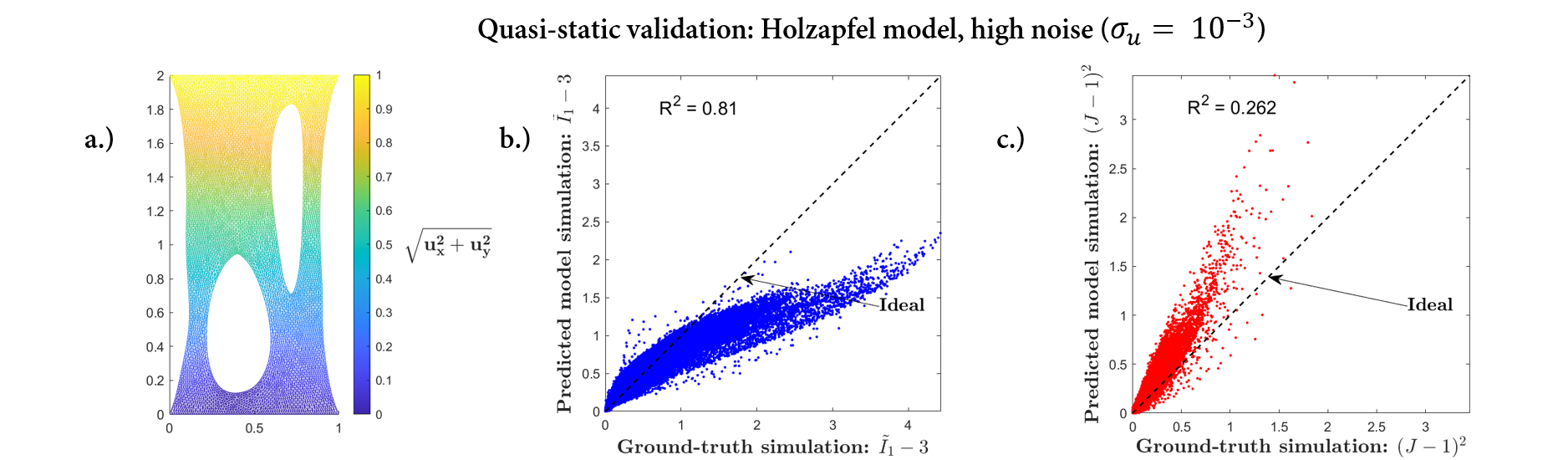}
    \caption{\RR{Quasi-static validation for the hidden Holzapfel benchmark model \eqref{eq:HZ} -- comparing simulation results from the ground-truth model and the mean of the energy density models predicted using the high-noise ($\sigma_u = 10^{-3}$) displacement data. (a) Deformed geometry at $\varphi=1$ obtained using the mean of the predicted models. (b) Predicted vs.~true strain invariant $(\Tilde{I}_1-3)$ across all quadrature points and loadsteps. (c) Predicted vs.~true strain invariant $(J-1)^2$ across all quadrature points and loadsteps.}}\label{fig:ValidHN}
\end{figure}

\subsection{Model discovery with dynamic data}

\begin{figure}[ht]
	\centering
	\begin{subfigure}[ht]{\textwidth}
	    \centering
	    \includegraphics[width=\textwidth]{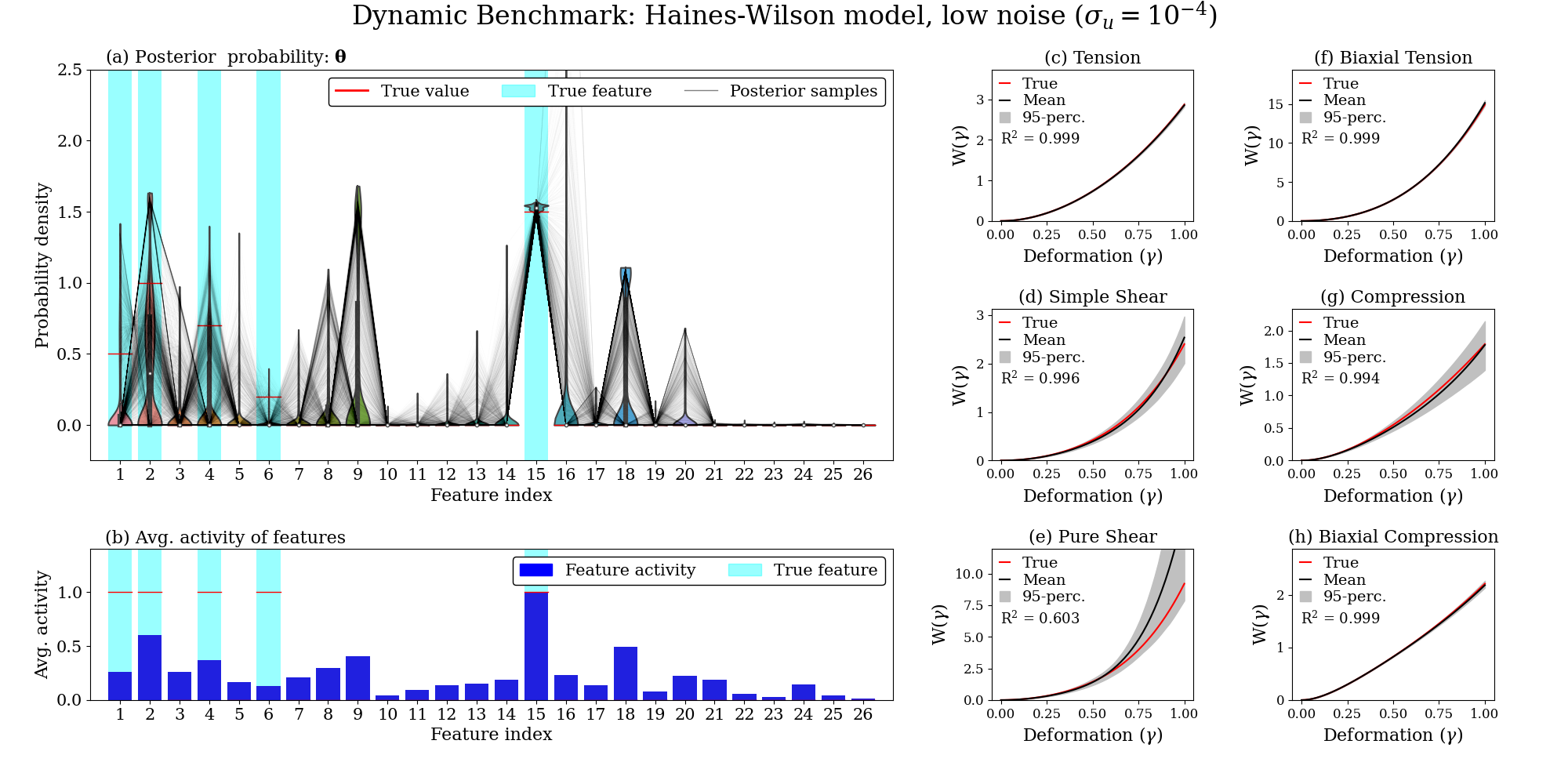}
	\end{subfigure}
	\hrule \vskip 15pt
	\begin{subfigure}[ht]{\textwidth}
	    \centering
	    \includegraphics[width=\textwidth]{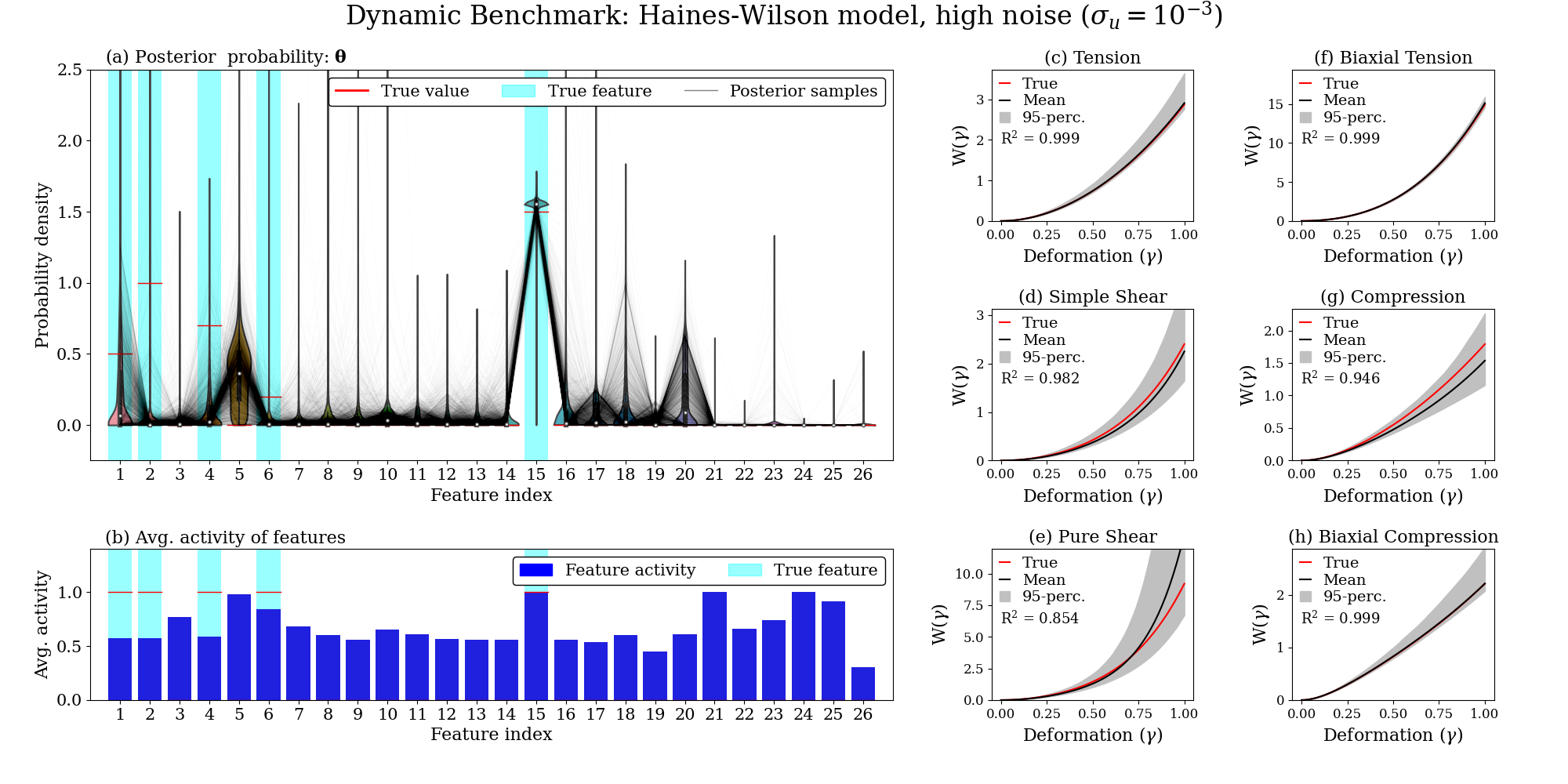}
	\end{subfigure}
    \caption{\customCaptionDynamic{Haines-Wilson}{HW}}\label{fig:HW-dyn}
\end{figure}

\begin{figure}[ht]
	\centering
	\begin{subfigure}[ht]{\textwidth}
	    \centering
	    \includegraphics[width=\textwidth]{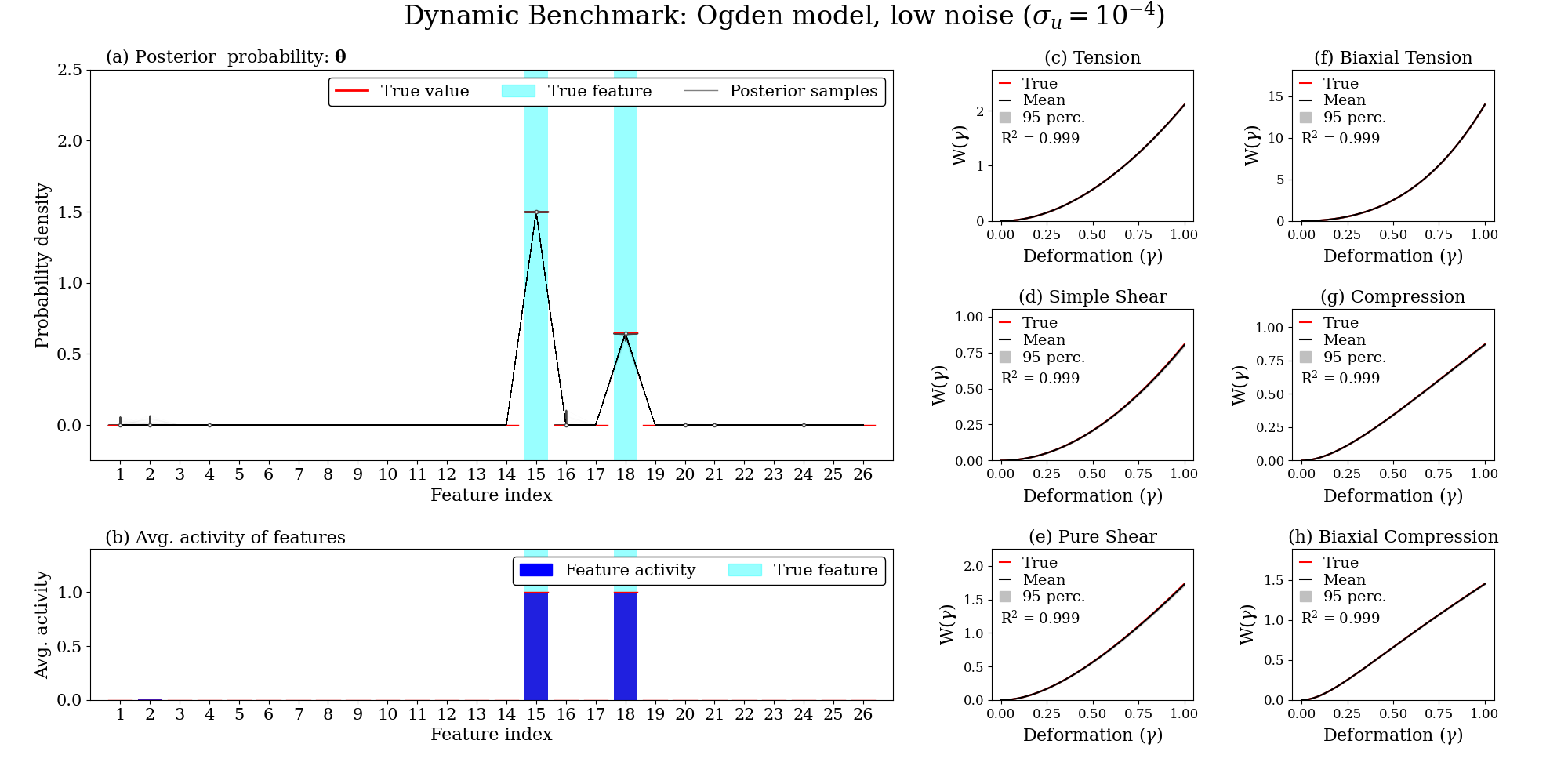}
	\end{subfigure}
	\hrule \vskip 15pt
	\begin{subfigure}[ht]{\textwidth}
	    \centering
	    \includegraphics[width=\textwidth]{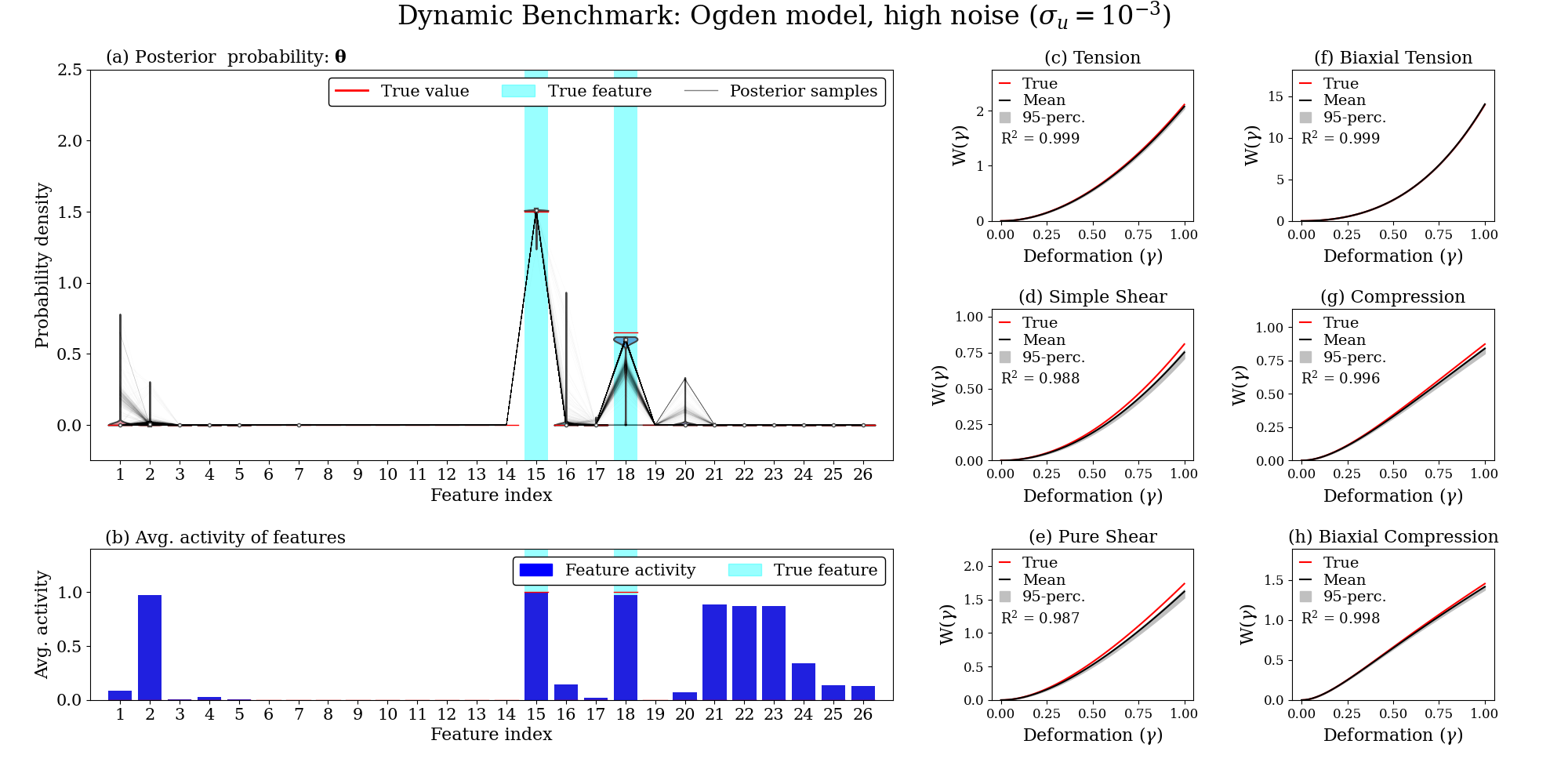}
	\end{subfigure}
    \caption{\customCaptionDynamic{Ogden}{OG}}\label{fig:OG-dyn}
\end{figure}

\begin{figure}[ht]
	\centering
	\begin{subfigure}[ht]{\textwidth}
	    \centering
	    \includegraphics[width=\textwidth]{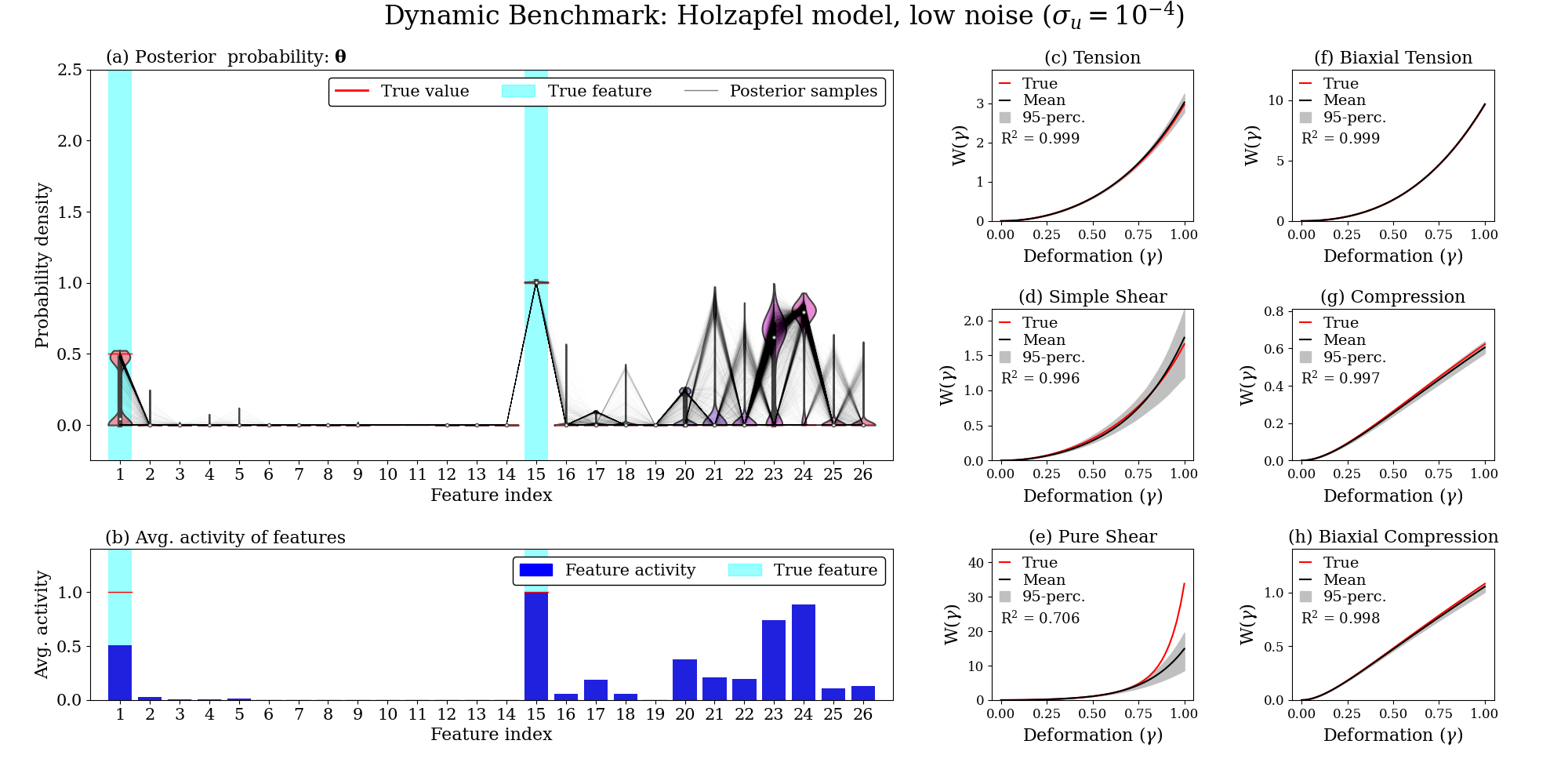}
	\end{subfigure}
	\hrule \vskip 15pt
	\begin{subfigure}[ht]{\textwidth}
	    \centering
	    \includegraphics[width=\textwidth]{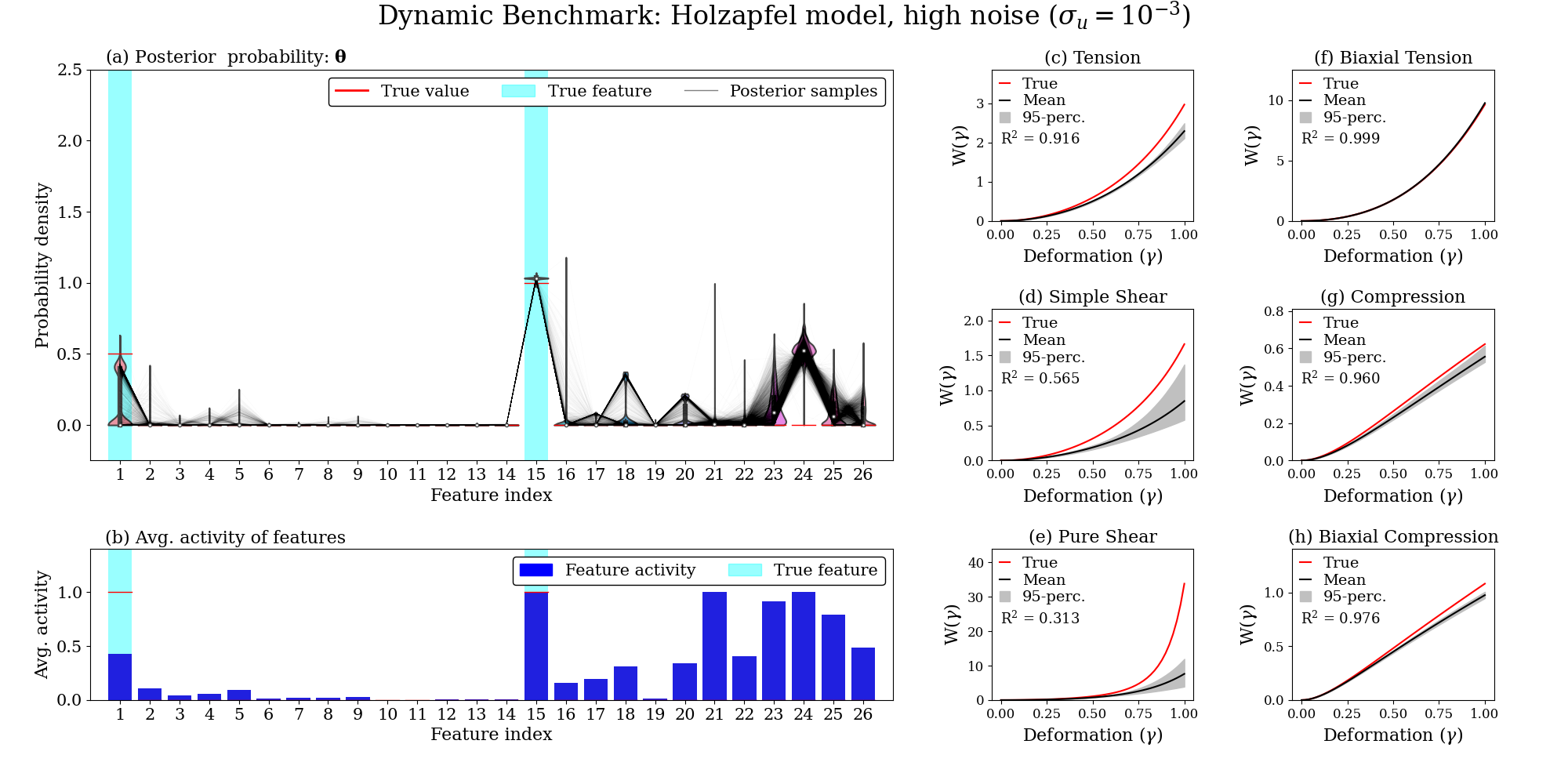}
	\end{subfigure}
    \caption{\customCaptionDynamic{anisotropic Holzapfel}{HZ}}\label{fig:HZ-dyn}
\end{figure}

The model discovery results with the dynamic data (which include the inertial forces) are summarized in Figures~\ref{fig:HW-dyn}-\ref{fig:HZ-dyn}. For the sake of brevity, a representative selection of Haines-Wilson \eqref{eq:HW}, Ogden \eqref{eq:OG}, and (anisotropic) Holzapfel \eqref{eq:HZ} benchmarks are presented. Similar observations with respect to accuracy, uncertainty, parsimony, multi-modality, and generalization can be made as in the quasi-static case.

\section{Conclusion}
\label{sec:conclusion}

We developed {Bayesian-EUCLID} -- a Bayesian framework for discovering interpretable and parsimonious hyperelastic constitutive models {with quantifiable uncertainties} in an unsupervised setting, i.e., without using any stress data and using only realistically obtainable displacement fields and global reaction force data. As opposed to calibrating an \textit{a priori} assumed parametric model, we use a large library of interpretable features inspired from several {physics-based} as well as phenomenological constitutive models{, which leverage domain knowledge accumulated} over the past decades. To {ensure parsimony and circumvent the lack} of stress labels, the hierarchical Bayesian {learning} approach {adopts} a {sparsity-inducing} spike-slab prior and a physics-constrained unsupervised likelihood based on conservation of linear momentum in the weak form. The efficacy of the Bayesian framework is tested on several benchmarks based on isotropic and anisotropic material {models} under quasi-static/dynamic loading, wherein the data is generated artificially with noise levels representative of contemporary DIC setups. The discovered constitutive models -- obtained as multi-modal posterior probability distributions --  accurately surrogate the true constitutive response with high confidence. Aleatoric uncertainties are automatically accounted for by hierarchically placing hyperpriors on the noise-related variables in the Bayesian model. The discovered models show good generalization under epistemic uncertainties (i.e., when {the} true features are unknown \textit{a priori} {and thus missing from the adopted feature library}) and automatically satisfy physical constraints via specially chosen model priors and features. {The interpretability of the approach enabled separately identifying the volumetric, deviatoric and direction-dependent (anisotropic) behavior of the material from a single experiment.}  {Future} work will include the extension to inelasticity as well as experimental validation, \RR{which will also entail adapting the EUCLID framework to work with a plane-stress assumption}.

\section*{Acknowledgments}

MF and LDL would like to acknowledge funding by SNF through grant N. $200021\_204316$ ``Unsupervised data-driven discovery of material laws''.

\section*{Declaration of competing interest}
The authors declare that they have no known competing financial interests or personal relationships that could have appeared to influence the work reported in this paper.

\section*{Code availability}
The codes developed in current study will be freely open-sourced at the time of publication.

\section*{Data availability}
The data generated in current study will be made freely available {at} the time of publication.

\bibliographystyle{elsarticle-harv}
\bibliography{references}

\appendix

\section{Protocols for data generation and benchmarks}\label{sec:protocols}

Table~\ref{tab:parameters} lists all the parameters used for the data generation and Bayesian learning. A consistent system of units for lengths, time and mass are used -- with each normalized based on the specimen side length, rate of loading, and material density. To generate realistic data artificially using FEM, we use a high-resolution mesh with $63,601$ nodes -- the same one used by \citet{flaschel_unsupervised_2021}. The noisy displacement data are also denoised using KRR following the same protocols. However, to demonstrate data efficiency of the proposed method, we project the denoised displacement field onto a coarser mesh with only $n_n=1,441$ nodes (each with two degrees of freedom). In the dynamic case, to avoid the computational expense of running a fine mesh for {a very large} number time steps, we use the coarser mesh with $n_n=1,441$ nodes to generate the data. For further data efficiency, we randomly sub-sample $n_\text{free}=100$ free degrees of freedom per snapshot for the Bayesian likelihood computation in both the quasi-static and dynamic cases (see discussion in Section~\ref{sec:likelihood}).

The parameter $\lambda_r$ is set to ensure similar importance to force balance at free and fixed degrees of freedom in \eqref{eqn:finaltheta}. For consistency, we choose $\lambda_r$ to be equivalent  to that used by \citet{flaschel_unsupervised_2021} (with the latter having a different definition of $\lambda_r$).

\begin{table}[]
\centering
\caption{Summary of parameters used for data generation and benchmarks of Bayesian-EUCLID.} 
\label{tab:parameters}
\begin{tabular}{lcc}
\hline
\multicolumn{1}{l}{\textbf{Parameter}} & \textbf{Notation} &\textbf{Value} \\ \hline
\textit{Quasi-static data generation:}\\
$\quad$ Number of nodes in mesh for FEM-based data generation  & - & $63,601$  \\
$\quad$ Number of nodes in data available for model discovery  & $n_n$ & $1,441$\\
$\quad$ Number of reaction force constraints & $n_{\beta}$   & $4$   \\ 
$\quad$ Number of data snapshots    & $n_{t}$     & $5$  \\
$\quad$ Number of time steps & $L_\text{static}$   & $5$  \\
$\quad$ Loading parameter    & $\varphi$  & $\{0.1 \times l: l=1,\dots,L_\text{static}\}$ \\ \hline
\textit{Dynamic data generation:}\\
$\quad$ Number of nodes in mesh for FEM-based data generation  & - & $1,441$  \\
$\quad$ Number of nodes in data available for model discovery  & $n_n$ & $1,441$\\
$\quad$ Number of reaction force constraints & $n_{\beta}$   & $4$   \\ 
$\quad$ Material density & $\rho$   & {$1$}   \\ 
$\quad$ Number of data snapshots & $n_{t}$     & $5$  \\
$\quad$ Number of time steps & $L_\text{dynamic}$ & 50000  \\
$\quad$ Loading rate for dynamic data  & $\dot\varphi$ & $0.1$ \\
$\quad$ Step size for explicit time integration in FEM & $\delta t$ & 0.0002 \\ \hline
\textit{Feature library:}\\
$\quad$ Number of features in library $\boldsymbol{Q}$ (see \eqref{eq:library} and Table~\ref{tab:features})  & $n_f$  & $26$  \\
$\quad$ Highest degree among Mooney-Rivlin polynomial features  & $N_\text{MR}$ & 4   \\
$\quad$ Number of volumetric features      & $N_\text{vol}$ & $1$   \\
$\quad$ Number of Ogden features       & $N_\text{Ogden}$    & $3$   \\
$\quad$ Highest degree among anisotropic polynomial features & $N_\text{aniso}$  & 4   \\ \hline
\textit{Bayesian learning:}\\
$\quad$ Number of free degrees of freedom randomly sampled per snapshot & $n_\text{free}$    & $100$\\
$\quad$ Weight parameter for reaction force balance   & $\lambda_r$    & $10$  \\
$\quad$ Hyperparameters for random variable $\nu_s$ & $(a_{\nu},b_{\nu})$ & $(0.5,0.5)$      \\
$\quad$ Hyperparameters for random variable $p_0$  & $(a_p,b_p)$    & $(0.1,5.0)$      \\
$\quad$ Hyperparameters for random variable $\sigma^2$     & $(a_{\sigma},b_{\sigma})$ & $(1.0,1.0)$\\
$\quad$ Number of burn-in samples per MCMC chain  & $N_\text{burn}$ & 250\\
$\quad$ Length of each MCMC chain after discarding the burn-in samples & $N_\text{G}$ & 750\\
$\quad$ Number of independent MCMC chains & $N_\text{chains}$ & 4
\end{tabular}
\end{table}

\section{{Pseudo-code for posterior distribution sampling}}
\label{sec:AppPSEUD}

A pseudo-code for  MCMC sampling of the posterior probability distribution \eqref{eq:posterior} is presented in Algorithm~\ref{alg:MCMC}.

\begin{algorithm}[t]
\caption{ MCMC sampling of the posterior probability distribution \eqref{eq:posterior}}\label{alg:MCMC}
{\begin{algorithmic}[1]
\State \textbf{Input:} observed displacement, acceleration and reaction force data (see Section~\ref{sec:available_data})
\For{$t=1,\dots,n_t$}
\State Randomly sample (without replacement) $n_\text{free}$ \textit{free} degrees of freedom from the $t^\text{th}$ data snapshot
\State Assemble $\bfA^{\text{free},t}\in\Rset^{n_\text{free}\times n_f} $ and $\bfb^{\text{free},t}\in\Rset^{n_\text{free}}$ (see Section~\ref{sec:likelihood}) at the above degrees of freedom
\State Assemble $\bfA^{\text{fix},t}\in\Rset^{n_\beta\times n_f} $ and $\bfb^{\text{fix},t}\in\Rset^{n_\beta}$ (see Section~\ref{sec:likelihood}) for all the reaction forces
\EndFor
\State Assemble $\bfA$ and $\bfb$ as indicated in \eqref{eqn:finaltheta}
\For{$k=1,\dots,N_\text{chains}$} \Comment{creating multiple chains.}
    \State Initialize an empty chain of size $(N_\text{burn}+N_G)$: $\calC_k$
    \State Initialize: $\theta_1\sim \calU(0.95,1.05)$, \ $\bftheta_{-1}=\bf{0}$ \Comment{$\calU$ denotes uniform distribution}
    \State Initialize: $\sigma^2\sim \calU(0.95,1.05)$
    \State Initialize: $\nu_s\sim \calU(0.95,1.05)$
    \State Initialize: $p_0\sim \calU(0.095,0.105)$
    \State Initialize: $z_1\sim \text{Bern}(0.5)$, $\bfz_{-1}=\bf{0}$ \Comment{$\text{Bern}(0.5)$ denotes Bernoulli trial with equally likely outcomes}
    \For{$q=1,\dots,(N_\text{burn}+N_G)$} \Comment{beginning Gibbs sampling}
        \State $\bftheta \leftarrow \bftheta$ sampled   using \eqref{eq:cond_0}
        \State $\sigma^2 \leftarrow \sigma^2$ sampled   using \eqref{eq:cond_1}
        \State$\nu_s \leftarrow \nu_s$ sampled   using \eqref{eq:cond_2}
        \State $p_0 \leftarrow p_0$ sampled   using \eqref{eq:cond_3}
        \State $\text{order}\leftarrow\text{Perm}\left([1,\dots,n_f]\right)$ \Comment{$\text{Perm}(\cdot)$ denote random permutation of an array}
        \For{$j=1,\dots,n_f$}
            \State $i \leftarrow \text{order}[j]$
            \State $z_i \leftarrow z_i$ sampled using \eqref{eq:cond_4}
        \EndFor
        \State Record updated states in the chain: $\calC_k[q]\leftarrow \{\bftheta,\sigma^{2},\nu_s,p_0,\bfz\}$ 
    \EndFor
    \State Discard first $N_\text{burn}$ samples of the chain: $\calC_k \leftarrow \calC_k[(N_\text{burn}+1):(N_\text{burn}+N_G)]$
\EndFor
\State \textbf{Return: } Concatenation of all chains, $\calC_1\frown\calC_2\frown\dots\frown\calC_{N_\text{chains}}$\Comment{$\frown$ denotes concatenation of sequences}
\end{algorithmic}
}
\end{algorithm}

\section{Performance under uncertainties related to anisotropy and high displacement noise.}
\label{sec:appEPIS}

\begin{figure}
	\centering
    \includegraphics[width=\textwidth]{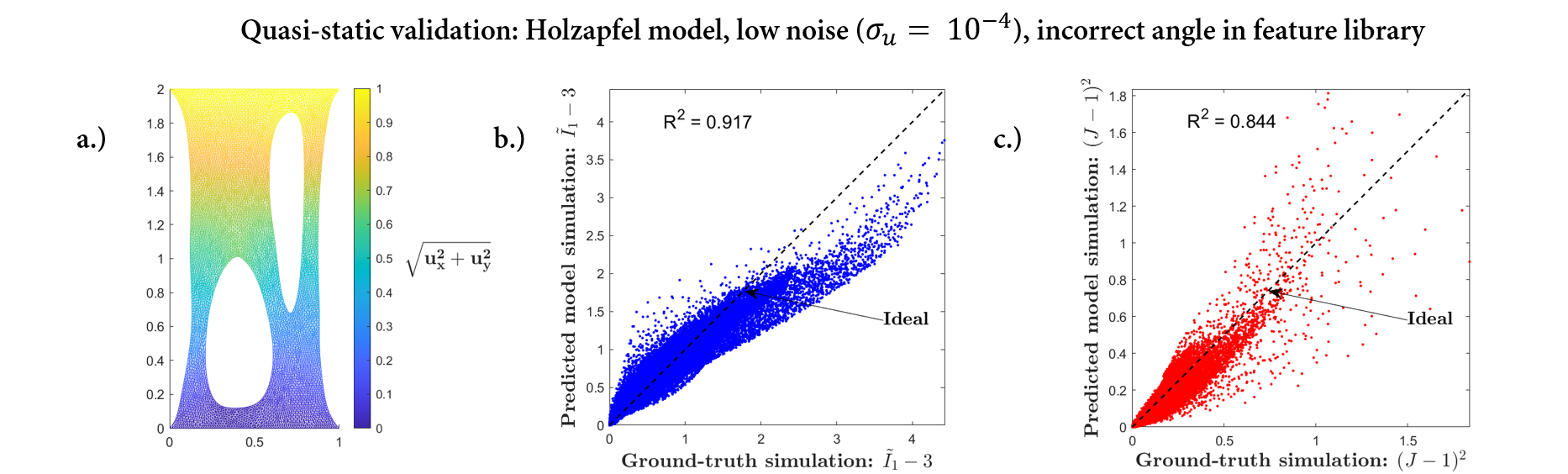}
    \caption{\RR{
    Quasi-static validation for the hidden Holzapfel benchmark model \eqref{eq:HZ} -- comparing simulation results from the ground-truth model and the mean of the energy density models predicted using the low-noise ($\sigma_u = 10^{-4}$) displacement data and assuming incorrect anisotropy directions $\alpha_1,\alpha_2=\pm45^{\circ}$ in the feature library (true directions are $\alpha_1,\alpha_2=\pm30^{\circ}$). (a) Deformed geometry at $\varphi=1$ obtained using the mean of the predicted models. (b) Predicted vs.~true strain invariant $(\Tilde{I}_1-3)$ across all quadrature points and loadsteps. (c) Predicted vs.~true strain invariant $(J-1)^2$ across all quadrature points and loadsteps.
    }}
    \label{fig:ValidLN_ICF}
    \vskip 10pt
	\centering
    \includegraphics[width=\textwidth]{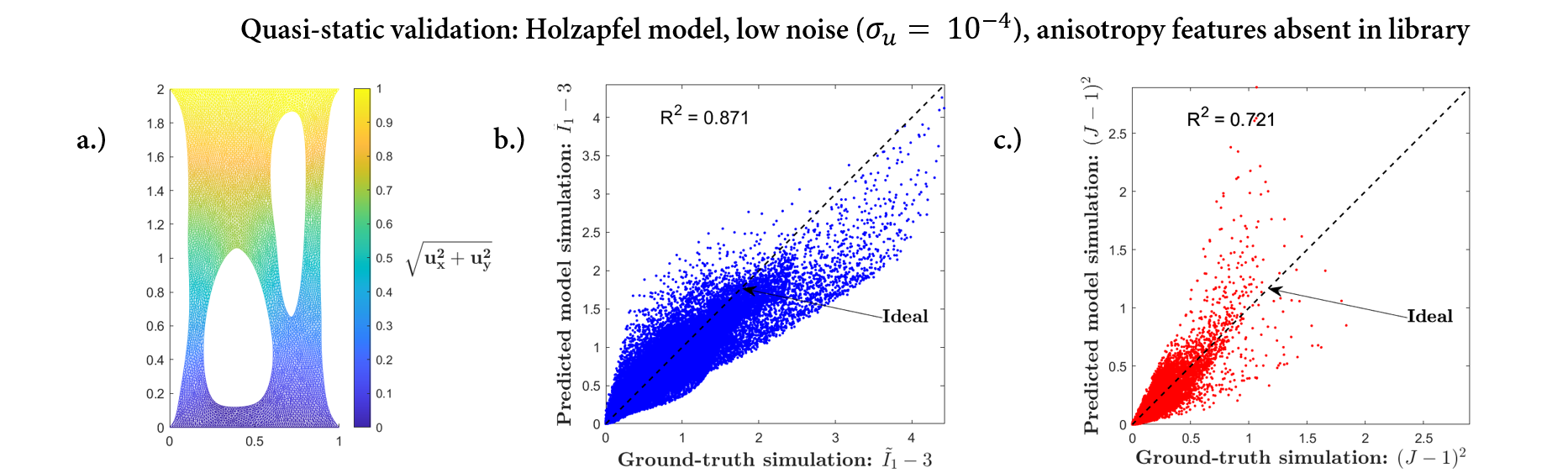}
    \caption{\RR{
    Quasi-static validation for the hidden Holzapfel benchmark model \eqref{eq:HZ} -- comparing simulation results from the ground-truth model and the mean of the energy density models predicted using the low-noise ($\sigma_u = 10^{-4}$) displacement data and anisotropic features excluded from the feature library. (a) Deformed geometry at $\varphi=1$ obtained using the mean of the predicted models. (b) Predicted vs.~true strain invariant $(\Tilde{I}_1-3)$ across all quadrature points and loadsteps. (c) Predicted vs.~true strain invariant $(J-1)^2$ across all quadrature points and loadsteps.
    }}
    \label{fig:ValidLN_NF}
    \vskip 10pt
	\centering
    \includegraphics[width=\textwidth]{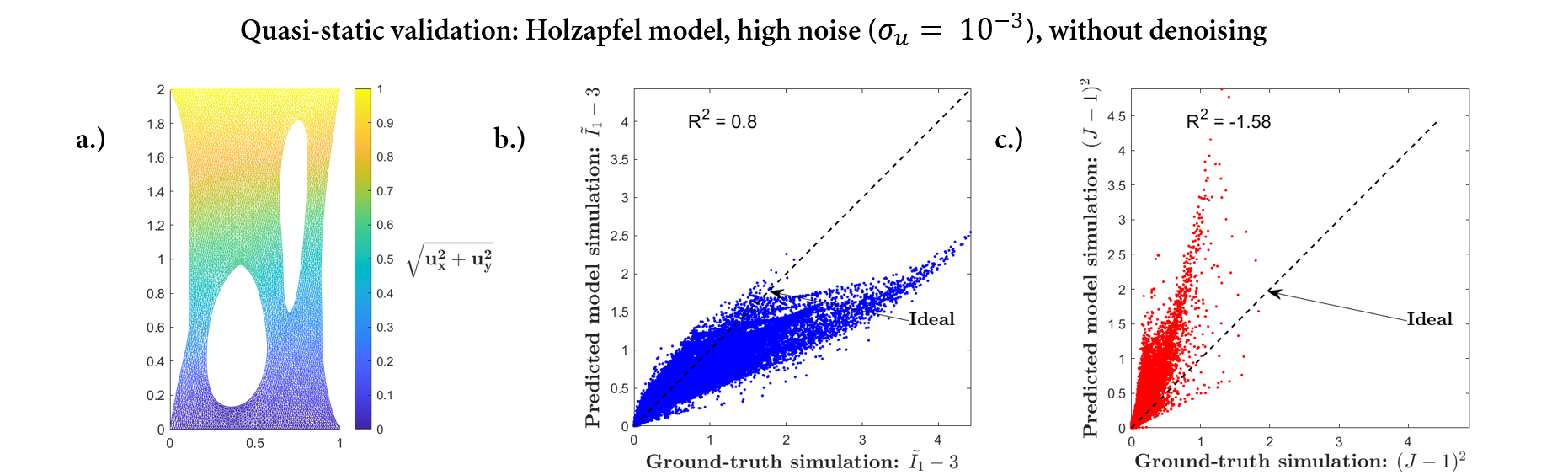}
    \caption{\RR{
    Quasi-static validation for the hidden Holzapfel benchmark model \eqref{eq:HZ} -- comparing simulation results from the ground-truth model and the mean of the energy density models predicted using the high-noise ($\sigma_u = 10^{-3}$) displacement data without any denoising. (a) Deformed geometry at $\varphi=1$ obtained using the mean of the predicted models. (b) Predicted vs.~true strain invariant $(\Tilde{I}_1-3)$ across all quadrature points and loadsteps. (c) Predicted vs.~true strain invariant $(J-1)^2$ across all quadrature points and loadsteps.
    }}
    \label{fig:ValidHNND}
\end{figure}

\RR{This section deals with evaluating the performance of the Bayesian-EUCLID framework under epistemic uncertainties related to anisotropy, i.e, due to selection of an incorrect/deficient library, and under severe displacement noise. For this purpose, we perform the FEM-based validation simulations on the hidden Holzapfel benchmark model specified by (\ref{eq:HZ}) with fibers along $\alpha_1,\alpha_2=\pm30^{\circ}$ directions. The results of the simulations employing the mean of predicted models are compared to those with the ground-truth Holzapfel model. Other details regarding the simulations are discussed in Section \ref{sec:aleatoricunc}. Figures \ref{fig:ValidLN_ICF}, \ref{fig:ValidLN_NF}, and \ref{fig:ValidHNND} present the validation results in the form of prediction accuracy of strain invariants across the specimen using: 
\begin{enumerate}[(i)]
\item low-noise ($\sigma_u = 10^{-4}$) displacement data with incorrectly assumed anisotropy directions:  $\alpha_1,\alpha_2=\pm45^{\circ}$ (true directions are $\alpha_1,\alpha_2=\pm30^{\circ}$) in the feature library \eqref{eq:library},
\item low-noise ($\sigma_u = 10^{-4}$) displacement data with anisotropic features not included in the feature library \eqref{eq:library}, and
\item high-noise ($\sigma_u = 10^{-3}$) displacement data without denoising.
\end{enumerate}
The $R^2$ scores shown in Figures \ref{fig:ValidLN_ICF} and \ref{fig:ValidLN_NF}, being considerably less than 1, indicate that these predicted models perform poorly when compared to the model predicted with correctly assumed anisotropy features (see Figure \ref{fig:ValidLN}).} \RR{Although not attempted in this work, it is therefore possible to iterate over different assumed angles (directions) to discover the in-plane material's anisotropy direction by maximizing the $R^2$ score for the plots.} \RR{Comparing Figures \ref{fig:ValidHN} and \ref{fig:ValidHNND} indicates that the displacement noise levels of $\sigma_u=10^{-3}$ (without denoising) are enough to induce spurious model predictions and that denoising the displacement data prior to learning the material model notably improves the prediction accuracy.}

\section{Correlation between the physics-constrained likelihood variance $\sigma^2$ and displacement noise variance $\sigma_u^2$}
\label{sec:appSIGS}

\begin{figure}[ht]
	\centering
	 \includegraphics[width=\textwidth]{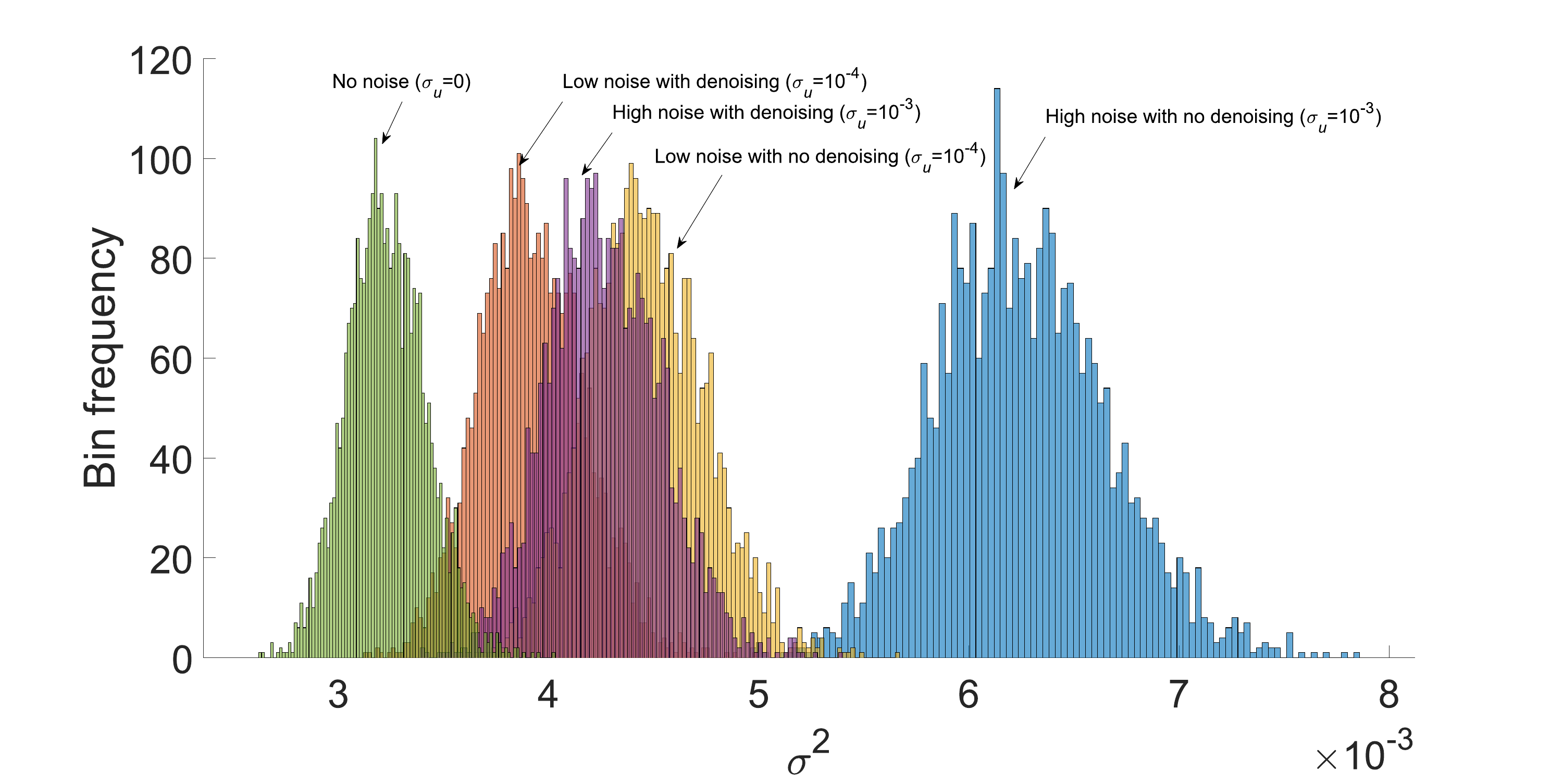}
    \caption{Distribution of the variance $\sigma^2$ for the Ogden benchmark \eqref{eq:OG} using quasi-static data with varying levels of $\sigma_u$ and with/without denoising of displacement data.}\label{fig:sig2}
\end{figure}

\figurename\ref{fig:sig2} shows the distribution of the variance $\sigma^2$ in the MCMC chain for  the Ogden model (\ref{eq:OG}) benchmark with quasi-static data. To demonstrate the correlation between $\sigma$ and $\sigma_u$, the following cases are considered:
\begin{enumerate}[(i)]
    \item $\sigma_u=0$,
    \item $\sigma_u=10^{-4}$ with denoising,
    \item $\sigma_u=10^{-3}$ with denoising,
    \item $\sigma_u=10^{-4}$ without denoising,
    \item $\sigma_u=10^{-3}$ without denoising.
\end{enumerate}
Both the average value and spread of $\sigma^2$ are the smallest for the noiseless data and increase with $\sigma_u$. Additionally, $\sigma^2$ reduces significantly with the denoising of the displacement data, \RR{which also corroborates the observations regarding the effects of denoising presented in \ref{sec:appEPIS}.}

\end{document}